\documentclass[11pt,a4paper]{oxarticle}
\pdfoutput=1

\usepackage{titlesec}
\titlespacing*{\subsubsection}
{0pt}{5pt plus 1pt minus 1pt}{5pt plus 1pt}

\usepackage[usenames, dvipsnames]{xcolor}
\usepackage{graphicx, float, array, xspace, amscd, amsthm, amssymb, latexsym, longtable, bbm, fancyhdr,slashed,pifont}
\usepackage[letterpaper, body={6.5in, 9.25in}, top=1.0in, left=1.5in]{geometry}
\usepackage[bbgreekl]{mathbbol}
\usepackage[utf8]{inputenc}
\usepackage[sort&compress, comma, square, numbers]{natbib}
\usepackage[bottom]{footmisc}
\usepackage{subcaption}
\usepackage{bm}
\usepackage{mathtools}
\usepackage{tikz}
\usepackage{tikz-cd}
\usepackage[pdfpagelabels=false]{hyperref}
\hypersetup{
    colorlinks,
    linkcolor={blue!50!black},
    citecolor={blue!50!black},
    urlcolor={blue!50!black}
}

\numberwithin{equation}{section}

\usepackage{tabularx, caption, boldline}
\usepackage{cellspace}
\DeclareSymbolFontAlphabet{\mathbb}{AMSb}
\DeclareSymbolFontAlphabet{\mathbbl}{bbold}

\let\SS=\S 

\newcommand{\g}{\gamma}\newcommand{\G}{\Gamma}

\newcommand{\z}{\zeta}

\newcommand{\Th}{\Theta}

\renewcommand{\l}{\lambda}

\newcommand{\n}{\nu}
\newcommand{\X}{\Xi}

\newcommand{\p}{\pi}\renewcommand{\P}{\Pi}\newcommand{\vp}{\varpi}
\renewcommand{\r}{\rho}
\newcommand{\s}{\sigma}\renewcommand{\S}{\Sigma}

\newcommand{\U}{\Upsilon}
\newcommand{\vph}{\varphi}

\DeclareFontFamily{OT1}{pzc}{}
\DeclareFontShape{OT1}{pzc}{m}{it}{<-> s * [1.200] pzcmi7t}{}
\DeclareMathAlphabet{\mathpzc}{OT1}{pzc}{m}{it}

\newcommand{\cA}{\mathcal{A}}
\newcommand{\cB}{\mathcal{B}}
\newcommand{\cC}{\mathcal{C}}
\newcommand{\cD}{\mathcal{D}}
\newcommand{\cE}{\mathcal{E}}
\newcommand{\cF}{\mathcal{F}}

\newcommand{\cH}{\mathcal{H}}

\newcommand{\cL}{\mathcal{L}}
\newcommand{\cM}{\mathcal{M}}
\newcommand{\cN}{\mathcal{N}}
\newcommand{\cO}{\mathcal{O}}

\newcommand{\cU}{\mathcal{U}}

\newcommand{\cW}{\mathcal{W}}

\newcommand{\cY}{\mathcal{Y}}

\DeclareFontFamily{U}{bbold}{}
\DeclareFontShape{U}{bbold}{m}{n}
{  <-5.5> s*[1.05] bbold5
	<5.5-6.5> s*[1.05] bbold6
	<6.5-7.5> s*[1.05] bbold7
	<7.5-8.5> s*[1.05] bbold8
	<8.5-9.5> s*[1.05] bbold9
	<9.5-11.5> s*[1.05] bbold10
	<11.5-16> s*[1.05] bbold12
	<16-> s*[1.05] bbold17
}{}

\newcommand{\IC}{\mathbbl{C}}

\newcommand{\IF}{\mathbbl{F}}

\newcommand{\II}{\mathbbl{I}}

\newcommand{\IK}{\mathbbl{K}}

\newcommand{\IP}{\mathbbl{P}}
\newcommand{\IQ}{\mathbbl{Q}}
\newcommand{\IR}{\mathbbl{R}}

\newcommand{\IT}{\mathbbl{T}}

\newcommand{\IY}{\mathbbl{Y}}
\newcommand{\IZ}{\mathbbl{Z}}

\newcommand{\Izero}{\mathbbl{0}}

\newcommand{\mtB}{\text{B}}

\newcommand{\mtE}{\text{E}}
\newcommand{\mtF}{\text{F}}
\newcommand{\mtG}{\text{G}}

\newcommand{\mtI}{\text{I}}

\newcommand{\mtU}{\text{U}}
\newcommand{\mtV}{\text{V}}
\newcommand{\mtW}{\text{W}}

\newcommand{\fr}{\mathfrak{r}}

\newcommand{\ft}{\mathfrak{t}}

\newcommand{\1}{\mathbf{1}}

\font\eightrm=cmr8 at 8pt
\font\csc=cmcsc10

\newcommand{\beq}{\begin{equation}}
\newcommand{\eeq}{\end{equation}}
\newcommand{\beqnn}{\begin{equation*}}
\newcommand{\eeqnn}{\end{equation*}}
\newcommand{\bea}{\begin{eqnarray}}
\newcommand{\eea}{\end{eqnarray}}
\newcommand{\bean}{\begin{eqnarray*}}
	\newcommand{\eean}{\end{eqnarray*}}

\newcommand{\cicy}[2]{\begin{matrix} #1\end{matrix}\!\left[\begin{matrix}#2 \end{matrix}\right]}

\newcommand{\defineas}{\buildrel\rm def\over =}

\newcommand{\place}[3]{\vbox to0pt{\kern-\parskip\kern-7pt
		\kern-#2in\hbox{\kern#1in #3}
		\vss}\nointerlineskip}
\newcommand{\capt}[3]{\parbox{#1}{\renewcommand{\baselinestretch}{1.0}
		\caption{\label{#2}\small\it #3}}}

\newcommand{\smallfrac}[2]{\frac{\scriptstyle #1}{\scriptstyle #2}}

\newcommand{\+}{\phantom{-}}
\renewcommand{\=}{\;=\;}

\newcommand{\tref}[1]{table~\ref{#1}}
\newcommand{\sref}[1]{\SS\ref{#1}}
\newcommand{\cref}[1]{Chapter \ref{#1}}

\DeclareFontFamily{U}{wncy}{}
\DeclareFontShape{U}{wncy}{m}{n}{<->wncyr10}{}
\DeclareSymbolFont{mcy}{U}{wncy}{m}{n}
\DeclareMathSymbol{\sha}{\mathord}{mcy}{"58}

\newcommand{\Tr}{\text{Tr}}

\newcommand{\tr}{\,\text{Tr}\,}
\newcommand{\Teich}{\text{Teich}}
\newcommand{\Frob}{\text{Frob}}

\newcommand{\HV}{\text{HV}}

\newcommand{\Fr}{\text{Fr}}

\newcommand{\ee}{\text{e}}
\newcommand{\me}{\text{e}}

\newcommand{\ii}{\text{i}}
\newcommand{\dd}{\text{d}}

\newcommand{\wt}[1]{\widetilde{#1}}
\newcommand{\wh}[1]{\widehat{#1}}
\makeatletter
\g@addto@macro\bfseries{\boldmath}

\def\blindfootnote{\xdef\@thefnmark{}\@footnotetext}
\makeatother

\newenvironment{smallarray}[1]
{\null\,\vcenter\bgroup\scriptsize
	\renewcommand{\arraystretch}{0.7}%
	\arraycolsep=.13885em
	\hbox\bgroup$\array{@{}#1@{}}}
{\endarray$\egroup\egroup\,\null}

\newcommand{\str}{\vrule height 12pt depth7pt width 0pt}

\newcommand{\strHV}{\vrule height 15pt depth7pt width 0pt}

\newcommand{\tablepreambleOctic}[1]{
	\vspace{-0.2cm}
	\begin{center}
		\begin{longtable}{| >{\footnotesize$~} c <{~$} | >{~\footnotesize} l <{~} |>{\centering\footnotesize $}p{3.7in}<{$}|}\hline
			\multicolumn{3}{|c|}{\vrule height 13pt depth8pt width 0pt \small $p=#1$}\tabularnewline[0.5pt] \hline
			\vrule height 13pt depth8pt width 0pt (\wh \varphi_1,\wh \varphi_2) & smooth/sing. & R_p(T) \tabularnewline[0.5pt] \hline\hline
			\endfirsthead
			\hline
			\multicolumn{3}{|l|}{\footnotesize\sl $p=#1$, continued}\tabularnewline[0.5pt] \hline
			(\varphi_1,\varphi_2) & smooth/sing. & R_p(T)\tabularnewline[0.5pt] \hline\hline
			\endhead
			\hline\hline 
			\multicolumn{3}{|r|}{{\footnotesize\sl Continued on the following page}}\tabularnewline[0.5pt] \hline
			\endfoot
			\hline
			\endlastfoot}
		
\newcommand{\tablepostamble}{\end{longtable}\end{center}}

\newcommand{\tablepreambleSplitQuintic}[1]{
	\vspace{-0.2cm}
	\begin{center}
		\begin{longtable}{| >{\footnotesize$~} c <{~$} | >{~\footnotesize} l <{~} |>{\centering\footnotesize $}p{3.7in}<{$}|}\hline
			\multicolumn{3}{|c|}{\vrule height 13pt depth8pt width 0pt \small $p=#1$}\tabularnewline[0.5pt] \hline
			\str (\varphi_1,\varphi_2) & smooth/sing. & R_p(T) \tabularnewline[0.5pt] \hline\hline
			\endfirsthead
			\hline
			\multicolumn{3}{|l|}{\footnotesize\sl $p=#1$, continued}\tabularnewline[0.5pt] \hline
			(\varphi_1,\varphi_2) & smooth/sing. & R_p(T)\tabularnewline[0.5pt] \hline\hline
			\endhead
			\hline\hline 
			\multicolumn{3}{|r|}{{\footnotesize\sl Continued on the following page}}\tabularnewline[0.5pt] \hline
			\endfoot
			\hline
			\endlastfoot}

\newcommand{\tablepreamble}[1]{
\vspace{-0.3cm}
\begin{center}
\begin{longtable}{| >{\footnotesize$~} c <{~$} | >{\footnotesize$~} c <{~$} | >{~\footnotesize} l <{~} 
| >{\centering\footnotesize$}p{1in}<{$} |>{\centering\footnotesize $}p{3.3in}<{$} |}
\hline
\strHV p & \vph & smooth/sing. & $singularity$ & R_p(T) \tabularnewline[0.5pt] \hline\hline
\endhead
\hline\hline 
\multicolumn{5}{|r|}{{\footnotesize\sl Continued on the following page}}\tabularnewline[0.5pt] \hline
\endfoot
\hline
\endlastfoot}

\usepackage{lmodern}
\usepackage{mmacells}
\mmaSet{uselistings=false}
\usepackage{tcolorbox}

\mmaSet{
	morefv={gobble=2},
	linklocaluri=mma/symbol/definition:#1,
	morecellgraphics={yoffset=1.9ex}
}

\newtcolorbox{defn}{colback=red!5!white,colframe=red!75!black}
\newtcolorbox{funcDefn}{colback=red!5!white,colframe=red!75!black}
\newtcolorbox{objDefn}{colback=blue!5!white,colframe=blue!75!black}
\newtcolorbox{optDefn}{colback=green!5!white,colframe=green!75!black}

\newtcolorbox[auto counter,number within=section]{funcDefnN}[2][]{%
	colback=red!5!white,colframe=red!75!black,fonttitle=\bfseries,title=Function ~\thetcbcounter: #2,#1}
\newtcolorbox[auto counter,number within=section]{optDefnN}[2][]{%
	colback=green!5!white,colframe=green!75!black,fonttitle=\bfseries,title=Option ~\thetcbcounter: #2,#1}
\newtcolorbox[auto counter,number within=section]{objDefnN}[2][]{%
	colback=blue!5!white,colframe=blue!75!black,fonttitle=\bfseries,title=Object ~\thetcbcounter: #2,#1}

\newcommand{\mmm}[1]{\mmaUnd{\(#1\)}}
\newcommand{\mmmd}[1]{\mmaDef{\(#1\)}}

\newcommand{\mmtd}[1]{\texttt{#1}}

\mmaDefineMathReplacement[→]{->}{\to}[2]

\begin{document}
	\proofmodefalse
	
	
	\thispagestyle{empty}  
    \begin{flushright} MITP--24--032\end{flushright}
	\begin{center}
		\null\vskip0.5in
		{\Huge Local Zeta Functions \\of Multiparameter Calabi--Yau Threefolds \\from the Picard--Fuchs Equations \\[12pt]}
\vskip1cm
{\csc Philip Candelas${}^{*1}$, Xenia de la Ossa${}^{*2}$, and Pyry Kuusela${}^{\dagger3}$.\\[2cm]}
\blindfootnote{\null\hskip-10pt \hfill $^1\,$candelas@maths.ox.ac.uk \hfill $^2\,$delaossa@maths.ox.ac.uk \hfill 
$^3\,$pyry.r.kuusela@gmail.com\hfill \kern10pt}
{${}^*$\it Mathematical Institute\\
University of Oxford\\
Andrew Wiles Building\\
Radcliffe Observatory Quarter\\
Oxford, OX2 6GG, UK\\}
\vskip20pt
{${}^\dagger$\it
PRISMA+ Cluster of Excellence \& Mainz Institute for Theoretical Physics\\
Johannes Gutenberg-Universit\"at Mainz\\
55099 Mainz, Germany\\
}
\vspace*{1cm}
{\bf Abstract}
\end{center}
\vskip-5pt
\begin{minipage}{\textwidth}
\baselineskip=15pt
\noindent 
The deformation approach of \cite{Candelas:2021tqt} for computing zeta functions of one-parameter Calabi--Yau threefolds is generalised to cover also multiparameter manifolds. Consideration of the multiparameter case requires the development of an improved formalism. This allows us, among other things, to make progress on some issues left open in previous work, such as the treatment of apparent and conifold singularities and changes of coordinates. We also discuss the efficient numerical computation of the zeta functions. As examples, we compute the zeta functions of the two-parameter mirror octic, a non-symmetric split of the quintic threefold also with two parameters, and the $S_5$ symmetric five-parameter Hulek--Verrill manifolds. These examples allow us to exhibit the several new types of geometries for which our methods make practical computations possible. They also act as consistency checks, as our results reproduce and extend those of \cite{Hulek2005,Kadir:2004zb}. To make the methods developed here more approachable, a \textit{Mathematica} package \mmtd{CY3Zeta} for computing the zeta functions of Calabi--Yau threefolds, which is attached to this paper, is presented.
\end{minipage}
\clearpage

\thispagestyle{empty}

{\baselineskip=15pt
	\tableofcontents} 


\newpage
\pagenumbering{arabic}

\section{Introduction}
\vskip-10pt
The local zeta function $\zeta_p(X,T)$ of an algebraic variety $X$ can be thought of as a generating function of the numbers of solutions over finite fields $\IF_{\! p^n}$ of the equations defining a manifold. Somewhat surprisingly, it turns out that these functions for Calabi--Yau threefolds have a direct connection to physics of string theory compactifications on these manifolds, for example encoding existence of supersymmetric flux vacua and rank-two attractor points \cite{Candelas:2019llw,Kachru:2020abh,Kachru:2020sio,Candelas:2023yrg,Bonisch:2022mgw,Bonisch:2022slo}. In addition, the zeta functions are widely-studied in connection with number theory, making them ideal and interesting objects to study in order to investigate connections between number theory and physics. As the more intricate aspects of the theory of zeta functions, such as their connection to modular forms, is still not fully developed, investigating concrete examples for a wide range of prime numbers $p$ to high numerical accuracy is important to obtain examples of interesting connections and to formulate and test conjectures concerning these. 

While the zeta functions can be computed using techniques such as direct counting of solutions to polynomial equations or evaluating Gauss sums, these techniques tend to be computationally complex, which greatly limits the range of primes $p$ for which the zeta functions can be evaluated with currently existing computational resources. This can, for instance, make it difficult to identify modular forms conjecturally related to supersymmetric flua vacua \cite{Kachru:2020sio}. For one-parameter Calabi--Yau manifolds, the state-of-the-art was improved in \cite{Candelas:2021tqt} where Candelas, de la Ossa, and van Straten presented a practical method for computing their zeta functions using series expansions for their periods near the large complex structure point. 

The aim of this paper is to generalise the methods of \cite{Candelas:2021tqt} to cover multiparameter manifolds. The main result of this paper, which we present in \sref{sect:generic_Frobenius_map}, is an efficient numerical method for computing the local zeta functions of multiparameter Calabi--Yau threefolds $X_{\bm \varphi}$. We do this, analogously to \cite{Candelas:2021tqt}, by finding a matrix $\mtU_p(\bm \varphi)$, which will determine the local zeta function $\zeta_p(X_{\bm \varphi},T)$ via the relation
\begin{align*} 
\zeta_p(X_{\bm \varphi},T) &\= \frac{R_p(X_{\bm \varphi},T)}{(1-T)(1-pT)^{h^{11}}(1-p^2T)^{h^{11}}(1-p^3T)}~,\\[-3pt]
\intertext{where} 
R_p(X_{\bm \varphi},T) &\= \det \left( \mtI - T \mtU_p(\bm \varphi) \right)~.\\
\end{align*}
\vskip-15pt
We are able to find a relatively simple expression for the matrix $\mtU_p(\bm \varphi)$, by developing further the formalism of \cite{Candelas:2023yrg}, where the periods of the Calabi--Yau manifold $X_{\bm \varphi}$ are expressed in terms of the generators of the homology algebra of its mirror manifold $\wt X_{\bm t}$ with a Kähler parameter $\bm t$ given by the mirror map. In particular, we are able to express the matrix $\mtU_p(\bm \varphi)$ in terms of a representation of the homology algebra and the periods of $X_{\bm \varphi}$. Additionally, we discuss several computational techniques which make evaluation of the matrices $\mtU_p(\bm \varphi)$ significantly faster, and subtleties that are not apparent in the case of threefolds, but are important for further generalisations \cite{Jockers:2023zzi}.

The approach we take here is computationally less intensive compared to several existing methods, such as evaluating Gauss sums, and thus allows us to compute the zeta functions to considerably higher values of the prime $p$ than has been previously possible. Additionally, the approach based on the Picard--Fuchs equations we develop here requires only simple geometric data: the number of complex structure parameters, triple intersection numbers, singular loci, and periods, making the method amenable to computer implementation.

In \sref{sect:Examples}, we present three examples of multiparameter manifolds whose zeta functions we have computed using the methods presented in this paper: the two-parameter family of mirror octic manifolds, a two-parameter split of the quintic threefold, and  the five-parameter family of mirror Hulek--Verrill manifolds. In addition to demonstrating the techniques developed in this paper, we use each of these examples to discuss a particular subtlety or a generalisation of these methods. The mirror octic example is used to discuss and demonstrate how the deformation method can be used to compute the local zeta function even for manifolds with conifold singularities, and how the choice of coordinates on the complex structure moduli space affects the computation. The split quintic computation demonstrates using different bases of the middle cohomology to deal with the \textit{apparent singularities} encountered in \cite{Candelas:2021tqt}. The case of Hulek--Verrill manifolds would in principle require making computations with $12 {\,\times} 12$ matrices whose components are five-parameter series. This is not computationally feasible on current hardware. However, we are able to develop techniques that allow us to consider various lines in the moduli space, and thus deal with series in one variable only.

These examples also work as highly-nontrivial consistency checks on our methods. The mirror octic has been studied previously in detail using different methods in \cite{Kadir:2004zb}, and the zeta functions of the Hulek--Verrill manifolds the `symmetric' line in the moduli space can be found using the direct point-counting methods \cite{Hulek2005}. We find complete agreement with the results of \cite{Kadir:2004zb} and \cite{Hulek2005}, as far as they overlap ours. The techniques developed in this paper allow extending these results to higher primes $p$. After discussing the examples, we include a very concise summary of the results and discuss the limitations of the presented methods as well as directions for future research.

In appendix \ref{app:p-adic_numbers}, we give a telegraphic review of the basic properties of the $p$-adic numbers that we utilise in the text. The appendix \ref{app:elliptic_curve} presents a computation of the zeta functions of the Legendre family of elliptic curves. This example helps to illustrate the deformation method we use without involving the technical details that are necessary for the more involved case of multiparameter Calabi--Yau threefolds. Some computational details are delegated to the appendices \ref{app:U(0)} and \ref{app:univariate_series}.

In the final appendix \ref{app:CY3Zeta}, we include documentation for a \textit{Mathematica} package, \texttt{CY3Zeta}, that provides user-friendly implementation of the algorithms discussed in this paper, with the aim of making the computation of local zeta functions more accessible. In particular, the package can be used to explicitly compute the polynomials $R_p(X_{\bm \varphi},T)$ determining the local zeta function, as well as the ancillary matrices, such as $\mtU_p(\bm \varphi)$ which are extensively used throughout the paper. As input only basic geometric data, such as the triple intersection numbers and the fundamental period, of the Calabi--Yau manifold is required.

Some of the material in this paper is adapted from one of the present authors' doctoral thesis \cite{Kuusela:2022hga}, and a very brief overview of the methods developed here has appeared before in a paper~\cite{Candelas:2023yrg} by the present authors and J. McGovern.

\subsection{Conventions and notation} \label{sect:conventions}
\vskip-10pt
Throughout the paper, we study families of Calabi--Yau threefolds $X_{\bm \varphi}$, with $m$ complex structure parameters $\bm \varphi = (\varphi_1,\dots,\varphi_m)$. We are interested in cases $X_{\bm \varphi}/\IQ$ when the manifolds are defined over rational numbers. That is, we require that the polynomials defining the manifold (at least locally) have coefficients in $\IQ$, or equivalently in $\IZ$ with compatible transition functions. Given the inclusions $\IF_{\! p} \hookrightarrow \IZ$, we can study the family of manifolds over these finite fields, and by considering field extensions, this can be extended to include manifolds defined over fields $\IF_{\! p^n}$. To avoid any confusion, where needed, we will denote the manifold $X_{\bm \varphi}$ defined over field $\IK$ by $X_{\bm \varphi}/\IK$.
 
We denote $m \times m$ matrices by symbols in blackboard bold font, and $m$-component vectors with symbols in bold font. We often also treat such vectors as multi-indices, and denote the sum of their components by $x_1 + \cdots + x_m = |\bm x|$. Unless otherwise stated we employ Einstein summation convention, with the indices $a,b,c,\dots$ from the beginning of the Latin alphabet taking values $0,\dots,m$, and the indices $i,j,k,\dots$ taking values $1,\dots,m$.

Some symbols that appear in multiple sections are collected, with their definitions, in~\tref{tab:notation}.
\vskip10pt
\begin{table}[H]
	\renewcommand{\arraystretch}{1.35}
	\centering
	\begin{tabularx}{0.99\textwidth}{|>{\hsize=.12\hsize\linewidth=\hsize}X|
			>{\hsize=0.79\hsize\linewidth=\hsize}X|>{\hsize=0.07\hsize\linewidth=\hsize}X|}
		\hline
		\textbf{Symbol} & \hfil \textbf{Definition/Description} & \hfil \textbf{Ref.}\\[3pt]
		\hline	\hline
		
		$\bm \varphi$
		& 
		The coordinates $(\varphi^{1},\dots,\varphi^{m})$ on the complex structure moduli space of a Calabi--Yau manifold $X_{\bm \varphi}$.
		&
		\sref{sect:conventions}
		\\[4pt] \hline

    	$\theta_i$
		& 
		The logarithmic derivative $\varphi^i \partial_{\varphi^i}$ (no sum implied) with respect to the complex structure modulus $\varphi^i$.
		&
        \sref{sect:Frobenius_map}
  	    \\[4pt] \hline

        $\vartheta^a, \vartheta_a$
		& 
		The linear combinations 0 to 3 of logarithmic derivatives defined by $ (\vartheta_0,\vartheta_{i}, \vartheta^{i},\vartheta^0) = \left(1,\theta_i, \wh Y^{jki} \theta_j \theta_k, \frac{1}{m} \wh Y^{ijk} \theta_i \theta_j \theta_k \right)$.
		&
        \eqref{eq:vartheta_definition}
  	    \\[4pt] \hline

  		$v_a, v^a$
		& 
		The basis vectors of the constant basis of $H^3(X_{\bm \varphi},\IC)$.
		&
        \eqref{eq:Omega_basis_expansion}
  	    \\[4pt] \hline

  		$\vartheta_a\Omega, \vartheta^a\Omega$
		& 
		The basis vectors of the derivative basis of $H^3(X_{\bm \varphi},\IC)$ corresponding to the logarithmic derivatives of the holomorphic $(3,0)$-form $\Omega$.
		&
        \eqref{eq:H^3_basis}
  	    \\[4pt] \hline              
		
		$\varpi$
		& 
		The period vector of $X_{\bm \varphi}$ in the Frobenius basis.
		&
		\eqref{eq:period_vector_Frob_basis_expansion}
		\\[4pt] \hline
		$\Pi$
		& 
		The period vector of $X_{\bm \varphi}$ expressed in the integral symplectic basis.
		&
		\eqref{eq:Pi_varpi_relation_general}
		\\[4pt] \hline

  		$\mtE(\bm \varphi)$
		& 
		The change-of-basis matrix from the constant basis to the derivative basis. Also known as the period matrix.
		&  
        \eqref{eq:E_matrix_definition}
  	    \\[4pt] \hline  

  		$\wt \mtE(\bm \varphi)$
		& 
		The logarithm-free period matrix defined by setting $\log \varphi^i = 0$ in $\mtE(\bm \varphi)$.
		&  
        \eqref{eq:Log-free_E_definition}
  	    \\[4pt] \hline        
		
		$\IF_{\! p^n}$
		& 
		The finite field with $p^n$ elements.
		&
        \sref{sect:Weil_conjectures}
		\\[4pt] \hline
		
		$\zeta_p(X_{\bm \varphi},T)$
		& 
		The local zeta function of a Calabi--Yau manifold $X_{\bm \varphi}$.
		&
		\eqref{eq:zeta_function_form}
		\\[4pt] \hline
		
		$R_p(X_{\bm \varphi},T)$
		& 
		The numerator of the zeta function $\zeta_p(X_{\bm \varphi},T)$.
		&
		\eqref{eq:zeta_function_form}
		\\[4pt] \hline
		
		$\mtU_p(\bm \varphi)$
		& 
		The matrix representing the action of the inverse Frobenius map $\Fr_p^{-1}$ on the middle cohomology.
		&
		\eqref{eq:R(T)_determinant}
		\\[4pt] \hline

  		$\alpha^i, \beta_i, \gamma$
		& 
		The prime-dependent coefficients appearing in the matrix $\mtU_p(\bm 0)$.
		& 
        \eqref{eq:U(0)_general}
  	    \\[4pt] \hline      

  		$S_n(\bm \varphi)$
		& 
		The denominator of the rational matrix $\mtU_p(\bm \varphi) \mod p^n$.
		&  
        \eqref{eq:U(varphi)_p-adic_expansion}
  	    \\[4pt] \hline    

  		$\IY_a$
		& 
		The $m \times m$ matrix whose components are given by the topological quantities $Y_{aij}$, $\left[\IY_a\right]_{ij} = Y_{aij}$
		&
        \eqref{eq:homology_algebra_rep}
  	    \\[4pt] \hline

  		$\epsilon_i, \mu_i, \eta$
		& 
		Matrices giving a representation of the (co-)homology algebra of the mirror manifold of $X_{\bm \varphi}$.
		&
        \eqref{eq:homology_algebra_rep}  
  	    \\[4pt] \hline

  		$\wh Y^{ijk}$
		& 
		The `inverse triple intersection numbers' that satisfy $Y_{ijk} \wh Y^{ijs} = \delta_k^s$.
		&
        \eqref{eq:Yhat_definition}
  	    \\[4pt] \hline   

  		$\II$
		& 
		The $m{\times}m$ unit matrix.
		&
        \sref{sect:conventions}
  	    \\[4pt] \hline 
  		$\Izero$
		& 
		The $m{\times}m$ zero matrix.
		&
        \sref{sect:conventions}
  	    \\[4pt] \hline 
  		$\bm 1$
		& 
		The $m$-component vector with all entries 1.
		&
        \sref{sect:conventions}
  	    \\[4pt] \hline        
  		$\bm 0$
		& 
		The $m$-component zero vector.
		&
        \sref{sect:conventions}
  	    \\[4pt] \hline  
  		$\bm \delta_i$
		& 
		The $m$-component vector with components $[\bm \delta_i]_j = \delta_{ij}$.
		&
        \eqref{eq:homology_algebra_rep}
  	    \\[4pt] \hline        
	\end{tabularx}
	\vskip10pt
	\capt{6.5in}{tab:notation}{Some quantities that are used throughout the paper with references to where they are first introduced. }
\end{table}
\newpage

\section{Review of Mirror Symmetry and Zeta Functions}
\vskip-10pt
We begin with a brief review of the salient aspects of the theory of local zeta functions and mirror symmetry in order to keep the paper self-contained, and to simultaneously introduce the notation. Most of the material appearing in this section is standard, although we have reformulated some of it in a language that is useful for discussing zeta functions of multiparameter threefolds. 

For more comprehensive introduction to the properties of zeta functions of Calabi--Yau threefolds, we refer the reader to \cite{Candelas:2021tqt,Candelas:2007mb} for an exposition aimed at physicists, or \cite{Koblitz:1609457} for a physicist-friendly mathematical treatment. The literature on mirror symmetry is extensive, but the aspects of mirror symmetry discussed here are presented in more detail for example in \cite{Hosono:1994ax,Hosono:1993qy,Candelas:1990pi,Candelas:1990rm} and references therein.
\subsection{The Local zeta function and Weil conjectures} \label{sect:Weil_conjectures}
\vskip-10pt
Let $X_{\bm \varphi}$ be a manifold that is defined as a zero locus of polynomials $P_i$ in some ambient space $\IP^n$, with the coefficients of $P_i$ rational. Such a manifold is said to be defined over $\IQ$, which is denoted by $X_{\bm \varphi}/\IQ$. It is then possible to consider solutions $P_i(x) = 0$ with $x \in \IQ^n \subset \IC^n$. The set of solutions is then denoted~$X_{\bm \varphi}(\IQ)$.

Given such a manifold, one can further clear the denominators of the polynomials $P_i$ to get polynomials with coefficients in $\IZ$. Using the natural projection $\IZ \to \IZ/p\IZ$, we can then consider the manifold to be defined over the finite field $\IF_{\! p} \simeq \IZ/p\IZ$. In practice, this amounts to studying the defining polynomials $P_i$ modulo $p$. As above, we say that the manifold is defined over $\IF_{\! p}$, denoting it $X_{\bm \varphi}/\IF_{\! p}$, and denote the finite set of solutions $P_i(x) \equiv 0 \mod p$ by $X_{\bm \varphi}(\IF_{\! p})$. One can similarly consider $X_{\bm \varphi}/\IF_{\! p^n}$ and the finite set $X_{\bm \varphi}(\IF_{\! p^n})$ for any finite field with $p^n$ elements.

A significant amount of interesting geometric information is in fact encoded in the sets $X_{\bm \varphi}(\IF_{\! p^n})$. We denote the number of $\IF_{\! p^n}$ points on a manifold $X_{\bm \varphi}/\IF_{\! p^n}$ by $N_{p^n}(X_{\bm \varphi})$, that is, $N_{p^n}(X_{\bm \varphi}) = \cN(X_{\bm \varphi}(\IF_{\! p^n}))$, where $\cN(A)$ denotes the number of elements of the set $A$. It turns out to be useful to define a generating function for these quantities.  The \textit{local zeta function} or the \textit{Hasse-Weil zeta function} of the manifold $X_{\bm \varphi}$ at the prime $p$ is defined as
\begin{align} \label{eq:Zeta_function_definition}
\zeta_p(X_{\bm \varphi}, T) \= \exp \left( \sum_{n=1}^\infty \frac{N_{p^n}(X_{\bm \varphi}) T^n}{n} \right).
\end{align}
The Weil conjectures, originally due to Weil \cite{Weil}, and later proved by Dwork \cite{Dwork}, Grothendieck~\cite{Grothendieck}, and Deligne \cite{Deligne1974,Deligne1980}, can be stated as:
\begin{enumerate}
	\item \textbf{Rationality}: $\zeta_p(X_{\bm \varphi},T)$ is a rational function of $T$ of the form
	\begin{align} \notag
	\zeta_p(X_{\bm \varphi},T) \= \frac{R^{(1)}_p(X_{\bm \varphi},T)R^{(3)}_p(X_{\bm \varphi},T)\cdots R^{(2d-1)}_p(X_{\bm \varphi},T)}{R^{(0)}_p(X_{\bm \varphi},T)R^{(2)}_p(X_{\bm \varphi},T)\cdots R^{(2d)}_p(X_{\bm \varphi},T)}~,
	\end{align}
	where $R^{(i)}_p(X_{\bm \varphi},T)$ is a polynomial in $T$ with integer coefficients. The degree of $R_p^{(i)}(X_{\bm \varphi},T)$ is given by the Betti number $b_i(X_{\bm \varphi})$ of the manifold $X_{\bm \varphi}$.
	\item \textbf{Functional equation}: $\zeta_p(X_{\bm \varphi},T)$ satisfies the functional equation
	\begin{align} \label{eq:Weil_conjectures_functional_equation}
	\zeta_p\left(X_{\bm \varphi},p^{-d} T^{-1}\right) \= \pm p^{\frac{d}{2}\chi(X_{\bm \varphi})} T^\chi \zeta_p(X_{\bm \varphi},T)~,
	\end{align}
	where $\chi(X_{\bm \varphi})$ is the Euler characteristic of $X_{\bm \varphi}$.
	\item \textbf{Riemann hypothesis}: The polynomials $R^{(i)}_p(X_{\bm \varphi},T)$ factorise over $\IC$ as 
	\begin{align} \notag
	R^{(i)}_p(X_{\bm \varphi},T) \= \prod_{j=1}^{b_i}\big(1-\lambda_{ij}(X_{\bm \varphi})T\big)~,
	\end{align}
	where the $\lambda_{ij}(X_{\bm \varphi})$ are algebraic integers of complex modulus $|\lambda_{ij}(X_{\bm \varphi})| = p^{i/2}$.
\end{enumerate}
When the Picard group of $X_{\bm \varphi}$ is generated by divisors that are defined over $\IF_{\! p}$, the polynomials corresponding to the cohomology groups $H^2(X_{\bm \varphi},\IC)$ and $H^4(X_{\bm \varphi},\IC)$ factorise into linear factors, thus being given by
\begin{align} \notag
R^{(2)}_p(X_{\bm \varphi},T) \= (1-pT)^{h^{11}}~, \qquad R^{(4)}_p(X_{\bm \varphi},T) \= (1-p^2T)^{h^{11}}~.
\end{align}
Therefore, in this case, the zeta function is completely determined by a single degree-$(2m{+}2)$ polynomial $R_p(X_{\bm \varphi},T) \defineas R^{(3)}_p(X_{\bm \varphi},T)$.
\begin{align} \label{eq:zeta_function_form}
\zeta_p(X_{\bm \varphi},T) = \frac{R_p(X_{\bm \varphi},T)}{(1-T)(1-pT)^{h^{11}}(1-p^2T)^{h^{11}}(1-p^3T)}.
\end{align}
This polynomial can be computed explicitly using the periods of the Calabi--Yau manifold. Roughly speaking, the reason for the relation to the periods is due to the fact that the polynomial $R_p(T,X_{\bm \varphi})$ can be related to the \textit{Frobenius map} acting on the third cohomology. 

Denote by $\Frob_p$ the Frobenius map that acts on the coordinates $\bm x$ of the ambient space $\IK^k$ by
\begin{align} \notag
\IK^k \to \IK^k \; : \;  \bm x = (x_1,\dots,x_k) \mapsto (x_1^p,\dots,x_k^p) \= \bm x^p~. 
\end{align}
Recall that Fermat's little theorem shows that $a^{p} = a \mod p$ for any $a \in \IF_{\! p}$, implying that the Frobenius map fixes any element of $\IF_{\! p}$. In fact, since the polynomial $x^p-x$ has at most $p$ roots in any field extension, the elements fixed by the Frobenius map are exactly those in $\IF_{\! p}$. 

If $X_{\bm \varphi}$ is a variety defined over $\IF_{\! p}$ and we study the solutions over the algebraic closure $\overline{\IF_{\! p}}$, $\Frob_p$ defines a self-map $\Frob_p: X_{\bm \varphi} \to X_{\bm \varphi}$. To see this, note that, if $X_{\bm \varphi}$ is defined as the vanishing locus of the polynomial $P$ which has coefficients in $\IF_{\! p}$, it follows that
\begin{align} \notag
P\big(x^{p}\big) \= P(x)^{p} \= 0 \mod p~.
\end{align}
Therefore the fixed points of $\Frob_p$ are exactly those counted in $N_{p}(X_{\bm \varphi})$ appearing in the definition of the zeta function.

It turns out that it is possible to define so-called \textit{$p$-adic cohomology theories} $H^k(X_{\bm \varphi},\IQ_p)$, such that one can pull back the Frobenius map to get an automorphism 
\begin{align} \label{eq:Frobenius_map_Fr_definition}
\Fr_p \defineas (\Frob_{p})_*: H^k(X_{\bm \varphi},\IQ_p) \to H^k(X_{\bm \varphi},\IQ_p)
\end{align}
The $H^k(X_{\bm \varphi},\IQ_p)$ are finite dimensional vector spaces over the field $\IQ_p$ of $p$-adic numbers (for a brief introduction to $p$-adic numbers, see appendix \ref{app:p-adic_numbers} and references therein). The Lefschetz fixed-point theorem can be applied for this cohomology theory, giving a simple relation between the point counts and the action of $\Fr_p$:
\begin{align} \notag
N_{p^n}(\bm \varphi) \= \sum_{m=0}^{6} (-1)^m \, \Tr \Big(\Fr_{p^n} \big| H^m(X_{\bm \varphi},\IQ_p) \Big)~.
\end{align}
From this formula it can be seen that the characteristic polynomial of the inverse Frobenius map acting on the middle cohomology $H^3(X_{\bm \varphi},\IQ_p)$ is exactly the polynomial $R_p(X_{\bm \varphi},T)$:
\begin{align} \label{eq:R(T)_determinant}
R_p(X_{\bm \varphi},T) \= \det \left(\mtI - T \, \Fr_{p}^{-1} \big| H^3(X_{\bm \varphi},\IQ_p)\right) \= \det \left(\mtI - T \mtU_p(\bm \varphi)\right)~,
\end{align}
where $\mtU_p(\bm \varphi)$ is a matrix representing the action of $\Fr_{p}^{-1}$ on $H^3(X_{\bm \varphi},\IQ_p)$.

It is important to note that the field of $p$-adic numbers has characteristic $0$, in contrast to $\overline{\IF_{\! p}}$, which has characteristic $p$. This is necessary for the Lefschetz fixed-point formula to hold, as otherwise one would only obtain the correct result $\!\!\! \mod p$, which is not enough to compute the full zeta function. To construct $p$-adic cohomology theories, the variety defined over $\IF_{\! p}$ needs thus to be lifted to a variety over the $p$-adic integers $\IZ_p$, which can be then studied over $\IQ_p$ (see for instance \cite{Monsky1970a,vanderPut1986a}). A key step in constructing the lift is to consider the embedding of the finite field $\IF_{\! p}$ into $\IQ_p$ given by the Teichmüller lift $\Teich: \IF_{\! p} \hookrightarrow \IZ_p \subset \IQ_p$ (for definitions, see appendix \ref{app:p-adic_numbers}). The full construction of a $p$-adic cohomology theory is an involved process. However, many of the properties we need to find the action of the Frobenius map are luckily essentially independent of the choice of the cohomology theory \cite{Candelas:2019llw}. In practice, this means that we can perform many of the computations, chiefly the power series expansions of the periods and their derivatives in a more familiar cohomology, such as the de Rham cohomology, and in the end interpret the result as power series whose coefficients are $p$-adic integers. All we have to do to take lifting into account is that when computing quantities associated to the manifold $X_{\bm \varphi}$ with $\bm \varphi \in \IF_{\! p}^m$, we must at the end substitute
\begin{align} \notag
\bm \varphi \mapsto \Teich(\bm \varphi) = (\Teich(\varphi_1),\dots,\Teich(\varphi_m))~.
\end{align}

In \cite{Candelas:2019llw} an effective practical method for computing the polynomials $R_p(X_{\bm \varphi},T)$ for one-parameter families of Calabi--Yau manifolds was developed. In this paper, we generalise this method to multiparameter manifolds.
\subsection{Mirror symmetry and Calabi--Yau periods}
\vskip-10pt
We are interested in families of manifolds parametrised by the complex structure parameters $\bm \varphi$, so we wish to find the action of the Frobenius map on families of cohomologies. To do this efficiently, finding a convenient basis of the third cohomology $H^3(X_{\bm \varphi})$ is a key.\footnote{As explained above, the computations we do are essentially valid for both $p$-adic and Dolbeault cohomology, and we are free to switch between the two cohomology theories by essentially just taking the field we are working over to be either that of $p$-adic or complex numbers. For this reason, when a statement is essentially valid for both cohomologies or when we are treating the both theories simultaneously, we do not distinguish between the two, and talk simply about the cohomology $H^3(X_{\bm \varphi})$, leaving the choice unspecified.} An ideal tool for this purpose is Dwork's deformation theory \cite{MR0159823,MR0188215}, the idea being to first find the action of the Frobenius map on a simple manifold, which we take here to be a manifold with a large complex structure, and then study how the action changes as the complex structure of the manifold is varied. 

Deformations of the complex structure of Calabi--Yau manifolds lie also at the heart of mirror symmetry, which is a conjectural relation that can be used to relate a Calabi--Yau threefold $X_{\bm \varphi}$ to another threefold $\wt X_{\bm t}$, the \textit{mirror of $X_{\bm \varphi}$}.\footnote{The subscript $\bm t$ here denotes the complexfied Kähler class of $\wt X$, which is related to $\bm \varphi$ by the mirror map \eqref{eq:mirror_map}.} In particular, this mapping relates the middle cohomology $H^3(X_{\bm \varphi},\IC)$ of $X_{\bm \varphi}$ to the even cohomology $H^{2*}(\wt X_{\bm t},\IC)$ of its mirror $\wt X_{\bm t}$. This relation, and the universal structures that follow from mirror symmetry considerations, allow us to find a convenient basis of the middle cohomology for any multiparameter Calabi--Yau threefold $X_{\bm \varphi}$ in terms of topological data of its mirror $\wt X_{\bm t}$. We give here a brief review of the aspects of mirror symmetry essential for motivating and understanding the construction we present in \sref{sect:generic_Frobenius_map}.

Griffiths transversiality (see for example \cite{Candelas:1990pi}) implies that differentiating the holomorphic three-form with respect to the complex structure moduli gives three-forms that are no longer holomorphic. Rather, by taking enough derivatives, the whole of $H^3(X_{\bm \varphi},\IC)$ can be spanned by the following forms
\begin{align} \label{eq:Griffiths_transversiality} 
\begin{split} 
&\Omega \in H^{(3,0)}(X_{\bm \varphi},\IC)~, \\
\partial_{\varphi^i} &\Omega \in H^{(3,0)}(X_{\bm \varphi},\IC) \oplus H^{(2,1)}(X_{\bm \varphi},\IC)~,\\
\partial_{\varphi^i} \partial_{\varphi^j}& \Omega \in  H^{(3,0)}(X_{\bm \varphi},\IC) \oplus H^{(2,1)}(X_{\bm \varphi},\IC) \oplus H^{(1,2)}(X_{\bm \varphi},\IC)~,\\
\partial_{\varphi^i} \partial_{\varphi^j} \partial_{\varphi^k} &\Omega \in  H^{(3,0)}(X_{\bm \varphi},\IC) \oplus H^{(2,1)}(X_{\bm \varphi},\IC) \oplus H^{(1,2)}(X_{\bm \varphi},\IC) \oplus H^{(0,3)}(X_{\bm \varphi},\IC)~.
\end{split}
\end{align}
We can therefore always choose a basis for the $2m+2$-dimensional space $H^3(X_{\bm \varphi},\IC)$ among these derivatives. Any additional derivatives can then be expressed in terms of the basis, leading to a system of differential equations that $\Omega$ satisfies, called the \textit{Picard--Fuchs equations}. These can be solved to find $\Omega$ in practice.  

Picking a basis of $2m{+}2$ functions among the space of solutions to the Picard--Fuchs equations is equivalent to choosing a particular (constant) basis $\{ v_a, v^a \}$ of $H^3(X_{\bm \varphi},\IC)$. The holomorphic three-form can be expanded as
\begin{align} \label{eq:Omega_basis_expansion}
\Omega \= \sum_{a=0}^m \left(\varpi^a v_a - \varpi_a v^a \right)~.
\end{align}
Equivalently, the three-form $\Omega$ can be expressed in terms of a vector of \textit{periods} $\varpi^a$, $\varpi_b$:
\begin{align} \label{eq:period_vector_Frob_basis}
\varpi \= \left(\begin{matrix}\varpi^0\\\varpi^{i}\\ \varpi_{j}\\ \varpi_{0} \end{matrix}\right).
\end{align}
We will shortly fix the basis of solutions by imposing boundary conditions on $\varpi^a$ and $\varpi_b$.
\subsection{Indicial algebra and the Frobenius basis} \label{sect:Frobenius_basis}
\vskip-10pt
The \textit{indicial algebra} of the Picard--Fuchs equation satisfied by the periods \cite{Hosono:1994ax,Braun:2007vy} is defined by the relations satisfied by the (co-)homology ring elements $\epsilon_i, \mu^i,$ and $\eta$, viewed as elements of $H^2(\wt X_{\bm t},\IZ)$, $H^4(\wt X_{\bm t},\IZ)$, and $H^6(\wt X_{\bm t},\IZ)$, respectively, or their homology duals
\begin{align} \label{eq:indicial_algebra_definition}
\epsilon_i \epsilon_j \= \epsilon_j \epsilon_i~, \qquad \epsilon_i \epsilon_j = Y_{ijk} \, \mu^k~, \qquad \epsilon_i \epsilon_j \epsilon_k \= Y_{ijk} \, \eta~, \qquad \epsilon_i \mu^j \= \delta_i^j \eta~, \qquad \epsilon_i \eta \= 0~,
\end{align}
where $Y_{ijk}$ are the classical triple intersection numbers of $\wt X_{\bm t}$, expressed in terms of the generators $\me_i$ of the second cohomology $H^2(\wt X_{\bm t}, \IZ)$ as
\begin{align} \notag
Y_{ijk} \= \int_{\wt X_{\bm t}} \me_i \wedge \me_j \wedge \me_k~, \qquad \me_i,\me_j,\me_k \in H^2(\wt X_{\bm t}, \IZ)~. 
\end{align}
Near a large complex structure point, or equivalently a point of maximal unipotent monodromy, the periods can be extracted as coefficients in the expansion of
\begin{align} \label{eq:all-period_epsilon_expansion}
\begin{split}
\varpi(\bm \varphi,\bm \epsilon) \= \bm \varphi^{\bm \epsilon}f(\bm \varphi, \bm \epsilon) \,  &\defineas \varpi^0 + \varpi^{i} \epsilon_i  + \varpi_{i} \mu^i + \varpi_0 \eta~,\\
&\= f + (\ell^i f + f^i) \epsilon_i + \frac{1}{2!} \left(\ell^i \ell^j f  + 2 \ell^i f^j + f^{ij} \right) Y_{ijk} \mu^k\\
&\hskip28pt +\frac{1}{3!} \left(\ell^i \ell^j \ell^k f + 3 \ell^i \ell^j f^k + 3 \ell^i f^{jk} + f^{ijk} \right)Y_{ijk} \eta~,
\end{split}
\end{align}
where $\bm \varphi = (\varphi^1,\dots,\varphi^m)$ is a vector of complex structure parameters, and we have used the shorthand $\ell^i = \log \varphi^i$. We have also denoted $\partial_{\epsilon_i} f = f^i$, $\partial_{\epsilon_i} \partial_{\epsilon_j} f = f^{ij}$, and $\partial_{\epsilon_i} \partial_{\epsilon_j} \partial_{\epsilon_k} f = f^{ijk}$. The coordinates are chosen so that the large complex structure point is at $\bm \varphi = \bm 0$ and
\begin{align} \notag
f(\bm \varphi,\bm \epsilon) \= \sum_{\bm k \in \IZ_{\geq 0}^m} A_{\bm k}(\bm \epsilon) \, \bm \varphi^{\bm k}
\end{align}
is a power series in $\bm \varphi$ and $\bm \epsilon$ with rational coefficients. We also require that the mirror map is given by the standard expression
\begin{align} \label{eq:mirror_map}
t^i \= \frac{1}{2\pi \ii}\frac{\varpi^i}{\varpi^0}~.
\end{align}
In particular, this choice fixes the freedom to rescale the coordinates by rational coefficients, and it is the choice we will be using throughout the paper. Scaling $\varphi$ by a rational constant amounts to a non-rational change of basis of $H^3(X_{\bm \varphi},\IC)$, which would be reflected in the $p$-adic cohomology (see for example the discussion in \sref{sect:mirror_octic}).  There is a residual freedom corresponding to the choice of the integral basis of $H^{2*}(\wt X_{\bm t}, \IZ)$. This latter freedom will not affect our discussion in this paper. In all examples known to us, this choice of scaling makes the coefficients of the series expansion of the fundamental period $p$-adic integers for all but finitely many $p$. However, we are not aware of a theorem guaranteeing this in~general.

We also remark that for computational purposes, instead of using $f^{i}$, $f^{ij}$, and $f^{ijk}$, it is often useful to define the combinations 
\begin{align} \label{eq:tilded_f_definition}
\wt f_i = \frac{1}{2!} Y_{ijk} f^{jk} \qquad \text{and} \qquad \wt f = \frac{1}{3!} Y_{ijk} f^{ijk}~,
\end{align}
which naturally enter the logarithm-free quantities we define in \eqref{eq:Log-free_E_definition}. 

The expansion \eqref{eq:all-period_epsilon_expansion} implicitly fixes the boundary conditions of the Picard--Fuchs equations, giving us the \textit{period vector} $\varpi$ in the \textit{(arithmetic) Frobenius basis}.
\begin{align} \label{eq:period_vector_Frob_basis_expansion}
\varpi \= \begin{pmatrix}
f \\[3pt]
f^i + f \ell^i \\[3pt]
\frac{1}{2!}Y_{ijk}\left(f^{jk} + 2 f^j \ell^k + f \ell^j \ell^k \right)  \\[3pt]
\frac{1}{3!}Y_{ijk}\left(f^{ijk} + 3 f^{ij} \ell^k + 3 f^i \ell^j \ell^k + f \ell^i \ell^j \ell^k \right)  \\[3pt]
\end{pmatrix}~.
\end{align}

We introduce the `\textit{inverse triple intersection numbers}' $\wh Y^{ijk}$ as a set of constants that satisfy the relation
\begin{align} \label{eq:Yhat_definition}
Y_{ijk} \wh Y^{ijs} \= \delta_k^s~.
\end{align}
This does not define the quantities $\wh Y^{ijs}$ uniquely. Rather, one can shift $\wh Y^{ijs}$ by any $A^{ijs}$ which is `orthogonal' to the triple intersection numbers, that is
\begin{align} \notag
Y_{ijk}A^{ijs} \= 0~.
\end{align}
When we use the quantities $\wh Y^{ijs}$ to form a basis of $H^3(X_{\bm \varphi})$, different choices of the constants $A^{ijs}$ amount simply to choosing a different basis, which explains the apparent ambiguity. The freedom to choose these constants can always be used to make $\wh Y^{ijs}$ symmetric in the first two indices, which fixes the antisymmetric part $A^{[ij]s}$. Otherwise we leave $A^{ijs}$ unspecified in general, and choose these conveniently on a case-by-case basis. Any ambiguities resulting from this will not affect the following discussion, and we will always specify the choice we have made in particular examples. 

With the help of $\wh Y^{ijk}$, the elements $\mu^k$ can be expressed as
\begin{align} \notag
\mu^k \=  \wh Y^{ijk} \epsilon_i \epsilon_j~,
\end{align}
Arguing as in \cite{Candelas:2023yrg} that the indicial algebra elements can be related to the monodromy matrices around the loci $\varphi^i = 0$, we can find an explicit representation where the indicial algebra elements are given by
\begin{align} \label{eq:homology_algebra_rep}
\epsilon_i \=  \begin{pmatrix}
0 & \+\!\bm 0^T & \hskip3pt\bm 0^T & 0\\[3pt]
\bm \delta_i &  \Izero & \hskip-3pt\Izero & \bm 0\\[3pt]
\bm 0 & \,\,\mathbb{Y}_i & \hskip-3pt\Izero & \bm 0 \\[3pt]
0 & \+\!\bm 0^T & \+\!\bm \delta_i^T & 0
\end{pmatrix}, \quad \mu^i \= \begin{pmatrix}
0 & \+\!\bm 0^T  & \hskip3pt\bm 0^T & 0\\[3pt]
\bm 0 & \Izero & \hskip-3pt\Izero & \bm 0\\[3pt]
\bm \delta_i  & \Izero & \hskip-3pt\Izero & \bm 0 \\[3pt]
0 & \+\!\bm \delta_i^T & \hskip3pt\bm 0^T & 0
\end{pmatrix}, \quad \eta \= \begin{pmatrix}
0 & \+\!\bm 0^T & \hskip3pt\bm 0^T & 0\\[3pt]
\bm 0 &  \Izero &  \hskip-3pt\Izero & \bm 0\\[3pt]
\bm 0  &  \Izero &  \hskip-3pt\Izero & \bm 0 \\[3pt]
1 &  \+\!\bm 0^T & \hskip3pt\bm 0^T & 0
\end{pmatrix}.
\end{align}
Here $\IY_i$ denotes the symmetric $m \times m$ matrix whose components are given by the triple intersection numbers $Y_{ijk}$, $1 \leqslant j,k \leqslant m$, $\Izero$ is the constant zero matrix, $\bm 0$ is the constant zero vector and $\bm \delta_i$ is an $m$-component vector whose components are given by $\delta_{ij}$, $1 \leqslant j \leqslant m$.
\subsection{The integral basis} \label{sect:From_Frobenius_to_Integral}
\vskip-10pt
Choosing a symplectic integral basis $\alpha_i,\beta^i$ of $H^3(X,\IZ)$, we can expand the holomorphic three-form~as
\begin{align} \notag
\Omega \= z^a \alpha_a - \cF_b(z) \beta^b~, \qquad \cF_b \;\defineas \; \frac{\partial \cF(z)}{\partial z^b}~,
\end{align}
where $\cF$ is the prepotential \cite{Candelas:1990pi}. We denote by $\Pi$ the corresponding vector
\begin{align} \label{eq:Pi_vector}
\Pi \= \begin{pmatrix}
\Pi_{0}\\[3pt]
\Pi_{i}\\[3pt]
\Pi^{0}\\[3pt]
\Pi^{i}
\end{pmatrix} \= \begin{pmatrix}
\frac{\partial}{\partial z^0} \cF\\[3pt]
\frac{\partial}{\partial z^i} \cF\\[3pt]
z^0\\[3pt]
z^i
\end{pmatrix}~.
\end{align}
Near a point of maximal unipotent monodromy, we can find a change-of-basis matrix $\rho$ from the Frobenius basis to the integral symplectic basis by comparing the asymptotics and monodromies of the vectors $\Pi$ and $\varpi$. This matrix is given by
\begin{align} \notag
\rho \= \begin{pmatrix}
-\frac{1}{3} Y_{000} & -\frac{1}{2} \bm Y_{00}^T & \;\;\+\bm 0^T &\hskip12pt 1\\[3pt]
- \frac{1}{2} \bm Y_{00} & -\bm \IY_{0}  & - \II &\hskip12pt \bm 0 \\[3pt]
1 & \+\bm 0^T & \;\;\+\bm 0^T &\hskip12pt 0 \\[3pt]
\bm 0 & \; \II & \+\Izero &\hskip12pt \bm 0
\end{pmatrix}~.
\end{align}
The components of this matrix are as follows: $\IY_0$ is an $m{\times}m$ matrix whose components are $(\IY_0)_{ij} = Y_{0ij}$. Requiring that $Y_{0ij} \in \{0, 1/2\}$, the matrix $\rho$ is uniquely fixed. These constants can be found by requiring that $\Pi$ has integral monodromy around the point of maximal unipotent monodromy \cite{Hosono:1994ax}.\footnote{The constants $Y_{0ij}$ are conjecturally given by $Y_{0ij} = -\frac{1}{2}Y_{iij} \mod \IZ$ \cite{Mayr:2000as}.} The symbol $\bm Y_{00}$ denotes the $m$-component vector whose $i$'th component is given by $Y_{i00}$, $1 \leqslant i \leqslant m$. The constants $Y_{i00}$ and $Y_{000}$ are given by
\begin{align} \notag
Y_{i00} \= -\frac{1}{12} \int_{\wt{X}_{\bm t}} c_2(\wt{X}_{\bm{t}}) \wedge \ee_i~, \qquad Y_{000}\;= -3\chi(\widetilde X_{\bm t})\frac{\z(3)}{(2\p\ii)^3} \;= +3\chi(X_{\bm \varphi})\frac{\z(3)}{(2\p\ii)^3}~,
\end{align}
where $\ee_i$ are the generators of $H^2(\wt{X}_{\bm t},\IZ)$, $c_2(\wt{X}_{\bm{t}})$ denotes the second Chern class of $\wt{X}_{\bm{t}}$, and $\chi(X_{\bm \varphi})$ and $\chi(\widetilde X_{\bm t})$ the Euler characteristics of $X_{\bm \varphi}$ and $\widetilde X_{\bm t}$, respectively.

In writing this, we have separated the diagonal matrix $\nu$, which affects the normalisation of the periods, giving the transformation between the period vectors $\varpi$ and $\wh \varpi$ in what were termed in \cite{Candelas:2021tqt} the \textit{arithmetic Frobenius basis} and the \textit{complex Frobenius basis}. 
\begin{align} \label{eq:complex_Frobenius_basis_varpi}
\wh \varpi \= \nu^{-1} \varpi~.
\end{align}
The relation between the vectors in the Frobenius and integral bases is given by
\begin{align} \label{eq:Pi_varpi_relation_general}
\Pi \= \rho \nu^{-1} \varpi~, \qquad \text{with} \qquad \nu = \text{diag}\left(1,2\pi\ii \, \bm 1, (2\pi \ii)^2 \, \bm 1,(2\pi\ii)^3 \right)~.
\end{align}
\subsection{Relation between the rational and Frobenius bases} \label{sect:frobenius_to_rational}
\vskip-10pt
Let us return to the first of eqns.~\eqref{eq:all-period_epsilon_expansion} and rewrite this in the form
\beq\begin{split}\label{eq:varpi_expansion}
\varpi(\bm \varphi,\bm \epsilon) &\= \varpi^0 (\mtI - \g\,\eta) + \varpi^{i} \epsilon_i  + \varpi_{i} \mu^i + (\varpi_0 +\g\,\vp^0) \eta~,\\[3pt]
&\= (\mtI-\g\,\eta)\big(\varpi^0 + \varpi^{i} \epsilon_i  + \varpi_{i} \mu^i + {\widetilde\vp}^0 \eta\big)~,
\end{split}\eeq
where, in the last relation, we have written
\beq\notag
{\widetilde\vp}_0 \= \vp_0 + \g\,\vp^0~.
\eeq

We can gather the periods, in this new basis, into a vector similar to that in \eqref{eq:period_vector_Frob_basis}:
\beq\notag
\widetilde\vp \= \begin{pmatrix}\vp^0\\ \vp^i \\ \vp_j \\[2pt] {\widetilde\vp}_0\\
\end{pmatrix}~.
\eeq

Abusing notation, we may write
\beq
(\mtI + \g\eta)\,\vp \=  \widetilde\vp \qquad\text{or equivalently}\qquad \vp \= (\mtI - \g\,\eta)\, \widetilde\vp ~.
\notag\eeq
The abuse of notation is that, in this relation we may think of $\vp$ and $\widetilde\vp$ as either vectors or matrices. In either case, $1 \mp  \gamma \,\eta$ is a matrix, with $\eta$ as in \eqref{eq:homology_algebra_rep}. 

In the above, the quantity $\g$ has appeared as an arbitrary parameter. We now choose  
\beq\label{eq:gammadef}
\g \= \chi(\wt X_{\bm t})\, \z(3)~.
\eeq
The virtue of this choice is that $\mtI + \gamma \eta$ is now a matrix that converts the complex Frobenius basis to a rational basis, as we see by the following relations
\beq\notag
\P \= {\tilde \r}\,\n^{-1}{\widetilde\vp} \qquad \text{where} \qquad \r \= \tilde \r \left(\mtI + \chi(\wt{X}_{\bm t}) \frac{\z(3)}{(2\p\ii)^3}\,\eta\right)~.
\eeq
The matrix $\tilde\r$ has the same form as the matrix $\r$ apart from the element $-Y_{000}/3$, which is replaced by zero:
\beq \notag
\tilde\rho \= 
\begin{pmatrix}
0& -\frac{1}{2} \bm Y_{00}^T & \;\;\+\bm 0^T &\hskip12pt 1\\[3pt]
- \frac{1}{2} \bm Y_{00} & -\bm \IY_{0}  & - \II &\hskip12pt \bm 0 \\[3pt]
1 & \+\bm 0^T & \;\;\+\bm 0^T &\hskip12pt 0 \\[3pt]
\bm 0 & \; \II & \+\Izero &\hskip12pt \bm 0
\end{pmatrix}~.
\eeq
Thus $\tilde\r$ is a matrix with rational entries and $\n^{-1} \widetilde\vp$ is a rational basis of periods.
 
The matrix $\left(\mtI - \chi(\wt{X}_{\bm t}) \frac{\z(3)}{(2\p\ii)^3}\,\eta\right)$ has an interesting relation to the Todd and $\Gamma$-classes (see \cite{Libgober:1999a,Hosono:2004jp,Iritani:2007a,Iritani2009a,Hori:2013ika,Halverson:2013qca}). To see this, we set
\newcommand{\Td}{\hbox{Td}}
\beq
\Td(z) \= \frac{z}{1-\ee^{-z}}
\notag\eeq
and note the identity
\beq\label{eq:characteristic classes}
\frac{\G\left(1+\frac{z}{2\p\ii}\right)}{\sqrt{\Td(z)}} \= \ee^{\ii\,\X(z) - z/4}~,
\eeq
with
\beq
\X(z) \;= \g_E \frac{z}{2\p} + \ii \sum_{k=1}^\infty \frac{\z(2k+1)}{2k+1} \left(\frac{z}{2\p\ii}\right)^{2k+1}~,
\notag\eeq
and $\g_E$ is Euler's constant. We are interested in multiplicative characteristic classes based on $\Td(z)$ and $\G\left(1 + \frac{z}{2\p\ii}\right)$. To proceed, we replace $z$, in the identity \eqref{eq:characteristic classes}, by $\wt \Th$, the curvature matrix of $\wt X_{\bm t}$. We denote the eigenvalues of $\wt \Th$ by $\l_k,\,k{\=}1,2,3$ and the $m$'th symmetric polynomial in the $\l_k$ by $\s_m$. The curvature matrix is a matrix-valued two-form, so the $\l_k$ are two forms and $\s_m$ is a $2m$-form.

We have
\beq
\prod_{k=1}^3 \ee^{\ii \, \X(\l_k) - \frac14 \l_k} 
\= \exp\left\{ -\s_1\left( \frac{\g_E}{2\p\ii} + \frac14 \right) - \frac{\z(3)}{3(2\p\ii)^3} \sum_k \l_k^3\right\} 
\= \mtI - \chi(\wt{X}_{\bm t})\,\frac{\z(3)}{(2\p\ii)^3}\,\eta~,
\notag\eeq
where, in passing to the second expression we have used the fact that $\s_1$ is the first Chern class and so vanishes, together with the fact that
\beq
\sum_k \l_k^3 \= \s_1^3 - 3 \s_1\s_2 + 3\s_3 \= 3\chi(\wt X_{\bm t})\eta,
\notag\eeq
when $\s_1{\=}0$.

Thus we have shown that
\beq \label{eq:complex_Gamma_hat}
 \frac{{\widehat\G}_{\! \wt X_{\bm t}}}{\sqrt{{\widehat\Td_{\wt X_{\bm t}}}}} \= \mtI - \chi(\wt{X}_{\bm t})\,\frac{\z(3)}{(2\p\ii)^3}\,\eta~,
\eeq
where ${\widehat\G}_{\!\wt X_{\bm t}}$ and ${\widehat\Td}_{\wt X_{\bm t}}$ denote the multiplicative characteristic classes. 

We will later be interested in the $p$-adic $\Gamma$-class and in relation to this we make a comment on the identity \eqref{eq:characteristic classes}. This follows from a standard identity for the $\G$-function, which we can write in the form
\beq\label{eq:Gamma identity}
\G\left(1 + \frac{z}{2\p\ii}\right)\= \exp\left\{ -\g_E\frac{z}{2\p\ii} + 
\sum_{n=2}^\infty (-1)^n\, \frac{\z(n)}{n}\left(\frac{z}{2\p\ii}\right)^n \right\}
\eeq
Now, recalling the reflection formula for the $\G$-function,we have
\beq\notag
\Td(z)\,\ee^{-\frac{z}{2}}\= \frac{z/2}{\sinh(z/2)} 
\= \G\left(1 + \frac{z}{2\p\ii}\right)\,\G\left(1 - \frac{z}{2\p\ii}\right)~.
\eeq
If we use the identity \eqref{eq:Gamma identity} to replace the product of $\G$-functions, and observe that the $\z$-functions of odd argument cancel, and then take a square root, we see that
\beq\label{eq:Todd identity}
\sqrt{\Td(z)}\,\ee^{-z/4} 
\= \exp\left\{\sum_{n=1}^\infty\frac{\z(2n)}{2n}\left(\frac{z}{2\p\ii}\right)^{2n}\right\}~.
\eeq
From \eqref{eq:Gamma identity} and \eqref{eq:Todd identity} we have
\beq\label{eq:characteristic classes two}
\frac{\G\left(1+\frac{z}{2\p\ii}\right)}{\sqrt{\Td(z)}\,\ee^{-\frac{z}{4}}} \=
\exp\left\{-\g_E \frac{z}{2\p\ii} - \sum_{k=1}^\infty \frac{\z(2k+1)}{2k+1} \left(\frac{z}{2\p\ii}\right)^{2k+1}\right\}~,
\eeq
from which we recover \eqref{eq:characteristic classes}.
Now, in $p$-adic analysis it is possible to define $p$-adic analogues of the $\G$- and $\z$-functions. It is interesting that these are related by a relation analogous to \eqref{eq:Gamma identity}
\beq\label{eq:pGamma identity}
\G_{\! p}(z) \= \exp\left( -\G_{\! p}'(1)\, z - \sum_{k=1}^{\infty}\frac{\z_p(2k+1)}{2k+1}\,z^{2k+1}\right)~,
\eeq
in which $\G_{\! p}'(1)$ is the $p$-adic analogue of $\g_E{\=}\G'(1)$. The $\z$ function terms with even argument are missing because, for the $p$-adic case, we have
\beq\notag
\z_p(2n) \= 0 \quad\text{for}\quad n\= 1,2,3\ldots~.
\eeq
The surprise is that the right hand side of \eqref{eq:pGamma identity} is equally analogous to the right hand side of \eqref{eq:characteristic classes two}. While we do not have a complete understanding of this, it is easily checked that there is no difference, in the complex context, between integrating 
\begin{align}
{\widehat\G}_{\! \wt X_{\bm t}}\quad  \text{and} \quad \frac{{\widehat\G}_{\! \wt X_{\bm t}}}{\sqrt{{\widehat\Td}_{\wt X_{\bm t}}}}
\end{align}
over a Calabi-Yau threefold. Therefore, for threefolds, the Frobenius and rational bases are related by
\begin{align} \label{eq:complex_change_of_basis}
\wt{\varpi} \= {\widehat\G}_{\! \wt X_{\bm t}}^{-1} \, \varpi~.
\end{align}
We note that we will see in \sref{sect:U(0)} that an analogous relation is true also in the $p$-adic context.
\newpage
\section{The Local Zeta Functions of Generic Threefolds} \label{sect:generic_Frobenius_map}
\vskip-10pt
\subsection{The Frobenius map} \label{sect:Frobenius_map}
\vskip-10pt
To compute the zeta function of a Calabi--Yau threefold $X_{\bm \varphi}$, we use the methods of \cite{Candelas:2021tqt}, which build on the work of Dwork and Lauder \cite{Dwork,Lauder2004a,Lauder2004b}. This approach is based on finding explicitly the matrix $\mtF_p(\bm \varphi)$ (or its inverse $\mtU_p(\bm \varphi)$) representing the action of the Frobenius map defined in \eqref{eq:Frobenius_map_Fr_definition} on the middle cohomology. This is done by first finding the matrix $\mtU_p(\bm \varphi_0)$ for a manifold $X_{\bm \varphi_0}$ for a simple manifold where the expressions can be relatively easily computed. We are hoping to find a universal expression for the initial value $\mtU_p(\bm \varphi_0)$, depending at most on some manifold-specific constants. From the theory of mirror symmetry, we know that such universal expressions exist for the form of the periods near the large complex structure points, which motivates taking the large complex structure point $\bm \varphi = \bm 0$ as the initial point. By studying how the matrix $\mtU_p(\bm \varphi)$ changes as we move in the complex structure moduli space, we are able to find an explicit expression for the matrices $\mtU_p(\bm \varphi)$ at other points in the moduli space. Then the relation \eqref{eq:R(T)_determinant} between the Frobenius map and the zeta function numerator $R_p(X_{\bm \varphi},T)$ can be used to arrive at the final result. 

Description of this method necessarily gets at times rather technical. For this reason, we have included analogous derivation of the zeta function of a family of elliptic curves in appendix \ref{app:elliptic_curve} that illustrates the techniques employed here in a simpler setting. We encourage the reader to consult the appendix \ref{app:elliptic_curve} alongside reading this section.

We can view the Frobenius map as acting on a vector bundle $\cH$, whose base is the complex structure moduli space $\cM_{\IC S}$ of the family $X_{\bm \varphi}$ of Calabi--Yau manifolds. The fibre over a point $\bm \varphi \in \cM_{\IC S}$ is the middle cohomology group $H^3(X_{\bm \varphi})$ of the manifold $X_{\bm \varphi}$ whose complex structure corresponds to the point $\bm \varphi$ in the moduli space.
\begin{figure}[H]
	\hskip155pt
	\begin{tikzcd}				
	& H^3(X_{\bm \varphi}) \arrow[hookrightarrow]{r}{}
	& \cH  \arrow{d}{\pi} \\		
	& 
	& \cM_{\IC S}
	\end{tikzcd}
	\vskip10pt
	\capt{6in}{fig:Cohomology bundle}{The vector bundle $\cH$.}	
\end{figure}

The action of the Frobenius map on this bundle can be defined by
\begin{align} \label{eq:Frobenius_map_bundle_action}
\left(\bm \varphi,H^3(X_{\bm \varphi}) \right) \mapsto \left(\Frob_{p^n}(\bm \varphi),\Fr_{p^n} H^3(X_{\bm \varphi}) \right) \= \left(\bm \varphi^{p^n},\Fr_{p^n} H^3(X_{\bm \varphi}) \right)~,
\end{align}
where $\Fr_{p^n}$ denotes the map $H^3(X_{\bm \varphi}) \to H^3(X_{\bm \varphi^{p^n}})$, induced by the action $\Frob_{p^n}: X_{\bm \varphi} \to X_{\bm \varphi^{p^n}}$.

There is also the canonical Gauss--Manin connection
\begin{align} \notag
\nabla: \Gamma(\cM_{\IC S},H^3(X_{\bm \varphi})) \to \Gamma(\cM_{\IC S},H^3(X_{\bm \varphi}) \otimes T^* \cM_{\IC S})
\end{align}
on $\cH$. It will be enough to consider the covariant derivatives along the vector fields  given by the logarithmic derivatives $\theta_i = \varphi^i \partial_{\varphi^i}$ (no sum implied). We denote the corresponding covariant derivatives by
\begin{align} \label{eq:Gauss--Manin_definition}
\nabla_i \defineas \nabla_{\theta_i}: \Gamma(\cM_{\IC S}, H^3(X_{\bm \varphi})) \to \Gamma(\cM_{\IC S},H^3(X_{\bm \varphi}))~.
\end{align}
The Frobenius map and these derivatives satisfy a compatibility relation, as well as Leibniz rule and linearity relation. For any section $v \in \Gamma(\cM_{\IC S},H^3(X_{\bm \varphi}))$ and any function $f: \cM_{\IC S} \to \IC$ on the moduli space, the following relations hold
\begin{align} \label{eq:Fr_Commutation}
\begin{split}
p \, \Fr_p ( \nabla_i v ) &\= \, \nabla_i ( \Fr_p \, v )~,\\
\nabla_i \left(f(\bm \varphi) v \right) &\= (\theta_i f)(\bm \varphi) \, v + f(\bm \varphi) \nabla_i v~,\\
\Fr \left( f(\bm \varphi) v \right) &\= f(\bm \varphi^p) \, \Fr(v)~. 
\end{split}
\end{align}
In addition, there is also a useful consistency condition of the Frobenius map with the intersection product $H^3(X_{\bm \varphi}) \times H^3(X_{\bm \varphi}) \to H^6(X_{\bm \varphi})$, which can be identified with the one given by the wedge~product
\begin{align} \label{eq:symplectic_compatibility}
\int_X \Fr \, \xi \wedge \Fr \, \eta \= p^3 \, \Fr \int_X \xi \wedge \eta~.
\end{align}

Following \cite{Candelas:2021tqt}, we wish to formulate these conditions explicitly as matrix equations on the matrices $\mtF_p(\bm \varphi)$ and $\mtB(\bm \varphi)$ corresponding to the Frobenius map and the Gauss--Manin connection, respectively. However, note that at a generic point in the moduli space, $\bm \varphi^{p^n} \neq \bm \varphi$. This implies that the action \eqref{eq:Frobenius_map_bundle_action} of the Frobenius map $\Fr_{p^n}$ cannot be reduced, in a natural way, to an action on the middle cohomology as the fibre is not kept fixed under $\bm \varphi \mapsto \bm \varphi^{p^n}$. Instead, the Frobenius map can be viewed as a map between distinct fibres $\Fr_{p^n}: \cH_{\bm \varphi} \to \cH_{\bm \varphi^{p^n}}$. Nevertheless, at fixed points of $\Fr_{p}$, that is, at the Teichm\"uller representatives $\Teich(\bm \varphi)$
of integral vectors $\bm \varphi \in \IF_{\! p}^m$, it is indeed possible to identify the action of $\Fr_{p^n}$ on the middle cohomology $H^3(X_{\Teich(\bm \varphi)})$. We denote the matrix describing this action, in the basis defined by \eqref{eq:H^3_basis} below, by $\mtF_p(\bm \varphi)$. The Teichmüller representatives $\Teich(\bm \varphi)$ provide a natural embedding of $\IF_{\! p}^m$ to $\IQ_p^m$. Hence we can identify $X_{\Teich(\bm \varphi)}$ with the manifold $X_{\bm \varphi}/\IF_{\! p}$ defined over the finite field~$\IF_{\! p}$.
\begin{figure}[H]
	\centering
	\begin{center}
		\includegraphics[width=14cm, height=7cm]{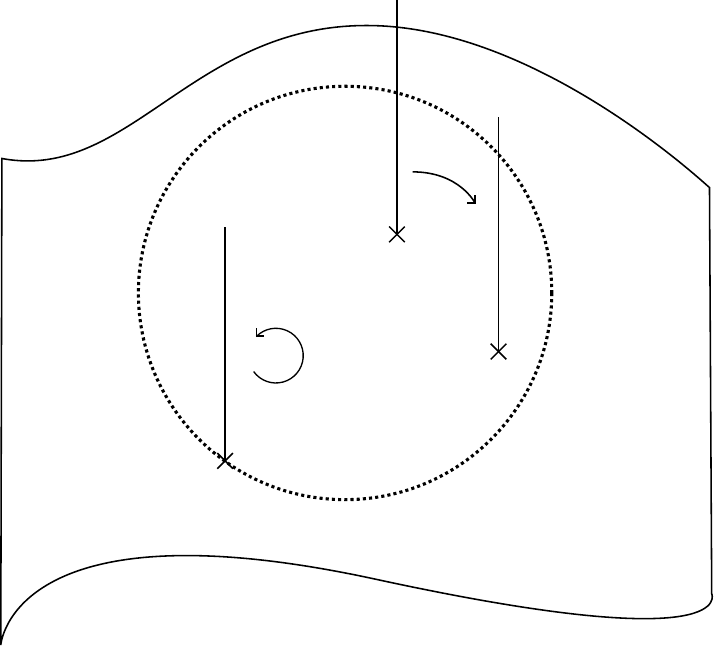}
	\end{center}
    \place{3.7}{2}{$\bm \varphi$} 
    \place{4.05}{1.45}{$\bm \varphi^p$}
    \place{0.75}{1.}{$\Teich(\bm n) = \Teich(\bm n)^p$}
    \place{2.3}{2}{$H^3(X_{\Teich(\bm n)})$}
    \place{2.9}{1.55}{$\Fr_p$} 
    \place{3.8}{2.45}{$\Fr_p$}    
    \place{2.95}{2.55}{$H^3(X_{\bm \varphi})$}  
    \place{4.45}{2.05}{$H^3(X_{\bm \varphi^p})$}     
	\vskip-10pt
	\capt{6in}{fig:Frobenius_map_fibration}{A heuristic sketch of the complex structure moduli space $\cM_{\IC S}$. Each point $\bm \varphi$ in the moduli space correspond to a Calabi--Yau manifold $X_{\bm \varphi}$. The fibre above each point is the middle cohomology group $H^3(X_{\bm \varphi})$ of the corresponding manifold. The dashed circle represents the $p$-adic unit disk $||x||_p < 1$, with $1$ also being the radius of convergence of the power series appearing in the expansions of the periods and their derivatives (the logarithms appearing in the periods drop out of the expression for the matrix $\mtU_p(\bm \varphi)$). At a generic point $\varphi$, the Frobenius map $\Fr_p$ acts between two distinct fibres $H^3(X_{\bm \varphi})$ and $H^3(X_{\bm \varphi^p})$, which in particular implies that the matrix $\mtU_p(\bm \varphi)$ is not well-defined. At Teichmüller representatives $\Teich(\bm n)$ of integral points $\bm n \in \IZ^m$ the fibres coincide, and the matrix $\mtU_p(\bm \varphi)$ is well-defined. The Teichmüller representatives lie at the boundary of the $p$-adic unit disk, where the period series do not converge, but the matrix $\mtU_p(\bm \varphi)$ does.}	
\end{figure}
In \cite{Candelas:2021tqt} it was noted that the solutions to the conditions laid out above turn out to be expressible in terms of solutions to Picard--Fuchs equations. In the remaining of this section, we will generalise this observation to the multiparameter case. 

The first step of this process is finding a convenient basis of sections of $\cH$ in which the identities we use to constrain the form of $\mtU_p(\bm \varphi)$ becomes tractable. Unlike in the one-parameter case, there is no clear canonical basis, and our choice is simply guided by the observation that our choice reduces to that used in \cite{Candelas:2021tqt} in the one-parameter case\footnote{The normalisation of periods used in \cite{Candelas:2021tqt} differs slightly from ours by factors of $Y_{111}$ and factorials.} and the fact that we find a simple expression for the matrix $\mtU_p(\bm 0)$ in terms of the matrices $\epsilon_i, \mu^i$ and $\eta$.

After the basis of sections is chosen, we can express the relations \eqref{eq:Fr_Commutation} and \eqref{eq:symplectic_compatibility} as matrix equations. These imply that the matrix $\mtU_p(\bm \varphi)$ can be expressed in terms a constant initial value matrix, which we take to be the matrix $\mtU_p(\bm 0)$ giving the Frobenius action at the large complex structure point, together with the change-of-basis matrix from the constant basis $\langle v_a \rangle$ to the basis given by the sections we chose. 

Using these relations in the large complex structure limit allows us to fix the initial matrix $\mtU_p(\bm 0)$, and thus the matrix $\mtU_p(\bm \varphi)$ up to a set of (prime-dependent) constants $\alpha_p^1,\dots,\alpha_p^{h^{1,2}}$, and $\gamma_p$. These can be fixed by requiring that $\mtU_p(\bm \varphi) \mod p^n$ can be expressed as a matrix of rational functions, as this turns out to single out unique values for the constants. We find that in all of the cases we have studied, it is possible to express these in terms of the Iwasawa logarithm and the $p$-adic zeta function (see appendix \ref{app:p-adic_numbers} for brief definitions).

In addition to these core ideas, we also briefly discuss certain properties of the matrices and the local zeta functions can be used to speed up the practical evaluation of the matrix $\mtU_p(\bm \varphi)$. This is especially important in the multiparameter case, as evaluating the matrix involves computing products of matrices whose entries are multiparameter series, which is highly demanding computationally.
\subsection{A basis of sections and the Gauss--Manin connection}
\vskip-10pt
To study the Frobenius map using the deformation methods, it is important to understand how it varies as we move in the complex structure moduli space $\cM_{\IC S}$. To do this, we find a convenient basis of sections of the vector bundle $\cH$. A natural choice of sections is given by the holomorphic 3-form $\Omega$ together with a suitable set of its logarithmic derivatives. Since we are computing the zeta functions by using the deformation method around the large complex structure limit $\bm \varphi \to 0$, it is important that this choice is made so that the Frobenius map in the given basis is regular in the large complex structure limit. 

Even with this condition, the choice of the basis is not unique. We make a particular choice which leads to a simple expression for the matrix $\mtU_p(\bm 0)$. However, other choices exist. These simply amounts to a change of basis. Indeed, often there does not exist a single convenient choice of basis of sections such that the corresponding set of vectors in $H^3(X_{\varphi},\IC)$ would be linearly independent for every value of $\varphi \in \cM_{\IC S}$ corresponding to a non-singular manifold. The points where the values of the chosen sections become linearly dependent are called, in analogy with \cite{Candelas:2021tqt} \textit{apparent singularities}. Studying the zeta functions at the apparent singularities requires a choosing a different basis. These are discussed further in \sref{sect:form_of_U(varphi)} and examples given in \sref{sect:non-symmetric_split_example}.

We choose a set of sections given by
\begin{align} \label{eq:H^3_basis}
\Omega~, \qquad \theta_i \Omega~, \qquad \wh Y^{jki} \theta_j \theta_k \Omega~, \qquad \wh Y^{ijk} \theta_i \theta_j \theta_k \Omega~.
\end{align}
It is useful to gather the combinations of derivatives appearing here into a vector
\begin{align} \label{eq:vartheta_definition}
(\vartheta_0,\vartheta_{i}, \vartheta^{i},\vartheta^0) \defineas \left(1,\theta_i, \wh Y^{jki} \theta_j \theta_k, \frac{\wh Y^{ijk}}{m} \theta_i \theta_j \theta_k \right)~,
\end{align}
Let us denote by $\mtE(\bm \varphi)$ the change-of-basis matrix from the constant basis $\{ v_a, v^a \}$ to this basis $\{ \vartheta_a \Omega, \vartheta^a \Omega \}$, which we call the \textit{derivative basis}.\footnote{This change of basis is required to have the Frobenius map take the simple form we use in this paper.} The components of this matrix are
\begin{align} \label{eq:E_matrix_definition}
\mtE(\bm \varphi)_a^{\,\, b} \= \left(\begin{matrix} \vartheta_a \varpi^b & \vartheta^a \varpi^b \\[3pt]
\vartheta_a \varpi_b & \vartheta^a \varpi_b
\end{matrix}\right)~.
\end{align}
To see that the basis vectors $\{ \vartheta_a \Omega, \vartheta^a \Omega \}$ are indeed linearly independent in the large complex structure limit, it is enough to note that the asymptotic form of $\mtE(\bm \varphi)$ in the large complex structure limit, $\bm \varphi \to \bm 0$, is not singular. In fact, it is given by
\begin{align} \label{eq:E_asymptotics}
\mtE(\bm \varphi) \=  \left(\begin{matrix}
~\+ 1 & \+\!\bm 0^T  & \+\!\! \bm 0^T & 0\\[3pt]
~\+ \bm \ell & \II & \Izero & \bm 0\\[3pt]
~\frac{1}{2!} \, \bm \ell^T \mathbb{Y}_{i} \bm \ell & \ell^i \bm \IY_i & \II & \bm 0 \\[3pt]
~\frac{1}{6!} \, Y_{ijk} \ell^i \ell^j \ell^k & \frac{1}{2!} \, \bm \ell^T \mathbb{Y}_{i} \bm \ell & \quad\bm\ell \quad{} & \quad1\quad{}
\end{matrix}\right) +\cO(\bm \varphi \log^3 \bm \varphi)\= \bm \varphi^{\bm \epsilon} +\cO(\bm \varphi \log^3 \bm \varphi)~.
\end{align} 
Since the basis corresponding to $\mtE(\bm \varphi)$ spans the third cohomology at a generic point $\bm \varphi$, the logarithmic derivatives of $\mtE(\bm \varphi)$ can be written in terms of the connection matrices $\mtB_i(\bm \varphi)$ of the Gauss--Manin connection
\begin{align} \notag
(\theta_i \mtE)(\bm \varphi) = \mtE(\bm \varphi) \mtB_i(\bm \varphi)~,
\end{align}
which is the first-order form of the Picard--Fuchs equations for the family. This relation could also be used to explicitly identify the matrices $B_i$, although these are not required for the purposes of computing the zeta function. Instead, only the asymptotic form of the matrices $B_i(\bm \varphi)$ in the large complex structure limit is needed, and this can be found by studying the asymptotic form of~$\theta_i E(\bm \varphi)$. 
\begin{align} \notag
\theta_i \mtE(\bm \varphi)  \; \sim \;  \begin{pmatrix}
\+\!0 & \!\bm 0  & \bm 0 & 0\\[3pt]
\+\bm \delta_i & \! 0 & 0 & \bm 0\\[3pt]
\+\!\bm \ell^T \bm Y_{ij} & \IY_i & 0 & \bm 0 \\[3pt]
\frac{1}{2!} \bm \ell^T \IY_{i} \bm \ell & \bm Y_{ij}^T \bm \ell& \,\bm \delta_i & 0
\end{pmatrix} \= \bm \varphi^{\bm \epsilon} \epsilon_i~.
\end{align}
Comparing this to the asymptotics of $\mtE(\bm \varphi)$, we deduce that
\begin{align} \label{eq:B_matrix_asymptotics}
\mtB_i(\bm 0) \= \epsilon_i~. 
\end{align}
\subsection{The identities satisfied by the matrix \texorpdfstring{$\mtU_p(\varphi)$}{U}}
\vskip-10pt
The identities \eqref{eq:Fr_Commutation}, using an argument completely analogous to the one-parameter case \cite{Candelas:2021tqt}, imply the following differential equation for the matrix of the Frobenius action\footnote{Note that here $B_i(\bm \varphi^p)$ denotes the matrix $\Fr_p(B_i(\bm \varphi)) = B_i(\bm \varphi)|_{\bm \varphi \to \bm \varphi^p}$, i.e. the connection matrix at $\bm \varphi$, where we have substituted $\bm \varphi^p$ for $\bm \varphi$. This matrix agrees with the connection matrix at $\bm \varphi^p$, which is why we do not distinguish between these two matrices.}
\begin{align} \label{eq:F_differential_equation}
\theta_i(\mtF_{\! p})(\bm \varphi) \= p \, \mtF_{\! p}(\bm \varphi) \mtB_i(\bm \varphi^p) - \mtB_i(\bm \varphi) \mtF_{\! p}(\bm \varphi)~.
\end{align}
The solution to these equations are given by
\begin{align} \notag
\mtF_{\! p}(\bm \varphi) \= \mtE^{-1}(\bm \varphi) \mtG_p(\bm 0) \mtE(\bm \varphi^p)~,
\end{align}
where $\mtG_p(\bm 0)$ is a fixed initial value. In terms of the inverse matrix $\mtU_p(\bm \varphi)$ this reads
\begin{align}  \label{eq:mtU_definition}
\mtU_p(\bm \varphi) \= \mtE^{-1}(\bm \varphi^p) \mtV_{\! p}(\bm 0) \mtE(\bm \varphi)~,
\end{align}
where $\mtV_{\! p}(\bm 0)$ is the matrix determining the inital value $\mtU_p(\bm 0)$. 

Note that $\mtE(\bm 0) \neq \mtI$. In fact, $\mtE(\bm 0)$ is not even defined owing to the presence of logarithms in $\mtE(\bm \varphi)$. So it is not a priori clear that the matrix $\mtU_p(\bm 0)$ exists or coincides with $\mtV_{\! p}(\bm 0)$. We will see presently that the logarithms in $\mtE(\bm \varphi)$ that are problematic cancel between $\mtE(\bm \varphi)$ and $\mtE(\bm \varphi^p)^{-1}$, at least in the case of threefolds, so $\mtV_{\! p}(\bm 0) = \mtU_p(\bm 0)$.
However, already in the case of fourfolds there are numerous examples for which $\mtV_{\! p}(\bm 0) \neq \mtU_p(\bm 0)$ (see for example \cite{Jockers:2023zzi}).

The symplectic product in \eqref{eq:symplectic_compatibility} is given in the constant Frobenius basis $\langle v_a, v^a \rangle$ where the period vector is given by $\varpi$, by the matrix
\begin{align} \notag
\sigma \= \nu^{-1} \rho^T \Sigma \rho \nu^{-1} \= \frac{1}{(2\pi \ii)^3} \begin{pmatrix}
0 		& \+\+\bm 0^T		& \+\bm 0^T 		& -1\\[3pt]
\bm 0 	& \+ \Izero			& \II 	& \+\bm 0\\[3pt]
\bm 0 	& -\,\II & \Izero			& \+\bm 0 \\[3pt]
1 		& \+\+ \bm 0^T 		& \+\bm 0^T			& \+0
\end{pmatrix}~,
\end{align}
where $\Sigma$ denotes the matrix giving the symplectic product in the symplectic integral basis
\begin{align} \notag
\Sigma \= \begin{pmatrix}
\+ 0 		& \+ \bm 0^T		& 1 		& \bm 0^T \\[3pt]
\+ \bm 0 	& \+ \Izero^{\phantom{T}}			& \bm 0 	& \II^{\phantom{T}} \\[3pt]
-1	        & \+ \bm 0^T        & 0  	    & \bm 0^T \\[3pt]
\+ \bm 0    & -\II^{\phantom{T}}              & \bm 0   & \Izero^{\phantom{T}}
\end{pmatrix}.
\end{align}
The compatibility condition \eqref{eq:symplectic_compatibility} is then written in matrix form as
\begin{align} \label{eq:symplectic_compatibility_matrix}
\mtV_{\! p}(\bm 0) \, \sigma \, \mtV_{\! p}(\bm 0)^T \= p^3 \sigma~.
\end{align}
We emphasise that this condition is indeed in principle imposed on $\mtV_{\! p}(\bm 0)$ and not on $\mtU_p(\bm 0)$, as it is $\mtV_{\! p}(\bm 0)$ that gives the action of the Frobenius map in the constant Frobenius basis, whereas $\mtU_p(\bm 0)$ gives the action in the derivative basis $\{ \vartheta^a \Omega, \vartheta_a \Omega \}$. In practice, for threefolds this difference does not matter as the matrices $\mtU_p(\bm 0)$ and $\mtV_{\! p}(\bm 0)$ are equal.
\subsection{The large complex structure limit of the Frobenius map} \label{sect:U(0)}
\vskip-10pt
Taking the limit $\bm \varphi \to \bm 0$ of \eqref{eq:F_differential_equation}, one obtains the following relations that constrain the form of the matrix $\mtU_p(\bm 0)$
\begin{align} \label{eq:U0_Commutation}
p \, \epsilon_i  \mtU_p(\bm 0) \= \mtU_p(\bm 0) \, \epsilon_i~, \qquad i = 1,\dots,m~.
\end{align}
As shown in detail in appendix \ref{app:U(0)}, the most general solution to these conditions can be written in a convenient form as
\begin{align} \label{eq:U(0)_general}
\begin{split}
&\mtU_p(\bm 0) = u_p \Lambda_p \left(\text{I} + \alpha_p^i \, \epsilon_i + \beta^{(p)}_i \, \mu^i + \wh \gamma_p \, \eta \right)~,\\
\Lambda_p \= \text{diag}(1&,p \,\bm 1,p^2 \, \bm 1,p^3)~, \qquad \wh \gamma_p \= \gamma_p + \frac{1}{3!}Y_{ijk} \alpha_p^i \alpha_p^j \alpha_p^k~,
\end{split}
\end{align}
where $\alpha_p^i, \beta_i^{(p)}$ and $\gamma_p$ are prime-dependent constants we will fix later. The reason for defining the coefficients $\gamma_p$ and $\wh \gamma_p$ in this way is so that we are able to use $\gamma$ to obtain a simple form \eqref{eq:U_matrix_exponential} later, while $\wh \gamma_p$ is often more convenient for numerical computations. The commutation relation \eqref{eq:U0_Commutation} also implies that $\mtV_p(\bm 0) = \mtU_p(\bm 0)$, as recalling \eqref{eq:E_asymptotics}, we can write
\begin{align} \notag
\mtU_p(\bm 0)  \= \lim_{\bm \varphi \to \bm 0} \bm \varphi^{-p \bm \epsilon} \mtV_p(\bm 0) \bm \varphi^{\bm \epsilon} \= \mtV_p(\bm 0)~,
\end{align}
where, in the last step, we have used the commutation relation \eqref{eq:U0_Commutation}.

The compatibility condition \eqref{eq:symplectic_compatibility_matrix} is equivalent to the conditions
\begin{align} \notag
u_p^2 \= 1~, \qquad \beta^{(p)}_i \= \frac{1}{2} \, Y_{ijk} \alpha_p^j \alpha_p^k~.
\end{align}
With this result, the matrix $\mtU_p(\bm 0)$ can be written in even more compact form as
\begin{align} \label{eq:U_matrix_exponential}
\mtU_p(\bm 0) \= u_p \Lambda_p \me^{\alpha_p^i \epsilon_i} (\mtI + \gamma_p \eta)~.
\end{align}
In the one-parameter case of \cite{Candelas:2021tqt} it was conjectured, supported with extensive numerical evidence that the correct solution is obtained by imposing (in the basis we are using)
\begin{align} \label{eq:alpha_gamma_coefficients}
u_p \= 1~, \quad \alpha_p^i \= 0~, \quad \gamma_p \= \chi(\wt X_{\bm t}) \zeta_p(3)~. 
\end{align}
Strictly speaking, $u_p=1$ is a particular choice which we conjecture to correspond to a manifold defined over $\IQ$ that is isomorphic to $X_{\bm \varphi}$ over $\IC$. By choosing different, prime-dependent values $u_p = \pm 1$, one can obtain zeta functions corresponding to varieties that are isomorphic over $\IC$ but not necessarily over $\IQ$. Such varieties are called \textit{twists} of the original variety (see appendix \ref{app:twists} for an example). 

We find that when using the basis of sections defined in \eqref{eq:H^3_basis} and the coordinates specified in \sref{sect:Frobenius_basis}, this result also applies in every multiparameter case we have studied. However, in different bases or using different coordinates, non-zero values of $\alpha_p^i$ can arise. We discuss this in more detail in the context of a specific example in section \ref{sect:mirror_octic}. The appearance of the $p$-adic zeta function values among these coefficients has also been proved in certain cases \cite{Beukers2023a}.

If we assume the above expression \eqref{eq:alpha_gamma_coefficients} for $\gamma_p$, it is perhaps interesting that, as first noted for one-parameter cases in \cite{Candelas:2021mwz}, by using the $\Gamma$-class computation from \sref{sect:frobenius_to_rational}, the matrix $\mtI - \gamma_p \eta$ can be expressed as
\begin{align} \label{eq:p-adic_gamma_hat}
\mtI - \gamma_p \eta \= \prod_{k=1}^3 \Gamma_p\left(\lambda_k\right)\; \defineas \; \wh{\Gamma}_p~,
\end{align}
where $\lambda_k$ are defined as in \sref{sect:frobenius_to_rational}. Notice that if we interpret the equations \eqref{eq:complex_Gamma_hat} and \eqref{eq:varpi_expansion} in the $p$-adic context, then the above relation \eqref{eq:p-adic_gamma_hat} provides us with the $p$-adic analogue of \eqref{eq:complex_Gamma_hat}, and we also have the analogue of the relation \eqref{eq:complex_change_of_basis}. Note also that from \eqref{eq:U_matrix_exponential} and \eqref{eq:alpha_gamma_coefficients}, together with the relation \eqref{eq:p-adic_gamma_hat}, it follows that we can write 
\begin{align}
\mtU_p(\bm 0) \= u_p \Lambda_p \me^{\alpha_p^i \epsilon_i} \, \wh{\Gamma}_p^{-1}~.
\end{align}

\subsection{Fixing the constants and the form of the matrix \texorpdfstring{$\mtU_p(\bm \varphi)$}{U}} \label{sect:form_of_U(varphi)}
\vskip-10pt
The matrix $\mtU_p(\bm \varphi) = \mtE^{-1}(\bm \varphi^p)\mtU_p(\bm 0)\mtE(\bm \varphi)$ is naturally expressed as a matrix of $p$-adically convergent series in variables $\bm \varphi$. However, the series converge only slowly, so it is useful to note \cite{Lauder2004a,Lauder2004b} that $\mtU_p(\bm \varphi)$ can conjecturally be expressed in terms of rational functions. To be specific, expanding the matrix $\mtU_p(\bm \varphi)$ as $p$-adic series, we can conjecturally write it as
\begin{align} \label{eq:U(varphi)_p-adic_expansion}
\mtU_p(\bm \varphi) \= \frac{\cU_p^{(0)}(\bm \varphi)}{S_0(\bm \varphi^p)} + \frac{\cU_p^{(1)}(\bm \varphi)}{S_1(\bm \varphi^p)} p + \frac{\cU_p^{(2)}(\bm \varphi)}{S_2(\bm \varphi^p)} p^2 + \dots~,
\end{align}
where $\cU_p^{(n)}(\bm \varphi)$ are matrices the entries of which are polynomials in $\bm \varphi$ whose coefficients are $p$-adic units. The $S_n(\bm \varphi)$, in this context, are polynomials in $\bm \varphi$.

This form also turns out to be the key to showing that the constants $\alpha_p^i$ and $\gamma_p$ indeed take the form~\eqref{eq:alpha_gamma_coefficients}. This is due to the fact that, at least in every case we have studied, the matrix $\mtU_p(\bm \varphi)$ takes the rational form above for only one set of values of $\alpha_p^i$ and $\gamma_p$. We conjecture that these values of the constant give the correct local zeta functions. This is corroborated by the examples we have studied, and further by the fact that in the one-parameter case this technique reproduces the conjectural values obtained in \cite{Candelas:2021tqt}.

To show that the matrices $\mtU_p(\bm \varphi)$ take this form, we first show that the logarithms appearing in the periods and their derivatives appearing in the matrix $\mtE(\bm \varphi)$ cancel. A straightforward but lengthy computation shows that
\begin{align} \notag
\mtE(\bm \varphi) \= \bm \varphi^{\bm \epsilon} \wt \mtE(\bm \varphi)~,
\end{align}
where the \textit{logarithm-free change-of-basis matrix} $\wt \mtE(\bm \varphi)$ is defined as the matrix we obtain by formally setting the logarithms to zero in the matrix $\mtE(\bm \varphi)$, after evaluating the derivatives appearing in the definition of this matrix.
\begin{align} \label{eq:Log-free_E_definition}
\wt \mtE(\bm \varphi) \defineas \mtE(\bm \varphi) \big|_{\log(\varphi_i) \mapsto 0}~.
\end{align}
This is a matrix of power series in $\bm \varphi$, whose columns are given by the \textit{logarithm-free period vectors}
\begin{align} \label{eq:logarithm_free_period_vectors}
\wt{\vartheta_a \varpi}(\bm \varphi) \defineas (\vartheta_a \varpi)(\bm \varphi) |_{\log \varphi^i \mapsto 0}~, \qquad \wt{\vartheta^a \varpi}(\bm \varphi) \defineas (\vartheta^a \varpi)(\bm \varphi) |_{\log \varphi^i \mapsto 0}~,
\end{align}
where the derivatives are first evaluated before setting $\log \varphi^i \mapsto 0$.

Recalling \eqref{eq:mtU_definition}, the matrix $\mtU_p(\bm \varphi)$  can be expressed as
\begin{align} \label{eq:Umatrix_Expanded_Expression}
\mtU_p(\bm \varphi) \= \wt \mtE(\bm \varphi^p)^{-1} \bm \varphi^{- p \bm \epsilon} \mtU_p(\bm 0) \bm \varphi^{\bm \epsilon} \wt \mtE(\bm \varphi)~.
\end{align}
Using the commutation relation \eqref{eq:U0_Commutation} it then immediately follows that
$\bm \varphi^{-p \bm \epsilon} \mtU_p(\bm 0) = \mtU_p(\bm 0) \bm \varphi^{-\bm \epsilon}$,
which means that we can in fact express the matrix $\mtU_p(\bm \varphi)$ manifestly as a matrix of power series in $\bm \varphi$ by writing it in terms of $\wt \mtE(\bm \varphi)$ as
\begin{align} \label{eq:U_varphi_form}
\mtU_p(\bm \varphi) \=  \wt \mtE(\bm \varphi^p)^{-1} \mtU_p(\bm 0)\wt \mtE(\bm \varphi)~.
\end{align}
In all cases we have studied, the polynomial $S_n(\bm \varphi)$ that gives the denominator of $\mtU_p(\bm \varphi) \mod p^n$, takes the form
\begin{align} \label{eq:U(varphi)_denominator}
S_n(\bm \varphi) \= \Delta(\bm \varphi)^{n-4} \cY(\bm \varphi)^{n-2} \cW(\bm \varphi)~,
\end{align} 
where $\Delta(\bm \varphi)$ gives the (hyper) conifold locus of the family of Calabi--Yau manifolds we are studying and $\cW(\bm \varphi)$ the denominator of the matrix $W^{-1}(\bm \varphi)$ which we will introduce in \eqref{eq:W_matrix_definition}). This factor contains the apparent singularities. The factor $\cY(\bm \varphi)$ represents additional factors in the discriminant that do not correspond to the (hyper) conifold locus nor to the divisors whose intersections give the large complex structure points. This factor contains, for instance, the K-points. Note that, somewhat curiously, when $n \leq 4$, the conifold discriminant $\Delta(\bm \varphi)$ does not appear in the denominator, and $\cU_p(\bm \varphi)$ is well-defined even when $\Delta(\bm \varphi) = 0$. This fact can be exploited to find the zeta function numerator even at the conifold loci, at least for one- and two-parameter manifolds. We will explore this in more detail in connection with the example of the mirror octic manifold in~\sref{sect:mirror_octic}. 

Even though for a large number of cases the denominator is of the form \eqref{eq:U(varphi)_denominator}, it is known that there are cases of (one-parameter) differential operators of Calabi--Yau type (see \cite{Almkvist:2005qoo} for the definition of this notion and examples of such operators) for which the denominator of the corresponding matrix $\mtU_p(\bm \varphi)$ takes on a slightly more general form. This can include, for instance, square roots of polynomials. We expect such cases to rise in connection with multiparameter Picard--Fuchs operators as well, although we are not aware precisely when such cases should arise, or if there is a `niceness' criterion that can be used to eliminate such examples.

Note that the form \eqref{eq:U(varphi)_p-adic_expansion} is not well-defined when $S_n(\bm \varphi) = 0 \text{ mod } p$. In particular, if one tries to compute the polynomial $R_p(X_{\bm \varphi},T)$ using the method described above, one would find that the expression does not in general converge. At the points that satisfy $\Delta(\bm \varphi)\cY(\bm \varphi) = 0 \text{ mod } p$ this is due to the fact that at such points the manifold $X_{\bm \varphi}/\IF_{\! p}$ is singular. Therefore, our assumptions which require that the manifold is a smooth Calabi--Yau threefold no longer hold. However, at the points where the discriminant does not vanish $\text{mod } p$, but $\cW(\bm \varphi) = 0 \text{ mod } p$, the manifold is smooth, and the zeta function can in principle be computed using the methods outlined above. At these the non-convergence is due to the fact that the set of sections \eqref{eq:H^3_basis} that we have chosen are not all linearly independent at such a point, and thus do not form a basis. Thus the convergence can be restored at these points by simply choosing a different set of sections. In the multiparameter case, this is often most conveniently done by choosing a different set of the `inverse' triple intersection numbers $\wh Y_{ijk}$, as we will illustrate in \sref{sect:non-symmetric_split_example}.

\subsection{Evaluating the matrix \texorpdfstring{$\mtU_p(\varphi)$}{U}}
\vskip-10pt
As remarked earlier, the action of the Frobenius map $\Fr_{p^n}$ is properly defined on a $p$-adic cohomology theory, the construction of which involves `lifting' the variety $X_{\bm \varphi}/\IF_{\! p}$ to a variety defined over the $p$-adic integers. The matrix $\mtU_p(\bm \varphi)$ should then be thought of as a matrix of power series with coefficients $p$-adic numbers in $\IQ_p$. To evaluate it for values $\bm \varphi \in \IF_{\! p}^m$, we need to use their Teichmüller representatives $\Teich(\bm \varphi) \in \IZ_p^m$, and evaluate the matrix $\mtU_p(\bm \varphi)$ at these points.

This gives rise to an additional subtlety: for the Teichmüller representatives $\bm \varphi^p = \bm \varphi$, which seems to imply that $\mtU_p(\bm \varphi)$ is a conjugate of a constant matrix. However, this would imply that the characteristic polynomial $R_p(X_{\bm \varphi},T)$ does not vary when we move in the moduli space, which we know to be incorrect. The problem with substituting the Teichmüller representatives directly in the matrix $\mtE(\bm \varphi^p)^{-1}$ lies in the fact that the matrix only converges inside the disk $||\varphi_i||_p<1$, but the Teichmüller representatives (apart from $0$) have $||\varphi_i||_p = 1$.

The correct way to proceed is that we must evaluate the product of matrices as power series in $\bm \varphi$ first, that is, for small $\bm \varphi$. This gives us a matrix $\mtU_p(\bm \varphi)$ that, owing to cancellations, converges in a larger region containing all $\bm \varphi \in \IQ_p^m$ with $||\varphi_i||_p \leq 1$. In this way, we have performed $p$-adic analytic continuation to a region containing the Teichmüller representatives, for which $||\varphi_i||_p = 1$. While the resulting series converge, they do so only slowly. This slow convergence is improved by noting that, as we have seen in \eqref{eq:U(varphi)_p-adic_expansion}, to a specific $p$-adic accuracy the series in $\mtU_p(\bm \varphi)$ can be summed to obtain a rational matrix.

\subsection{Fast inversion of \texorpdfstring{$\mtE(\varphi)$}{the matrix E}} \label{sect:inversion_of_E}
\vskip-10pt
To compute the matrix $\mtU_p(\bm \varphi)$ to high accuracy in practice, we need an efficient way of inverting the matrix $\mtE(\bm \varphi)$. A practical method of doing this is given in \cite{Thorne2019a}.\footnote{The basic idea of this method originates in unpublished work of Duco van Straten, and the details and the practical procedure were worked out in Thorne's thesis \cite{Thorne2019a}.} Using the Griffiths transversiality relations \eqref{eq:Griffiths_transversiality} and the Picard--Fuchs equations, it is straightforward to show that the inner products 
\begin{align} \notag
(\vartheta_a \Omega,\vartheta_b \Omega) = \int_X \vartheta_a \Omega \wedge \vartheta_b \Omega~, \quad (\vartheta^a \Omega,\vartheta^b \Omega) = \int_X \vartheta^a \Omega \wedge \vartheta^b \Omega~, \quad \text{and} \quad (\vartheta_a \Omega,\vartheta^b \Omega) = \int_X \vartheta_a \Omega \wedge \vartheta^b \Omega
\end{align}
satisfy a differential equations whose solutions are rational functions. These products are the components of the matrix
\begin{align} \label{eq:W_matrix_definition}
\mtW(\bm \varphi) \defineas \mtE(\bm \varphi)^T \sigma \mtE(\bm \varphi)~.
\end{align}
Using this matrix, the inverse matrix $\mtE^{-1}(\bm \varphi)$ can be expressed as
\begin{align} \notag
\mtE^{-1}(\bm \varphi) \= \left(\sigma \mtE(\bm \varphi) \mtW^{-1}(\bm \varphi)\right)^T~.
\end{align}
This is convenient to compute in practice because, as a matrix of rational functions, $\mtW(\bm \varphi)$ is easy to invert.
\subsection{Constraints from the Weyl conjectures} \label{sect:estimates}
\vskip-10pt
The functional equation \eqref{eq:Weil_conjectures_functional_equation} satisfied by the zeta function of a Calabi--Yau threefold implies that its numerator $R_p(X_{\bm \varphi},T)$ satisfies the following functional equation:
\begin{align} \notag
R_p(X_{\bm \varphi},T) \= p^{3(h^{12}+1)} \, T^{2(h^{12}+1)} R_p\left(X_{\bm \varphi}, p^{-3} T^{-1} \right)~.
\end{align}
Writing the polynomial $R_p(X_{\bm \varphi},T)$ as 
\begin{align} \notag
R_p(X_{\bm \varphi},T) \= 1 + \sum_{i=1}^{b^3} a_i T^i~,
\end{align}
this relation halves the number of independent parameters $a_i$, imposing the following relations among the coefficients
\begin{align} \label{eq:a_i_coefficients_relations}
a_{2(h^{12}+1)-i} \= p^{3(h^{12}+1-i)} a_{i}~, \qquad a_{2(h^{12}+1)} \= p^{3(h^{12}+1)}~.
\end{align} 
Our aim is to fix the remaining independent coefficients by using the relation \eqref{eq:R(T)_determinant} between the zeta function and the inverse Frobenius map acting on the third cohomology, which fully determines the zeta function of a Calabi--Yau threefold. The relation \eqref{eq:R(T)_determinant} could in principle be used directly to compute the zeta function numerator. However, in practice, finding the determinant of a matrix whose entries are multivariate series is computationally very taxing. In addition, these results would necessarily only give the coefficients $a_i$ up to certain $p$-adic accuracy. To resolve the first issue, it is more convenient to be able to express the independent coefficients $a_i$ as traces of powers of $\mtU_p(\bm \varphi)$. To address the second, using the Riemann hypothesis, we can derive bounds for the norms of $a_i$. This allows us to fix a $p$-adic accuracy to which the series need to be computed in order to obtain exact results for the $a_i$. This is also a crucial piece of information needed to accelerate convergence of the series in $\mtU_p(\bm \varphi)$, as only when working with a fixed $p$-adic accuracy and ignoring any coefficients of higher $p$-adic order, will we find that the \textit{a priori} infinite series appearing as elements of $\mtU_p(\bm \varphi)$ are actually rational functions with $\mtU_p(\bm \varphi)$ taking the form \eqref{eq:U(varphi)_p-adic_expansion}, which allows computing its values exactly to a given $p$-adic accuracy.

It is possible to find the coefficients of the characteristic polynomial by considering the following standard identities
\begin{align} \notag
\det \left(\mtI-T \mtU_p(\bm \varphi) \right) \= \exp \bigg( \Tr \log \big(\mtI- T \mtU_p(\bm \varphi)\big) \bigg) \= \exp \left(-\sum_{n=1}^{\infty} \frac{T^n}{n} \Tr \, \big(\mtU_p(\bm \varphi)^n\big)\right)~.
\end{align}
Expanding this in powers of $T$ gives each coefficient $a_i$ as a function of traces of powers of $\mtU_p(\bm \varphi)$. The entries of $\mtU_p(\bm \varphi)$ are series in $\bm \varphi$, and $\bm \varphi$ should be evaluated at the Teichm\"uller lifts ${\Teich(\bm \varphi) = (\Teich(\varphi_1),\dots,\Teich(\varphi_m))}$.

It is implicit in the Weil conjectures that the $p$-adic integers $a_i$ are also rational integers. This being so we can bound them in the following way: By the Riemann hypothesis, if we diagonalise the matrix $\mtU_p(\bm \varphi)$, the eigenvalues will be algebraic integers $\lambda_j$ of absolute value $|\lambda_j| = p^{3/2}$. The coefficients of the characteristic polynomial $R_p(X_{\bm \varphi},T)$ can be expressed as symmetric polynomials of these eigenvalues.
\begin{align} \notag
a_i \= (-1)^{i} \, \sigma_i(\lambda_1,\dots,\lambda_{2m+2})~,
\end{align}
where $\sigma_i$ is the $i$'th elementary symmetric polynomial, and no sum over $i$ is implied. From the Riemann hypothesis it immediately follows that every monomial in $\sigma_i$ has an absolute value of $p^{3i/2}$, whereas the number of monomials is $\binom{2m+2}{i}$. This gives a simple bound on the magnitude of the constant~$a_i$
\begin{align} \label{eq:a_i_bound}
|a_i| < \binom{2m+2}{i} \, p^{3i/2}~.
\end{align}
Thus, if we are interested in computing $a_i$, we only need to compute it modulo $p^n$, where $n$ is an integer such that $\binom{2m+2}{i} \, p^{3i/2} \leq p^n$. This gives the $p$-adic accuracy $n$ mentioned in \sref{sect:form_of_U(varphi)} to which we need to know $\mtU_p(\bm \varphi)$.
\newpage
\section{Examples} \label{sect:Examples}
\vskip-10pt
To illustrate the methods developed here in concrete cases, we study three different families of multiparameter Calabi--Yau manifolds: the two-parameter mirror manifold of the octic hypersurface in the weighted projective space $\IP^{(1,1,2,2,2)}$, the $S_5$ symmetric members of the five-parameter family of mirror Hulek--Verrill manifolds, and the non-symmetric split of the quintic threefold corresponding to the configuration matrix
\begin{align} \notag
\cicy{\IP^1\\\IP^4}{1 & 1\\4 & 1}~.
\end{align}
The first two examples act as checks of the methods presented here, as the zeta functions of these geometries have been studied using entirely different methods in \cite{Kadir:2004zb,Hulek2005}. 

In \cite{Kadir:2004zb}, Gauss sums were used to compute the zeta functions of the mirror octic manifold. Compared to the method presented in this paper, this technique benefits from more unified and simple treatment of singularities. However, as a drawback, it is computationally heavy, which is why we are able to easily extend the results of \cite{Kadir:2004zb} to greater primes. For the five-parameter family of Hulek--Verrill manifolds, the number of points on the manifolds over the finite fields $\IF_{\! p}$ was computed explicitly in \cite{Hulek2005}. This gives a simple closed-form formula, which can be used to compute the zeta function numerator, at least assuming the factorisation \eqref{eq:HV_R10_Factorisation}. However, this method relies on detailed knowledge of the toric geometry of the manifold, whereas the knowledge of the periods and the discriminant is enough for our method. In addition, it is not completely straightforward to generalise the direct counting technique for manifolds outside of the $S_5$ symmetric manifolds, where the factorisation \eqref{eq:HV_R10_Factorisation} no longer holds, and thus point counting over finite fields $\IF_{\! p^n}$ for $n>1$ is needed.

Treating these three examples also requires use of different strategies due to the computational complexity of various matrices involved, and the $p$-adic accuracy required to obtain exact results numerically. It is possible to study the two-parameter examples by expanding the periods as series in two parameters, which makes it easy to obtain results for any values of the parameters. However, for a larger number of parameters this becomes quickly computationally too cumbersome to be practical. These cases can be treated by studying lines inside of the multiparameter moduli space. This allows expressing the periods and their derivatives as series in one parameter, greatly reducing the computational complexity of the problem. By considering a suitable line, it is in principle possible to obtain the zeta function at any point in this~way.  
\subsection{A two-parameter example: the mirror octic} \label{sect:mirror_octic}
\vskip-10pt
We begin with a study of the mirror octic. Mirror symmetry for this manifold has been studied in detail in \cite{Candelas:1993dm}, and the zeta function has been computed in some cases in \cite{Kadir:2004zb}. This allows us to make highly non-trivial checks of the method presented above by comparing the results to those obtained~in~\cite{Kadir:2004zb}. 

The family of octic Calabi--Yau manifolds is given by resolving the singularities of varieties defined as degree-8 hypersurfaces in the weighted projective space $\IP^{(1,1,2,2,2)}$. 

The Hodge diamond of the family of octics is given by
\begin{align} \notag
h^{p,q} \= \begin{matrix}
& & & 1 & & & \\
& & 0 &  & 0 & & \\
& 0 &  & 2  &  & 0 & \\ 
1 &  & 86 &   & 86 &  & 1 \\ 
& 0 &  & 2  &  & 0 & \\
& & 0 &  & 0 & & \\
& & & 1 & & &  
\end{matrix}~.
\end{align}
Thus the mirror of this family gives a two-parameter model. These manifolds have been explicitly constructed in \cite{Candelas:1993dm}, where they were identified with the manifolds given by $\{P=0\}/\IZ_4^3$ with
\begin{align} \notag
P \= x_1^8 + x_2^8 + x_3^4 + x_4^4 + x_5^4 - 2 \phi x_1^4 x_2^4 - 8 \psi x_1 x_2 x_3 x_4 x_5 \= 0~,
\end{align}
and with the group $\IZ_4^3$ is realised as the group with generators
\begin{align} \notag
(s_1,s_2,s_3,s_4,s_5) \= (0,3,1,0,0),~(0,3,0,1,0),~(0,3,0,0,1)~,
\end{align}
which act on the coordinates as
\begin{align} \notag
(x_1,x_2,x_3,x_4,x_5) \mapsto (\alpha^{s_1} x_1, \alpha^{s_2} x_2, \alpha^{s_3} x_3, \alpha^{s_4} x_4, \alpha^{s_5} x_5)~,
\end{align}
with $\alpha$ a non-trivial eight root of unity.

The complex structure moduli space $\cM_{\IC S}$ of mirror octics can be identified as 
\begin{align} \notag
\cM_{\IC S} \= \text{Spec} \left[ \frac{\IC[\wh x, \wh y, \wh z]}{\langle \wh x \wh z-\wh y^2 \rangle} \right]~,
\end{align}
where the coordinates $\wh x$, $\wh y$, and $\wh z$ can be related to $\psi$ and $\phi$ by
\begin{align} \notag
\wh x \= \psi^8~, \qquad \wh y \= \psi^4 \phi~, \qquad \wh z \= \phi^2~.
\end{align}
The natural coordinates to use near the large complex structure point are 
\begin{align} \label{eq:LCS_coordinates_mirror_octic}
\varphi_1 \defineas \frac{1}{2^2 \phi^2}~, \qquad \varphi_2 \defineas -\frac{\phi}{2^{11} \psi^4}~,
\end{align}
in terms of which the large complex structure point is located as $\varphi_1=\varphi_2=0$.

The independent triple intersection numbers $Y_{ijk}$ are given by
\begin{align} \notag
Y_{111} \= 0~, \quad Y_{112} \= 0~, \quad Y_{122} \= 4~, \quad Y_{222} \= 8~,
\end{align}
where the index $1$ refers to the linear system of divisors inherited from the vanishing loci of degree-1 polynomials generated by the $x_1$ and $x_2$ in the ambient space, whereas the index $2$ refers to the linear system generated by the degree-$2$ polynomials. 

We take the quantities $\wh Y^{ijk}$ to be symmetric in the first two indices, with the independent values given~by
\begin{align} \notag
\wh Y^{121} \= - \frac{1}{4}~, \quad \wh Y^{221} \= \frac{1}{4}~, \quad \wh Y^{122} \= \frac{1}{8}~, \quad \wh Y^{111} \= 0~,  \quad \wh Y^{112} \= 0~, \quad \wh Y^{222} \= 0~.
\end{align}
The conifold locus $\Delta$ and the additional factor $\cY$ of the discriminant are given by
\begin{align} \notag
\Delta \= 1-2^{9} \varphi_2+2^{16}(1- 2^2 \varphi_1)\varphi_2^2~, \qquad \cY \= 2^2 \varphi_1-1~.
\end{align}
\subsubsection*{The Picard--Fuchs system}
\vskip-5pt
In the coordinates adapted to the large complex structure point, the fundamental period is
\begin{align} \notag
\varpi^0(\varphi_1,\varphi_2) \= \sum_{r,s=0}^\infty \frac{(8r+4s)!}{\left((2r+s)!\right)^3 (r!)^2 s!} \varphi_1^r \varphi_2^{2r+s}~.
\end{align}
In these coordinates, two of the differential operators giving the Picard--Fuchs system found in \cite{Candelas:1993dm} can be written as
\begin{align} \notag
\begin{split}
\cL_1 &\= (1-2^8 \varphi_2) \theta_2^3 - 2 \theta_1 \theta_2^2 - 3 \cdot 2^7 \varphi_2 \theta_2^2 - 11 \cdot 2^4 \varphi_2 \theta_2 - 3 \cdot 2^3 \varphi_2 ~,\\[5pt]
\cL_2 &\= (1-2^2 \varphi_1) \theta_1^2 + 2^2 \varphi_1 \theta_1 \theta_2 - \varphi_1 \theta_2^2  - 2 \varphi_1 \theta_1+ \varphi_1 \theta_2~.
\end{split}
\end{align}
Alternatively, a straightforward computation gives the Picard--Fuchs equations formulated as linear relations between the sections $\vartheta_a \Omega, \vartheta^a \Omega$ of the form
\begin{align} \notag
\begin{split}
\theta_1 \vartheta^0 \Omega &\= \frac{1}{\Delta \cY \cW} \sum_{a=0}^2 \Big(P^a(\varphi_1,\varphi_2) \, \vartheta_a \Omega + P_a(\varphi_1,\varphi_2) \, \vartheta^a \Omega \Big)~,\\
\theta_2 \vartheta^0 \Omega &\= \frac{1}{\Delta \cY \cW}\sum_{a=0}^2 \Big( Q^a(\varphi_1,\varphi_2) \, \vartheta_a \Omega +Q_a(\varphi_1,\varphi_2) \, \vartheta^a \Omega \Big)~,
\end{split}
\end{align}
where $P^a$, $P_a$, $Q^a$, and $Q_a$ are polynomials of multi-degree $(\text{deg}_{\varphi_1},\text{deg}_{\varphi_2}) \leq (3,3)$, which we will not display explicitly. We just note that all of these vanish at $\varphi_1 = \varphi_2 =0$, satisfying the asymptotic form \eqref{eq:B_matrix_asymptotics} of the connection matrix $\mtB(\varphi_1,\varphi_2)$. The factor $\cW = 2 \varphi_1+2^8 \varphi_2-1$ appearing in the denominator is the locus of apparent singularities for this choice of sections. We will see below that this is indeed the denominator of the matrix $\mtW^{-1}$.

To obtain series expressions for the periods around the large complex structure point, we use series ans\"atze for the functions $f^I$ appearing in the expression \eqref{eq:period_vector_Frob_basis_expansion} for the period vector:
\begin{align} \notag
f^{I} \= \sum_{\mu, \nu=0}^{\infty} c^{I}_{\mu \nu} \varphi_1^{\mu} \varphi_2^\nu~.
\end{align}
Here $I$ can be a one-, two, or three-component index. Substituting the Frobenius basis periods $\varpi_a, \varpi^a$ expressed in terms of these series into the first two differential equations, we obtain recurrence relations for the coefficients $c_{\mu \nu}^I$, which can be used to quickly compute the series $f^I$ to high order. To get a better idea of how this works in practice, let us consider the fundamental period. The recurrence relations can in this case be expressed as
\begin{align} \label{eq:c^0_recurrence}
c^0_{\mu \nu} \= \frac{(2\mu - \nu -2)(2\mu - \nu -1)}{\mu^2} \, c^0_{\mu-1,\nu}~, \qquad c^0_{0 \nu} \= \frac{8(2\nu-1)(4\nu-3)(4\nu-1)}{\nu^3} \, c^0_{0,\nu-1}~.
\end{align}
Together with the boundary conditions
\begin{align} \notag
c^0_{00} \= 1~, \qquad c^0_{\mu \nu} \= 0 \quad \text{if } \mu<0 \text{ or } \nu<0~,
\end{align}
these suffice to solve for all coefficients $c^0_{\mu \nu}$. The coefficients appearing in the other periods satisfy similar, albeit more lengthy, recurrence relations, which we refrain from displaying here.
\subsubsection*{Tme matrix $\mtU_p(\varphi_1,\varphi_2)$ and the zeta function}
\vskip-5pt
Having obtained the periods via recurrence relations, we only need to find the matrix $\mtW^{-1}(\varphi_1,\varphi_2)$ defined in \eqref{eq:W_matrix_definition} to be able to compute the matrix $\mtU_p(\varphi_1,\varphi_2)$. In the present case, it is not difficult to compute the required inner products of derivatives of periods to show that 
\begin{align} \notag
\mtW^{-1} = \smallfrac{(2\pi\ii)^3}{2048}\left(
\begin{smallarray}{cccccc}
 0 & \varphi_2 \mathcal{E} & \frac{\mathcal{A} \varphi_1 \varphi_2}{3 \mathcal{W}} & -\frac{\varphi _1 \mathcal{B}}{12 \mathcal{W}} & \frac{32 \mathcal{C} \varphi_2}{3 \mathcal{W}} & -\frac{\mathcal{D} \mathcal{E}}{12 \mathcal{W}} \\[7pt]
 -\varphi_2 \mathcal{E} & 0 & -\frac{8 \varphi_2 \mathcal{E}}{3} & \frac{1}{12} \left(1{-}256 \varphi_2\right) \mathcal{E} & -\frac{128 \varphi_2 \mathcal{E}}{3} & 0 \\[7pt]
 -\frac{\mathcal{A} \varphi_1 \varphi_2}{3 \mathcal{W}} & \frac{8 \varphi_2 \mathcal{E}}{3} & 0 & \frac{1}{6} \varphi_1 \left(1{-}512 \varphi_2\right) & ~~\frac{1}{12}{-}\frac{64}{3}\left(4 \varphi_1{+}1\right) \varphi_2~~ & 0 \\[7pt]
 \frac{\varphi_1 \mathcal{B}}{12 \mathcal{W}} & \frac{1}{12} \left(1{-}256 \varphi_2\right) \mathcal{E} & \frac{1}{6} \varphi_1 \left(1{-}512 \varphi_2\right) & 0 & 0 & 0
   \\[7pt]
 -\frac{32 \mathcal{C} \varphi_2}{3 \mathcal{W}} & \frac{128 \varphi_2 \mathcal{E}}{3} & ~\frac{64}{3} \left(4 \varphi_1{+}1\right) \varphi_2{-}\frac{1}{12}~ & 0 & 0 & 0 \\[7pt]
 \frac{\mathcal{D} \mathcal{E}}{12 \mathcal{W}} & 0 & 0 & 0 & 0 & 0 \\
\end{smallarray}
\right),
\end{align}
where we have denoted
\begin{align} \notag
\begin{array}{ll}
\cA \defineas 12 \varphi_1-256 \left(27-44 \varphi_1\right) \varphi_2+5~,& \cB \defineas 1-128 \varphi _2 \left(4 \varphi_1-256 \left(9-20 \varphi _1\right) \varphi _2+5\right)~,\\[5pt]
\cC \defineas 8 \varphi _1+256 \left(6 \varphi _1 \left(4 \varphi _1-3\right)+1\right) \varphi _2-1~,& \cD \defineas 512 \varphi_2 \left(128 \left(4 \varphi _1-1\right) \varphi_2+1\right)-1~,\\[5pt]
\cE \defineas 1-4 \varphi_1~,& \cW = 2 \varphi_1+2^8 \varphi_2-1~.
\end{array}
\end{align}
With this expression, computing the inverse matrix $\mtE^{-1}(\varphi_1^p,\varphi_2^p)$ is simple. Setting the coefficients
\begin{align} \notag
\alpha_p^i \= 0~, \qquad \gamma_p \= \chi(\wt X_{\bm t}) \zeta_p(3) \= -168 \zeta_p(3)~, 
\end{align}
it is also easy to compute the matrix $\mtU_p(\varphi_1,\varphi_2)$ for the first few primes $p$. As expected, we can check that the series appearing in this matrix indeed seem to converge to a rational function with the denominator \eqref{eq:U(varphi)_p-adic_expansion}. This is in practice seen as the coefficients of $\mtU_p(\varphi_1,\varphi_2)$ multiplied by \eqref{eq:U(varphi)_denominator} becoming finite polynomials.
\subsubsection*{The choice of coordinates}
\vskip-5pt
In many cases, it actually turns out that the numerical computations become slightly simpler when different coordinates are used --- which can result in significant reduction in computation times. We use the example of the mirror octic to address briefly the question of coordinate transformations. For simplicity, we restrict to a rescaling of coordinates, although analogous argument apply to other similar changes of coordinates as well. Let us therefore use the coordinates $\wh \varphi_i$ given~by
\begin{align} \notag
r_i \wh \varphi_i \= \varphi_i~, \qquad r_i \in \IQ
\end{align}
Under this transformation, the matrix $\mtE(\bm \varphi)$ can be written in terms of the new coordinates $\wh \varphi_i$ as
\begin{align} \notag
\mtE(\bm \varphi) \=  \wh{\bm \varphi}^{\bm \epsilon} \bm r^{\bm \epsilon} \, \wt \mtE(\wh{\bm \varphi})~,
\end{align}
where $\wt \mtE(\wh{\bm \varphi})$ is the logarithm-free matrix introduced in \eqref{eq:Log-free_E_definition}, where we have substituted in $\bm \varphi \mapsto \wh{\bm \varphi}$. The expression $\bm r^{\bm \epsilon}$ refers to the product $r_1^{\epsilon_1} \cdots r_m^{\epsilon_m}$. With this, the expression \eqref{eq:Umatrix_Expanded_Expression} for the matrix $\mtU_p(\bm \varphi)$ gives
\begin{align} \notag
\mtU_p(\bm \varphi) \= \wt \mtE(\wh{\bm \varphi}^p)^{-1} \bm r^{- \bm \epsilon} \wh{\bm \varphi}^{-\bm p \bm \epsilon} \, \mtU_p(\bm 0) \wh{\bm \varphi}^{\bm \epsilon} \bm r^{\bm \epsilon} \, \wt \mtE(\wh{\bm \varphi}) \= \wt \mtE(\wh{\bm \varphi}^p)^{-1} \wh{\mtU_p}(\bm 0) \, \wt \mtE(\wh{\bm \varphi})~.
\end{align}
To obtain the second equality, we have used the commutation property \eqref{eq:U0_Commutation} of the matrices $\mtU_p(\bm 0)$ and~$\epsilon_i$. Note that here the coefficients $r_i$ giving the scaling are not raised to the $p$'th power as the Frobenius map does not act on constants. This is essentially the reason we obtain a different matrix $\wh \mtU_p(\bm 0)$ when using the rescaled coordinates. From this expression we identify the matrix $\wh \mtU_p(\bm 0)$ associated to the coordinates $\wh \varphi_i$ as
\begin{align} \notag
\wh \mtU_p(\bm 0) \= \bm r^{-\bm \epsilon} \mtU_p(\bm 0) \bm r^{\bm \epsilon} \= \Lambda_p \bm r^{\frac{p-1}{p} \bm \epsilon} (\mtI - \gamma_p \eta) \= \Lambda_p \exp \left(\frac{p-1}{p} \log r_i \epsilon_i\right) \, (\mtI - \gamma_p \eta)~.
\end{align}
Comparing this to \eqref{eq:U_matrix_exponential}, identifies the coefficients $\alpha^i$ in this case as
\begin{align} \notag
\alpha_p^i \= \frac{p-1}{p} \log_p r^i~,
\end{align}
where we can use the $p$-adic Iwasawa logarithm (see appendix \ref{app:p-adic_numbers}) as this logarithm is based on the same series as the ordinary complex logarithm.\footnote{An astute reader might be puzzled by an issue of convergence: we wish to ultimately evaluate the logarithm at $p$-adic integers whose $p$-adic norm is outside of the region of convergence of the series corresponding to the Iwasawa logarithm. However, before lifting the series to $\IQ_p$, one can use the identities satisfied by the ordinary logarithm to write $\log \varphi = \log(\varphi^{p-1})/(p-1)$. This will converge for $p$-adic integers, and coincides with the definition of the Iwasawa logarithm.}
To make this discussion concrete, let us take, instead of the coordinates in \eqref{eq:LCS_coordinates_mirror_octic}, the following rescaled coordinates:
\begin{align} \label{eq:rescaled_coordinates}
\wh \varphi_1 \defineas \frac{1}{\phi^2} \= 2^2 \varphi_1~, \qquad \wh \varphi_2 \defineas \frac{\phi}{\psi^4} \= -2^{11} \varphi_2~.
\end{align}
Then the coefficients $\alpha_p^i$ should be, according to our previous discussion, given by
\begin{align} \label{eq:alpha_i_new}
\alpha_p^1 \= -2 \, \frac{p-1}{p} \log_p(2)~, \qquad \alpha_p^2 \= -11 \, \frac{p-1}{p} \log_p(2)~.
\end{align}
It is easy to check that with these coefficients, the resulting matrix $\mtU_p(\wt{\bm \varphi})$ takes indeed the required rational form. A numerical computation also reveals that the choice of these coefficients is the only one that results in such a rational matrix: Imposing the requirement that the entries of $\mtU_p(\varphi_1,\varphi_2)$ be rational functions with $p$-adic integers as coefficients of polynomials in the numerator and denominator, gives an overconstrained system, which can be used to solve the prime-dependent constants $\alpha_p^1$ and $\alpha_p^2$ to appropriate $p$-adic accuracies. From \eqref{eq:a_i_bound}, it follows that we need to compute $\mtU_p(\varphi_1,\varphi_2)$ to $p$-adic accuracy $p^6$, which in turn implies, that $\alpha_p^i$ need to be computed up to order $p^5$ or higher, although it is easily possible to compute these to higher accuracies as well. We gather the values computed in this way in \tref{tab:mirror_octic_alpha_gamma}. It is then an easy exercise to verify that these values agree with those in \eqref{eq:alpha_i_new}, up to the specified $p$-adic accuracy.

Using these values to compute the matrices $\mtU_p(\varphi_1,\varphi_2)$, one finds that these take exactly the form \eqref{eq:U(varphi)_p-adic_expansion}, which allows computing the independent coefficients $a_1$, $a_2$, and $a_3$ appearing in the zeta function numerators $R_p(X_{(\varphi_1,\varphi_2)},T)$.

\begin{table}[H]
	\begin{center}
		\renewcommand{\arraystretch}{1.3}
		\begin{tabular}{|l|l|l|}
			\hline
			$p$ & \hfil $\alpha_p^1$ 		& \hfil $\alpha_p^2$ 					\\ \hline \hline
			7 	& $161955+\cO(7^7)$ 	    & $478981+\cO(7^7)$  			\\[2pt] \hline
			11 	& $18516114+\cO(11^7)$ 	&  $4402772 +\cO(11^7)$ 	\\[2pt] \hline
			13  & $24288843 + \cO(13^7)$  &  $39465861 + \cO(13^7)$  	\\[2pt] \hline
			17  & $201574686 + \cO(17^7)$  &  $287983427 + \cO(17^7)$  	\\[2pt] \hline   
		\end{tabular}
		\vskip10pt
		\capt{5.3in}{tab:mirror_octic_alpha_gamma}{The values of the prime-dependent constants $\alpha_p^1$ and $\alpha_p^2$ appearing in the matrix $\mtU_p(0,0)$ when using coordinates \eqref{eq:rescaled_coordinates} for the cases $p=7,11,13,17$. These values agree with the values \eqref{eq:alpha_i_new} in terms of Iwasawa logarithms.}
	\end{center}
    \vskip-20pt
\end{table}

We have shown, in this example, that a non-zero $\alpha$ can be absorbed by a rescaling of the coordinates. Note however that this scale of the coordinates is fixed implicitly by the mirror map procedure, as in \eqref{eq:mirror_map}. A change in the scale of the parameters $\varphi_i$ corresponds to a shift in $t_i$:
\begin{align} \notag
\varphi^i \to \lambda \varphi^i~, \quad \text{corresponds to} \quad t^i \mapsto t^i + \frac{\log \lambda}{2\pi \ii}~, \quad \text{and so} \quad  q_i \defineas \exp(2 \pi \ii t^i) \mapsto \lambda q_i~,
\end{align}
and this would affect the calculation of instanton numbers that are independently known. In this example, we deliberately chose `incorrect' parameters and the non-zero value for $\alpha_p^i$ has corrected for this choice. We conjecture that with the correct choice of parameters, $\alpha_p^i$ are always zero, but we do not know this to be the case.
\subsubsection*{Conifold singularities}
\vskip-5pt
At conifold points, we expect, analogously to the observations made in \cite{Candelas:2021tqt}, that one of the roots of $R_p(X,T)$ vanishes, resulting in a degree five polynomial, which we expect further to contain a linear factor so that it takes the form
\begin{align} \label{eq:R_mirror_octic_conifold}
R_p(X,T) \= (1-\chi_p p T) (1- u_1 T + u_2 p T^2 - u_1 p^3 T^3 + p^6 T^4)~,
\end{align}
where $\chi_p$ is a character taking values $\chi_p = \pm 1$. Assuming that the polynomial takes this form, we can completely determine it by computing the coefficients $u_1$ and $u_2$, and the value of the character~$\chi_p$. The roots of the quartic factor of the polynomial $R_p(X,T)$ are expected to have the norm $p^{3/2}$ (see for instance \cite{Lauder2011a,Kadir:2004zb}). Therefore, like in \sref{sect:estimates}, we have the following bounds for $u_1$ and $u_2$:
\begin{align} \notag
|u_1| < 4p^{3/2}~, \qquad |u_2 p| < 6 p^{3}~.
\end{align}
This means that for primes $p \geq 7$, it is enough to compute these values modulo $p^4$. Recalling the form \eqref{eq:U(varphi)_denominator} of the denominator of the matrix $\mtU_p(\varphi_1,\varphi_2)$, we see that working modulo $p^4$, the factor $\Delta(\varphi_1,\varphi_2)$ corresponding to the conifold locus does not appear in the denominator. Therefore the matrix $\mtU_p(\varphi_1,\varphi_2) \mod p^4$ is well-defined even at the conifold locus, and we can use it to (conjecturally) evaluate the coefficients $u_1$ and $u_2$ appearing in the polynomial $R_p(X,T)$. To fix the remaining unknown, the value of the character $\chi_p$, we can compute the coefficient of $T^3$ in $R_p(X,T)$. Evaluating modulo $p^4$ is not enough to compute this, without using the form \eqref{eq:R_mirror_octic_conifold}, but using this form we know that this coefficient should be given by
\begin{align} \notag
-p^3 u_1 + p^2 u_2 \chi_p~.
\end{align}
The requirement that this agrees with the coefficient computed from the matrix $\mtU_p(\varphi_1,\varphi_2)$ modulo $p^4$ can be usually used to fix the value of $\chi_p$, with the only possible exceptions being the cases where $p^2 \mid u_2$. We display some numerators $R_p(X,T)$ for $p=11$, for both conifolds and smooth manifolds in \tref{tab:mirror_octic_numerators}. These values agree with those computed in \cite{Kadir:2004zb} using Gauss sums.

Note that the presence of singularities that are not of conifold type is reflected in the factor $\cY(\bm \varphi)$ in the conjectural expression \eqref{eq:U(varphi)_denominator} for the denominator $S_n(\bm \varphi)$ having non-trivial zeros. Since this factor appears with power $n-2$, it necessarily appears in the denominator $S_n(\bm \varphi)$ for the $p$-adic accuracies needed to compute the polynomials $R_p(X,T)$. Therefore the above technique cannot be used to evaluate the zeta function at these singularities. \vskip-5pt
\tablepreambleOctic{11}
\str (3,2) & conifold & -(p T-1) \left(p^6 T^4-40 p^3 T^3+9 p T^2-40 T+1\right)
 \tabularnewline[.5pt] \hline 
\str (3,3) & smooth & p^9 T^6+6 p^6 T^5+83 p^4 T^4+68 p^2 T^3+83 p T^2+6 T+1
 \tabularnewline[.5pt] \hline 
\str (3,6) & conifold & (p T+1) \left(p^6 T^4+32 p^3 T^3-80 p T^2+32 T+1\right)
 \tabularnewline[.5pt] \hline 
\str (4,5) & smooth & p^9 T^6+10 p^6 T^5+9 p^5 T^4+20 p^3 T^3+9 p^2 T^2+10 T+1
 \tabularnewline[.5pt] \hline 
\str (4,6) & smooth & p^9 T^6-2 p^6 T^5+235 p^4 T^4-452 p^2 T^3+235 p T^2-2 p T+1
 \tabularnewline[.5pt] \hline 
\str (5,5) & conifold & - (p T-1) \left(p^6 T^4+2 p^4 T^3+41 p T^2+2 p T+1\right)
 \tabularnewline[.5pt] \hline 
\str (5,8) & smooth & p^9 T^6-38 p^6 T^5+19 p^4 T^4+124 p^2 T^3+19 p T^2-38 T+1
 \tabularnewline[.5pt] \hline 
\tablepostamble
\vskip-40pt
\begin{table}[H]
\begin{center}
\capt{5.9in}{tab:mirror_octic_numerators}{Some denominators of the zeta function $\zeta_{11}$ of the mirror octic manifold with $p=11$ for various values of the coordinates $\varphi_1$ and $\varphi_2$. We have displayed both smooth and conifolds points.}
\end{center}
\vskip-30pt
\end{table}
\subsection{A five-parameter example: the Hulek--Verrill manifold}
\vskip-10pt
As an example of a case where it is beneficial to study zeta functions on complex lines in the moduli space, consider the five-parameter Hulek--Verrill manifolds, focusing on the $S_5$-symmetric complex lines on the patch $\varphi^0 = 1$ with $\bm \varphi = (\varphi, \varphi, \varphi, \varphi, \varphi)$, $\varphi \in \IC$. This example has been studied in the context of supersymmetric flux vacua by the present authors in \cite{Candelas:2023yrg}.

The five-parameter Hulek--Verrill manifolds \cite{Hulek2005} are mirrors to the complete intersection Calabi--Yau described by the CICY matrix
\begin{equation}\notag
\cicy{\IP^1\\\IP^1\\\IP^1\\\IP^1\\\IP^1}{1 & 1\\1 & 1\\1 & 1\\ 1 & 1\\ 1 & 1}_{\chi\,=-80~}.
\end{equation}
These manifolds $\text{HV}_{\bm \varphi}$ can be realised as the toric compactification of the hypersurface in $\IT^{5}=\IP^5 \setminus \{\prod_{\mu = 0}^5 X_\mu = 0\}$ given by the vanishing of the two polynomials
\begin{equation}\label{eq:HV_TwoPolynomials} 
P^1(\mathbf{X}) \= \sum_{\mu=0}^5 X_\mu~, \hskip30pt P^2(\mathbf{X};\bm{\varphi}) \= \sum_{\mu=0}^5 \frac{\varphi^\mu}{X_\mu}~.
\end{equation}
The six $\varphi^{\mu}$ furnish projective coordinates for the complex structure moduli space $\cM_{\IC S}$ of this manifold. From the defining equations \eqref{eq:HV_TwoPolynomials}, it is clear that interchanging the complex structure parameters $\varphi^\mu$ gives a biholomorphic manifold. Due to the symmetry, we can, without loss of generality, work exclusively in the patch $\varphi^{0}=1$ in which the five remaining $\varphi^{i}$ are the affine coordinates of $\cM_{\IC S}$.

The manifolds obtained in this way are smooth outside the conifold locus $\Delta = 0$ with
\begin{align} \label{eq:discriminant_HV}
\Delta \= \prod_{\eta_i \in \{ \pm 1 \}} \Big(\sqrt{\varphi^0} + \eta_1 \sqrt{\varphi^1} + \eta_2 \sqrt{\varphi^2} + \eta_3 \sqrt{\varphi^3} + \eta_4 \sqrt{\varphi^4} + \eta_5 \sqrt{\varphi^5}\Big)~,
\end{align}
which implies that
\begin{align} \notag
\cY \= 1~.
\end{align}
The triple intersection numbers $Y_{ijk}$ and the other topological quantities $Y_{abc}$ of the mirror Hulek--Verrill manifolds can be computed from their description as complete intersection varieties, and are given by
\begin{equation}\label{eq:Yijk}
Y_{ijk}\=\begin{cases}2\qquad i,j,k \text{ distinct.}\\ 0\qquad \text{otherwise,}\end{cases}\qquad Y_{ij0} \= 0~, \qquad Y_{i00} \= -2~, \qquad  Y_{000} \= 240 \frac{\zeta(3)}{(2\pi \ii)^3}~.
\end{equation}
The fundamental period is given by
\begin{align} \label{eq:Fundamental_Period_Series}
\varpi^0 \= \sum_{p_i=0}^{\infty} \left(\frac{(p_1+\cdots+p_5)!}{p_1! \cdots p_5!} \right)^{2} \varphi_1^{p_1} \cdots \varphi_5^{p_5}~,
\end{align}
The other periods are derivable from this by usual methods. In principle, the derivatives of the periods can then be computed as series in the five parameters $\varphi^i$, after which specialising to the symmetric case $\varphi^i = \varphi$ would give the expressions for periods and their derivatives the following computations require. However, this would mean working to high order with five-parameter series, which is computationally very expensive. Luckily, there is an alternative method which utilises recurrence relations to directly find univariate series expressions for the derivatives of the periods. We explain this in detail in appendix \ref{app:univariate_series}.

Recalling the values \eqref{eq:Yijk} that the triple intersection numbers $Y_{ijk}$ take, we can choose the quantities $\wh Y^{ijk}$ so that the derivative basis has the natural $S_5$ symmetry:
\begin{align} \label{eq:Y^{ijk}}
\wh Y^{iii} \= \wh Y^{iji} \= \wh Y^{jii} \= -\frac{1}{24}~, \qquad \wh Y^{ijk} \= \frac{1}{24}~,
\end{align}
On this line the discriminant \eqref{eq:discriminant_HV} becomes
\begin{align} \notag
\Delta_5 \= (1-\varphi)^{10}(1-9\varphi)^{5}(1-25\varphi)~.
\end{align}
However, for the purposes of computing the zeta function on the line, we do not need to account for the multiplicities, so we can here take the discriminant to be
\begin{align} \notag
\Delta = (1-\varphi)(1-9\varphi)(1-25\varphi)~.
\end{align}

\subsubsection*{Inversion of $\mtE(\varphi)$}
\vskip-5pt
To compute the inverse matrix $\mtE^{-1}(\varphi)$, we need the inner products $(\vartheta_a \Omega,\vartheta_b \Omega)$, $(\vartheta_a \Omega,\vartheta^b \Omega)$, and $(\vartheta^a \Omega,\vartheta^b \Omega)$. Due to the symmetries, there are only five independent such inner products that do not vanish:
\begin{align} \notag
\begin{split}
(\vartheta_{0} \Omega, \vartheta^0 \Omega) &\= - \frac{1}{(2\pi\ii)^3} \frac{5 (18 \varphi -1)}{12 \Delta}~, \qquad (\vartheta_{i} \Omega, \vartheta^i \Omega) \= \frac{1}{(2\pi\ii)^3} \frac{18 \varphi -1}{2 \Delta}~,\\[5pt]
(\vartheta_{i} \Omega, \vartheta^j \Omega) &\= - \frac{1}{(2\pi\ii)^3} \frac{7 \varphi }{2 \Delta}~, \qquad \qquad \quad \hskip2pt (\vartheta^{i} \Omega, \vartheta^0 \Omega) \= \frac{1}{(2\pi\ii)^3} \frac{1215 \varphi ^4-415 \varphi ^3+16 \varphi ^2}{72 \Delta^2}~,\\[5pt]
\qquad (\vartheta_{i} \Omega, \vartheta^0 \Omega) &\= - \frac{1}{(2\pi\ii)^3} \frac{1800 \varphi ^4+1915 \varphi ^3-462 \varphi ^2+11 \varphi }{12 \Delta^2}~.
\end{split}
\end{align}
With these, it is easy to compute the matrix $\mtW^{-1}(\varphi)$, which we refrain from giving here due to its size. However, by computing the matrix, its denominator can be identified as
\begin{align} \notag
\cW \= (10\varphi+1)(18\varphi-1)~.
\end{align}
\subsubsection*{The matrix $\mtU_p(\varphi)$ and the zeta function}
\vskip-5pt
With the information above, it is straightforward to construct the matrices $\mtE(\varphi)$ and $\mtE(\varphi)^{-1}$. The $S_5$ symmetry of the manifolds we are studying corresponds to permutations of the coordinates $\varphi^i \to \varphi^{\varsigma(i)}$, $\varsigma \in S_5$, and acts on the periods as
\begin{align} \notag
\varpi_0 \mapsto \varpi_0~, \qquad \varpi_i \mapsto \varpi_{\varsigma(i)}~, \qquad \varpi^i \mapsto \varpi^{\varsigma(i)}~, \qquad \varpi^0 \mapsto \varpi^0~,
\end{align}
but keeps the period vector invariant, implying that there are only four independent periods. The choice \eqref{eq:Y^{ijk}} of the quantities $\wh Y^{ijk}$ guarantees a similar symmetry property for the derivative vector $\vartheta$ under the action of the permutations $\varsigma$:
\begin{align} \notag
\vartheta_0 \mapsto \vartheta_0~, \qquad \vartheta_i \mapsto \vartheta_{\varsigma(i)}~, \qquad \vartheta^i \mapsto \vartheta^{\varsigma(i)}~, \qquad \vartheta^0 \mapsto \vartheta^0~.
\end{align}
As a consequence, the matrices $\mtE(\varphi)$ and $\mtE^{-1}(\varphi)$ have a corresponding symmetry property: applying simultaneously the same permutation $\varsigma \in S_5$ to the columns and rows $2,\dots,6$ and $7,\dots,11$ keeps the matrices invariant. Due to these symmetries, $\mtU_p(0)$ has the~form
\begin{align} \notag
\mtU_p(0) = u_p \Lambda_p \left(\text{I} + \alpha_p \, \epsilon_i + 12 \alpha_p^2 \, \mu^i + \chi(\wt X_{\bm t}) \zeta_p(3) \, \eta \right)~,
\end{align}
which can be verified, at least to the $p$-adic accuracy we are working to, by starting with the most general form \eqref{eq:U(0)_general} of $\mtU_p(0)$, and imposing the requirement that the coefficients in the series appearing in $\mtU_p(\varphi)$ are $p$-adic integers. Further, we can verify that $\alpha_p = 0$, to the given accuracy. 

With this form of $\mtU_p(0)$, the matrix $\mtU_p(\varphi)$ has also the same $S_5$ symmetry as the matrices $\mtE(\varphi)$ and $\mtE^{-1}(\varphi)$. This symmetry implies that its characteristic polynomials factorises over integers~as
\begin{align} \label{eq:HV_R10_Factorisation}
\det(\mtI- \mtU_p(\varphi)T) \= R_2(T)^4 R_4(T)~,
\end{align}
where $R_2(T)$ is a quadratic and $R_4(T)$ a quartic polynomial. Moreover, from the Weil conjectures it follows that these polynomials take the form
\begin{align} \notag
R_2(T) \= 1 + a_1 p T + p^3 T^2~, \qquad R_4(T) \= 1 + b_1 T + b_2 p T^2 + b_1 p^3 T^3 + p^6 T^4~,
\end{align}
leaving just three undertermined coefficients. It turns out that in all cases we have studied the polynomial $R_4(T)$ is exactly the numerator of the zeta function of the $\IZ_5$ quotient of the manifolds on the completely symmetric line. This was studied in \cite{Candelas:2019llw}, for example.
\begin{table}
\begin{center}
\begin{tabular}{|c|c|c|} \hline
{\vrule height 12pt depth7pt width 0pt}$p$ & $\varphi$ & $R_p(\HV_{(\varphi,\dots,\varphi)},T)$ \\ \hline \hline
\strHV 7 & 3  & $\left(p^3 T^2-2 p T+1\right)^4\left(p^6 T^4+2 p^3 T^3-54 p T^2+2 T+1\right)$ \\ \hline 
 \strHV 7 & 5 & $\left(p^3 T^2+4 p T+1\right)^4\left(p^3 T^2-34 T+1\right) \left(p^3 T^2+4 p T+1\right)$ \\ \hline
 \strHV 11 & 10 & $\left(p^3 T^2+1\right)^4\left(p^6 T^4-22 p^3 T^3+2 p T^2-22 T+1\right)$ \\ \hline
\strHV 13 & 4 & $\left(p^3 T^2-2 p T+1\right)^4\left(p^3 T^2+42 T+1\right) \left(p^3 T^2-2 p T+1\right)$ \\ \hline
 \strHV 13 & 11 & $\left(p^3 T^2+4 p T+1\right)^4\left(p^3 T^2-18 T+1\right) \left(p^3 T^2+4 p T+1\right)$ \\ \hline
 \strHV 17 & 14 & $\left(p^3 T^2+1\right)^4\left(p^3 T^2+1\right) \left(p^3 T^2-134 T+1\right)$ \\ \hline             
\end{tabular}
\vskip10pt
\capt{6.1in}{tab:mirror_HV_numerators}{Some denominators of the zeta function of the mirror Hulek--Verrill manifolds with various values of $p$ and $\varphi$. More complete data for the smooth manifolds can be found in the appendices of \cite{Candelas:2023yrg}.}
\end{center}
\end{table}
The polynomials $R_4(T)$ can be found in the tables of \cite{Candelas:2021tqt}. Thus leaves us with one parameter, $a_1$, to fix. Apart from using the matrix $\mtU_p(\varphi)$, this coefficient can be found by using the relation of the coefficients of the zeta function numerator to the number of points on the manifold defined over finite fields. The number of points on Hulek--Verrill manifolds over the field $\IF_{\! p}$ was computed already by Hulek and Verrill \cite{Hulek2005}. On the non-singular manifolds this number is given by
\begin{align} \notag
N_p\left(\HV_{(\varphi^1,\varphi^2,\varphi^3,\varphi^4,\varphi^5)} \right) &\= 48p^2 + 46p + 14 + \sum_{x,y,z=1}^{p-1} \left( {\frac{\U}{p}} \right) + \rho(\varphi^1) N_p(\cE)~,
\end{align}
where
\begin{align} \notag
\Upsilon &\= \left[(1+x+y+z)\left(\frac{\varphi^2}{x}+\frac{\varphi^3}{y}+\frac{\varphi^4}{z}+\varphi^5 \right)-\varphi^1-1 \right]^2 - 4\varphi^1~,
\end{align}
and here $\left(\frac{\Upsilon}{p}\right)$ denotes the Kronecker symbol. In this context
\begin{align} \notag
\rho(\varphi^1) \= \begin{cases}
p \qquad \text{if } \varphi^1 \equiv 1 \text{ mod } p~,\\
0 \qquad \text{otherwise}~,
\end{cases}
\end{align}
and $N_p(\cE)$ denotes the number of points over $\IF_{\! p}$ of the elliptic curve
\begin{align} \notag
(x+y+z)\left( \frac{\varphi^2}{x} +\frac{\varphi^3}{y} + \frac{\varphi^4}{z} \right) \= \varphi^5~.
\end{align}
Using the relation between the series expansion of the zeta function and the number of points on the manifold, we find that the coefficients $a_1$ are given by
\begin{align} \notag
a_1 \= \frac{1}{4}\Bigg(b_1 + 1 + 45p + 45p^2 + p^3 -N_p\left(\HV_{(\varphi,\varphi,\varphi,\varphi,\varphi)} \right)\Bigg)~.
\end{align}
Thus the zeta function for the $S_5$-symmetric family of Hulek--Verrill manifolds can be computed in two ways: using the point-counting formula together with the tables in \cite{Candelas:2021tqt}, and alternatively by computing the traces of powers of $\mtU_p(\varphi)$. Comparing the results gives yet another set of intricate consistency checks, and we indeed find a perfect agreement. We give some examples of the zeta function numerators in \tref{tab:mirror_HV_numerators}.
\subsection{A last example: the mirror of a non-symmetric split of the quintic} \label{sect:non-symmetric_split_example}
\vskip-10pt
Our last example is the mirror of the ``non-symmetric'' split quintic given by the configuration matrix
\begin{align} \notag
\cicy{\IP^1\\\IP^4}{1 & 1\\4 & 1}_{\chi=-168}~.
\end{align}
From this complete intersection description, it is easy to work out the triple intersection numbers, which are given by
\begin{align} \notag
Y_{111} \= 0~, \qquad Y_{112} \= 0~, \qquad Y_{122} \= 4~, \qquad Y_{222} \= 5~,
\end{align}
as well as the fundamental period, which can be expressed as
\begin{align} \label{eq:split_quintic_fundamental_period}
\varpi^0(\bm \varphi) \= \sum_{m_1,m_2=0}^{\infty} \frac{(m_1+m_2)!\,(m_1+4m_2)!}{(m_1!)^2(m_2!)^5} \, \varphi_1^{m_1} \varphi_2^{m_j}~.
\end{align}
The other periods can be worked out from this one either by deriving the Picard--Fuchs system, or alternatively by the familiar recipe of replacing the factorials by $\Gamma$-functions, deforming by $m_i \to m_i + \epsilon_i$, and expanding in the nilpotent matrices $\epsilon_i$. The periods can then be identified  by comparing to the expansion \eqref{eq:all-period_epsilon_expansion}.

The conifold locus can be identified as
\begin{align} \notag
\Delta \= \left(1-\varphi_1\right)^5-\varphi_2 \left(512+2816 \varphi_1-320 \varphi_1^2+144 \varphi_1^3-27 \varphi_1^4\right)+65536 \varphi_2^2~,
\end{align}
and $\cY(\bm \varphi)$, which is defined in \eqref{eq:U(varphi)_denominator}, is in this case a constant
\begin{align} \notag
\cY(\bm \varphi) \= 1~.
\end{align}
\subsubsection*{Apparent singularities --- choosing the \texorpdfstring{$\wh Y^{ijk}$}{inverse triple intersection numbers}}
\vskip-5pt
We work with two different choices of the coefficients $\wh Y^{ijk}$ to illustrate how these different choices affect the computation of the zeta function. In particular, for the two choices we make, the apparent singularities will be different. This is just a reflection of the fact that different bases of sections of the vector bundle $\cH$ corresponding to different choices of the constants $\wh Y^{ijk}$ will become degenerate at different points in the moduli space. The value of the zeta function does not depend on the choice of the basis of $H^3(X)$, so both choices can be used to compute the polynomials $R_p(X,T)$.

For both bases we take
\begin{align} \notag
\wh Y^{111} \= \wh Y^{222} \= 0~, \quad \wh Y^{121} \= -\frac{5}{32}~, \quad \wh Y^{122} \= \frac{1}{8}~, \quad \wh Y^{221} \= \frac{1}{4}~, 
\end{align}
but we take the last independent coefficient to be
\begin{align} \notag
\wh Y^{112} \= \begin{cases}
\hskip11pt 0 \hskip3pt~, \qquad \text{in case 1}~,\\ 
-\frac{5}{32}~, \qquad \text{in case 2}~.
\end{cases}\end{align}
The first choice turns out to be the simpler of the two.

\subsubsection*{The matrix $\mtU_p(\varphi_1,\varphi_2)$ and the zeta function}
\vskip-5pt
In both cases considered above, the inversion of the matrix $\mtE$ proceeds in complete analogy to the two previous cases. The only significant difference between the two cases is that the denominator of $W^{-1}$ takes a slightly different form in each:
\begin{align} \notag
\cW = \begin{cases}
2 \left(1{-}\varphi_1\right) \left(32 {-} 96 \varphi_1 {-} 8192 \varphi_2 {+}20480 \varphi_1  \varphi_2 {-}529 \varphi_1^2 {-}16128 \varphi_1^2 \varphi_2 {+} 593 \varphi_1^3   \right), \hskip38pt \text{ in case 1,}\\[5pt]
16 \left(4{-}29 \varphi_1\right) \left(64 {-} 392 \varphi_1 {-} 16384 \varphi_2 {+} 51200 \varphi_1 \varphi_2 {-} 1283 \varphi_1^2  {-} 43776 \varphi_1^2 \varphi_2 {+} 1611 \varphi_1^3 \right), \text{ in case 2.}
\end{cases}
\end{align} 
Therefore, for instance, the point $(\varphi_1,\varphi_2) {=} (6,1)$ has an apparent singularity for $X/\IF_7$ in the first case, but not in the second. By contrast, the point $(4,1)$ does not have an apparent singularity in the first case, even though it does in the second case. We can use this observation to compute the zeta function at the apparent singularities.

In both cases, the coefficients $\alpha_pi$ and $\gamma_p$ are given by
\begin{align} \label{eq:alphagamma_coefficients_split_quintic}
u_p \= 1~, \quad \alpha_p^i \= 0~, \quad \gamma_p \= \chi(\wt X_{\bm  t}) \,\zeta_p(3) \; = -168 \zeta_p(3)~. 
\end{align}
These, together with the matrix $\mtE(\varphi_1,\varphi_2)$ and the denominator \eqref{eq:U(varphi)_denominator} with $\Delta$, $\cY$, and $\cW$ as above permits the computation of the zeta function. As required, the results do not depend on the choice of the coefficients $\wh Y^{ijk}$, except at the apparent singularities, where the rational matrix $\mtU_p(\varphi_1,\varphi_2)$ is computable only in one of the cases. We display some representative results for $p=7$ in \tref{tab:non-symmetric_split_numerators}.

\tablepreambleSplitQuintic{7}
\str (1,2) & smooth & p^9 T^6+5 p^6 T^5-66 p^2 T^3+5 T+1
\tabularnewline[.5pt] \hline 
\str (1,4) & apparent (case 1) & p^9 T^6+12 p^6 T^5-11 p^4 T^4-200 p^2 T^3-11 p T^2+12 T+1
\tabularnewline[.5pt] \hline
\str (1,6) & apparent (case 2) & p^9 T^6-25 p^6 T^5+29 p^4 T^4+50 p^2 T^3+29 p T^2-25 T+1
 \tabularnewline[.5pt] \hline
\str (2,2) & smooth & p^9 T^6-10 p^6 T^5+25 p^4 T^4+36 p^2 T^3+25 p T^2-10 T+1
\tabularnewline[.5pt] \hline 
\str (4,5) & smooth & p^9 T^6-8 p^6 T^5+57 p^4 T^4-64 p^2 T^3+57 p T^2-8 T+1
\tabularnewline[.5pt] \hline 
\str (4,6) & apparent (case 2) & p^9 T^6+36 p^6 T^5+113 p^4 T^4+328 p^2 T^3+113 p T^2+36 T+1
\tabularnewline[.5pt] \hline 
\str (5,5) & smooth & p^9 T^6+20 p^6 T^5+73 p^4 T^4+152 p^2 T^3+73 p T^2+20 T+1
\tabularnewline[.5pt] \hline 
\str (6,3) & smooth &\left(p^3 T^2+1\right) \left(p^6 T^4+19 p^3 T^3-p^3 T^2+67 p T^2+19 T+1\right)
\tabularnewline[.5pt] \hline 
\tablepostamble
\vskip-40pt
\begin{table}[H]
	\begin{center}
		\capt{6in}{tab:non-symmetric_split_numerators}{Some numerators of the zeta function $\zeta_{7}(X_{\varphi_1,\varphi_2},T)$ of the non-symmetric split of the quintic studied in this section. Note that we are able to compute the values of the numerators at the apparent singularities where the apparent singularity only appears for one of the two choices of $\wh Y^{ijk}$.}
	\end{center}
\end{table}

\newpage
\section{Summary and Outlook}
\vskip-10pt
In this paper, we have generalised the deformation methods of \cite{Candelas:2021tqt} to encompass multiparameter Calabi--Yau manifolds. We can express the polynomial $R_p^{(3)}(X_{\bm \varphi},T)$ (see \eqref{eq:zeta_function_form}) that determines the zeta function of a Calabi--Yau threefold as
\begin{align} \notag
R_p^{(3)}(X_{\bm \varphi},T) \= \det \left(\mtI - T \mtU_p(\bm \varphi)\right)~.
\end{align}
Our main result is an explicit expression for the matrix $\mtU_p(\bm \varphi)$ in terms of the periods and a constant matrix $\mtU(\bm 0)$. To be specific,
\begin{align} \notag
\mtU_p(\bm \varphi) \=  \wt \mtE(\bm \varphi^p)^{-1} \mtU_p(\bm 0)\,\wt \mtE(\bm \varphi)~,
\end{align}
where $\wt{\mtE}(\bm \varphi)$ is the logarithm-free period matrix in the Frobenius basis, defined in \eqref{eq:E_matrix_definition}, and the matrix $\mtU_p(\bm 0)$ is given by
\begin{align} \notag
\mtU_p(\bm 0) \=  \text{diag}(1&,p \,\bm 1,p^2 \, \bm 1,p^3) \, \me^{\alpha_p^i \epsilon_i} (\mtI + \gamma \eta)~.
\end{align}
Here the matrices $\epsilon_i$ and $\eta$ satisfy the (co-)homology algebra of the mirror $\wt{X}_{\bm t}$ 
\begin{align} \notag
\epsilon_i \epsilon_j \= \epsilon_j \epsilon_i~, \qquad \epsilon_i \epsilon_j \epsilon_k \= Y_{ijk} \, \eta~, \qquad \epsilon_i \eta \= 0~,
\end{align}
with the explicit expression for these matrices in our chosen basis given in \eqref{eq:homology_algebra_rep}. In all of the cases we have studied, the coordinates $\bm \varphi$ of the complex structure moduli space of $X_{\bm \varphi}$ can be chosen so~that
\begin{align} \notag
\alpha_p^i \= 0~, \qquad \gamma_p \= \chi(\wt X_{\bm t}) \, \zeta_p(3)~. 
\end{align}
To speed up the convergence of the series appearing as the elements of $\mtU_p(\bm \varphi)$, we note that, at least in all of the multiparameter cases we have studied, the matrix $\mtU_p(\bm \varphi) \mod p^n$ takes on the rational form
\begin{align} \notag
\mtU_p(\bm \varphi) \=  \frac{\cU_p(\bm \varphi)}{S_n(\bm \varphi^p)} \mod p^n~,
\end{align}
with \begin{align} \notag
S_n(\bm \varphi) \= \Delta(\bm \varphi)^{n-4} \, \cY(\bm \varphi)^{n-2} \, \cW(\bm \varphi)~.
\end{align} 
In this expression, $\Delta(\bm \varphi)$ gives the (hyper) conifold locus of the family of Calabi--Yau manifolds we are studying, $\cW(\bm \varphi)$ corresponds to the apparent singularities, and the factor $\cY(\bm \varphi)$ represents additional singularities that are neither of the (hyper) conifold nor large complex structure type. We wish to emphasise, however, that there are known one-parameter cases where the above form of $S_n(\bm \varphi)$ needs to be slightly generalised, and we expect this to be true of some multiparameter cases as~well.

Even though we are able to study the arithmetic properties of many Calabi--Yau threefolds with the techniques presented here, there still remain some open questions and limitations to these methods. Perhaps the most significant shortcoming of the deformation method, based on the series expansions for the periods, is that series expansions in multiple parameters become quickly cumbersome as the number of parameters increases. Although we have managed to study a particular example with five parameters by specialising to lines in the moduli space, it is not clear how to derive efficiently the required univariate series for the periods and their derivatives in general. In addition, even if one can treat any line in moduli space, using such lines to compute the zeta functions for all possible values of moduli in $\IF_p^m$, in this way, quickly becomes very time-consuming at higher $p$. Another area where further developments could prove useful is the treatment of singularities. We are still unable to compute the matrix $\mtU_p(\bm \varphi)$ for Calabi--Yau threefolds with conifold singularities if the manifold has more than two parameters. In addition to this, we do not know yet how to treat other types of singularities, such as K-points.

It would also be interesting to study the further generalisation of the techniques presented here to cover higher-dimensional Calabi--Yau manifolds (the horizontal part of the middle cohomology of one-parameter Calabi--Yau fourfolds has been studied in \cite{Jockers:2023zzi}). This might shed more light on the relation of the matrix $\mtU_p(\bm 0)$ to the (co-)homology algebra and the $\Gamma$-class of the mirror. We can also ask whether the deformation theory can be developed around other regular singularities, apart from the large complex structure points, in a natural way. While this is certainly possible in special cases, with deformations around the Fermat quintic going back to the work of Dwork, the difficulty in developing such a method for general Calabi--Yau threefolds may be in finding a sufficiently universal expression for $\mtU_p(\bm 0)$ analogous to \eqref{eq:U(0)_general}.

Having an effective numerical procedure for computing local zeta functions of Calabi--Yau threefolds opens up many exciting possibilities for further study. The techniques presented in this paper have already been used by the present authors together with J. McGovern in \cite{Candelas:2023yrg} to study supersymmetric flux vacua in IIB compactifications on Calabi--Yau threefolds, and to verify, refine and extend the flux modularity conjectures set out in \cite{Kachru:2020abh,Kachru:2020sio}. The solutions presented in \cite{Candelas:2023yrg} were obtained using the symmetry properties of the compactification manifolds. With the effective method for computing zeta functions of a wide variety of Calabi--Yau threefolds, it should be possible to find much larger families of threefolds whose Hodge structure splits analogously to that for the supersymmetric flux vacua studied in \cite{Kachru:2020abh,Kachru:2020sio,Candelas:2023yrg}, or rank-two attractor points studied for example in \cite{Candelas:2019llw,Candelas:2021mwz}. The structure of the zeta function has also a connection to other properties of the Calabi--Yau manifold. For some manifolds, for example, the zeta function is related to the existence of complex multiplication, which is itself conjecturally related to the existence of rational conformal field theories \cite{Gukov:2002nw,Kidambi:2022wvh,Okada:2022jnq}. 

In section \ref{sect:frobenius_to_rational}, we discussed briefly the relation of the $\Gamma$-class to both the rational basis of $H^3(X_{\bm \varphi},\IC)$ and an analogous construction the matrix $\mtU_p(\bm 0)$ representing the action of the inverse Frobenius map on $H^3(X_{\bm \varphi},\IQ_p)$. We find it intriguing that in these relations both the complex gamma function $\Gamma(z)$ and the $p$-adic gamma functions $\Gamma_p(z)$ appear in such an analogous fashion, albeit in slightly different contexts. This raises the question of whether it is possible to define a $p$-adic $\Gamma$-class that would explain the appearance of $\chi(\wt X_{\bm t}) \zeta_p(3)$ in both the change-of-basis matrix $\rho$ (with $p=\infty$) and $\mtU_p(\bm 0)$, and put the complex and $p$-adic computations on a similar footing. Ultimately, one would like to find an adelic formulation of this process.

The fact that the matrix $\mtU_p(\bm 0)$ has a natural expression in terms of the (co-)homology algebra of the mirror manifold and is connected to a $p$-adic analogue of the $\Gamma$-class may give some hints of possible relevance of mirror symmetry to the zeta function. In light of this, we would also like to briefly revisit the speculation originally made in \cite{Candelas:2004sk} regarding the possible role of mirror symmetry in relation to the local zeta functions. In this reference, it was speculated that it may be useful to defined a `quantum' zeta function $\zeta_p^Q(X_{\bm \varphi},T)$ that would satisfy a natural mirror symmetry property
\begin{align} \notag
\zeta_p^Q(X_{\bm \varphi},T) \= \frac{1}{\zeta_p^Q(\wt X_{\bm \psi},T)}~,
\end{align}
where $\wt X_{\bm \psi}$ denotes the mirror manifold of $X_{\bm \varphi}$ with complex structure parameter $\bm \psi$. It was noted that such a function could be obtained, for instance, by defining 
\begin{align} \notag
\zeta_p^Q(X_{\bm \varphi},\wt X_{\bm \psi},T) \= \frac{\text{numerator } \zeta_p(X_{\bm \varphi},T)}{\text{numerator } \zeta_p(\wt X_{\bm \psi},T)} \= \frac{R_p(X_{\bm \varphi},T)}{R_p(\wt X_{\bm \psi},T)} \= \frac{\det \left(\mtI - T \mtU_p(\bm \varphi)\right))}{\det \left(\mtI - T \wt \mtU_p(\bm \psi)\right)}~.
\end{align}
It is perhaps interesting to note that, in the cases where the Picard group of $X_{\bm \varphi}$ is generated by divisors\footnote{Recall, however, that the denominator of the local zeta function $\zeta_p(X_{\bm \varphi},T)$ can take on a slightly more complicated form in case the Picard group of $X_{\bm \varphi}$ is not generated by divisors defined over $\IF_p$.} defined over $\IF_p$, by formally taking the large complex structure limit and setting the complex structure parameter $\bm \psi$ corresponding to the mirror manifold to zero, and computing the characteristic polynomial of $\wt \mtU_p(\bm 0)$, one recovers the usual zeta function \eqref{eq:zeta_function_form} :
\begin{align} \notag
\zeta_p^Q(X_{\bm \varphi}, \wt X_{\bm 0},T) \= \frac{R_p(X_{\bm \varphi},T)}{(1-T)(1-pT)^{h^{11}}(1-p^2T)^{h^{11}}(1-p^3T)} \= \zeta_p(X_{\bm \varphi},T)~.
\end{align}
In this way, it may be tempting to view the polynomial $R_p(\wt X_{\bm \psi},T)$ as a `quantum-corrected' version of the denominator of $\zeta_p(X_{\bm \varphi},T)$. However, it seems that there is no natural way of taking the large complex structure limit $\bm \psi \to \bm 0$ $p$-adically, as every other point $\bm \psi$ we study has $||\Teich(\bm \psi)||_p = 1$. In addition to which it seems that, when the Picard group of $X_{\bm \varphi}$ is not generated by divisors defined over $\IF_p$, the definition of $\zeta_p^Q(X_{\bm \varphi},\wt{X}_{\bm 0},T)$ would have to be further refined. Most importantly, we still lack a compelling enumerative interpretation for the coefficients of $\zeta_p^Q(X_{\bm \varphi}, \wt X_{\bm 0},T)$. Nevertheless, it would be interesting to study whether these connections and analogues with mirror symmetry could be further developed.

\vfill
\section*{Acknowledgements}
\vskip-10pt
We wish to thank Joseph McGovern for collaboration at the early stages of this work, for many useful remarks, particularly regarding the computational implementation of the algorithms presented, and for various interesting discussions. We are also grateful to Hans Jockers and Sören Kotlewski for interesting conversations regarding the multiparameter computations. We thank also Duco van Straten for illuminating and instructive discussions related to the Frobenius map and the form of the matrix $\mtU_p(\bm 0)$. We thank the anonymous referees for their useful comments. PK is supported in part by the Cluster of Excellence Precision Physics, Fundamental Interactions, and Structure of Matter (PRISMA+,
EXC 2118/1) within the German Excellence Strategy (Project-ID 390831469). This work is partially based on the doctoral thesis of PK \cite{Kuusela:2022hga} which was funded by the Osk.~Huttunen Foundation and Jenny ja Antti Wihurin rahasto.
\newpage
\appendix
\section{A Lightning Introduction to \texorpdfstring{$p$}{p}-Adic Numbers} \label{app:p-adic_numbers}
\vskip-10pt
Here we briefly review some aspects of $p$-adic numbers that are necessary to understand the present article. For a careful discussion, we refer the interested reader to \cite{Koblitz:1609457}, and for a shorter treatment in the style of this appendix, see \cite{Candelas:2000fq}.

The construction of the field of $p$-adic numbers is analogous to the way in which the field $\IR$ of real numbers is constructed from the field of rational numbers $\IQ$. Recall that $\IR$ is essentially the topological completion of $\IQ$, i.e. to obtain $\IR$, we add to $\IQ$ all limits of Cauchy sequences in $\IQ$. In this construction, we have implicitly made a choice to use the (Archimedian) norm $|*|$ given by the absolute value. However, there are other inequivalent norms $||*||_p$ that are defined as follows: given any $r \in \IQ$ and a prime $p$, we can write $r$ uniquely as
\begin{align} \notag
r \= \frac{m}{n} p^i~,
\end{align}
where $m,n,p \in \IZ$ and $m,n,p$ are mutually prime. The $p$-adic norm $||r||_p$ is then given~by
\begin{align} \notag
||r||_p \= p^{-i}~, \qquad \text{with } ||0||_p \= 0~.
\end{align}
If $p$ is a prime, this satisfies the properties of a norm, that is
\begin{align} \notag
\begin{split}
||r||_p &\geq 0~, \qquad \text{with equality if and only if }  r = 0~, \\[5pt]
||rs||_p &= ||r||_p ||s||_p~, \\[5pt]
||r+s||_p &\leq ||r||_p + ||s||_p~.
\end{split}
\end{align}
In fact, the $p$-adic norm $||r+s||_p$ satisfies a stronger bound than that provided by the triangle inequality 
\begin{align} \label{eq:non-Archimedian_property}
||r + s||_p \leq \max \big( ||r||_p, ||s||_p \big)~.
\end{align}
Thus the norm is \textit{non-Archimedian}. This fact has important consequences for the main text. In the usual Archimedian case, if we are given numbers $x$ and $y \neq 0$ with $|x| > |y|$, then there is an integer $N$ such that 
\begin{align} \notag
|Ny| > |x|~.
\end{align}
However, for $p$-adic numbers $X$ and $Y \neq 0$ with $||X||_p > ||Y||_p$, we have $||NY||_p < ||X||_p$ for all $N \in \IZ$. In a similar way, we also have 
\begin{align} \notag
||X+NY||_p = ||X||_p \qquad \text{for all $N \in \IZ$.} 
\end{align}

Ostrowski’s theorem (see \cite{Koblitz:1609457} for details) states that any non-trivial norm is equivalent to either $|*|$ or $||*||_p$ for some prime $p$, and that these are inequivalent with each other. Thus, if we complete $\IQ$ with respect to $||*||_p$, we obtain the field $\IQ_p$ of $p$-adic numbers, which is different from $\IR$.

Analogously to the decimal expansion in $\IR$, in $\IQ_p$ every $p$-adic number $\eta$ can be represented by infinite series of the form
\begin{align} \label{eq:p-adic_representation}
\eta \= \sum_{n=n_0}^\infty a_n p^n~, \qquad \text{where } n_0 \in \IZ~, \text{ and } 0 \leq a_n \leq p-1~. 
\end{align}
Note that $||a_n p^n||_p = p^{-n}$ so the terms in the series are increasingly small in the $p$-adic norm. 

Numbers $\eta$ such that $n_0 \geq 0$, i.e. $||\eta||_p \leq 1$ are called \textit{$p$-adic integers}. The ring of $p$-adic integers is denoted\footnote{This should not be confused with the field $\IZ/p\IZ$ of integers modulo $p$.} by $\IZ_p$. A number $\eta$ such that both $\eta$ and $1/\eta$ are $p$-adic integers is a \textit{$p$-adic unit}. If $\eta$ is a unit, then necessarily $||\eta||_p = 1$, that is, $\eta$ has $n_0 = 0$, $a_0 \neq 0$. The set of $p$-adic integers
\begin{align} \notag
\IZ_p \= \{x \in \IQ_p \,\,|\,\, ||x||_p \leq 1\}
\end{align}
plays a role analogous to the unit disk. We have sometimes referred to this disk as $D$ in the main~text.

If $x \in \IF_p$, then we have the relation
\begin{align}
x^p - x \= 0~,
\end{align}
which is satisfied exactly. However, if $x \in \IZ_p$, then
\begin{align} \notag
x^p - x \= p x_1~, \qquad \text{for some $x_1 \in \IZ_p$.}
\end{align}
However, if we can choose the \textit{Teichmüller representative} of $x$,
\begin{align} \notag
\Teich(x) \defineas \lim_{n \to \infty} x^{p^n}~,
\end{align}
it can be shown that the limit above exist in $\IQ_p$. This satisfies the equation
\begin{align} \notag
\Teich(x)^p - \Teich(x) \= 0 
\end{align}
exactly. In fact, the Teichmüller representative defines a multiplicative character $\Teich: \IF_{\! p}^* \to \IQ_p$, which embeds $\IF_{\! p}^*$ as a multiplicative group of $(p-1)$'th roots of unity.\footnote{Note that, for instance, $\Teich(p) = 0$, and $\Teich(x+p) = \Teich(x)$, so the Teichmüller representative does not define a bijective correspondence $\IQ_p \to \IQ_p$.}

Since $\IQ_p$ is complete and has a norm, we have available all the processes of analysis. We can discuss limits and continuity in a manner analogous to analysis over $\IR$. We can also develop the theory of special functions. However, due to the non-Archimedian property of the $p$-adic norm, many of these concepts are somewhat different in the $p$-adic case. Of particular interest is the convergence of series. Consider the partial sums
\begin{align} \notag
S_n \= \sum_{i=0}^n \alpha_i~, \quad \alpha_i \in \IQ_p~ \quad \text{of the series} \quad S \= \sum_{i=0}^\infty \alpha_i~.
\end{align}
The sequence $\{S_n\}_{n=0}^\infty$ is Cauchy if and only if $||\alpha_i||_p \to 0$ as $i \to \infty$. The property \eqref{eq:non-Archimedian_property} guarantees that such a sequence converges in the $p$-adic norm. In particular, there is no possibility of many small terms adding up to give a large contribution to the sum. This property is used extensively in the main text to evaluate $p$-adic sums exactly to certain accuracy: if we are interested in the value of the limit $S$ to accuracy $p^{m}$, that is, we wish to compute $S_\infty \! \mod p^{m+1}$, we can ignore any terms with $||\alpha_n||_p < p^{-m}$, that is any $\alpha_n$ such that $\alpha_n \equiv 0 \! \mod p^{m+1}$. If the sum is convergent, this implies that we can evaluate $S \! \mod p^{m+1}$ as a finite sum.

This also has an interesting implication for convergence of power series. Let
\begin{align} \notag
f(\eta) \= \sum_{n=0}^\infty \alpha_n \eta^n~.
\end{align}
The function $f(\xi)$ is well-defined for values $\xi \in \IQ_p$ such that $||\alpha_n \xi^n||_p \to 0$ as $n \to \infty$. One can define the radius of convergence $r$ by
\begin{align} \notag
\frac{1}{r} \= \lim \sup ||\alpha_n||_p^{1/n}~,
\end{align}
where $\lim \sup ||\alpha_n||_p^{1/n}$ denotes the least real number $x$ such that for any $X>x$ there are only finitely many $\alpha_n$ such that $||\alpha_n||_p^{1/n}>X$. One can show that the above series is convergent if and only if $||x||_p < r$ \cite{Koblitz:1609457}. In particular, there is no notion of conditional convergence, as this condition only depends on the norm of $x$.

An important example of a function defined by power series is the $p$-adic logarithm that is defined, in analogy to the usual case, via
\begin{align} \notag
\log_p(x+1) \= \sum_{n=1}^\infty (-1)^{n+1}\frac{x^n}{n}~,
\end{align}
which converges for $||x||_p < 1$. This satisfies the usual property $\log_p(xy) = \log_p(x) + \log_p(y)$, since this follows directly from the series expansion when this converges. Requiring that this property holds for all $x,y \in \IQ_p^*$, one can define a logarithm that is defined on the whole of $\IQ^*_p$ (or even the algebraically closed field $\IC^*_p$ containing $\IQ_p^*$). To fully fix this function, one needs to fix the value of $\log_p(p)$. The \textit{Iwasawa logarithm} is obtained by choosing $\log_p(p) = 0$. To evaluate this explicitly for $p$-adic integers, one can use the following observation (see for example \cite{Cohen2007a}): For any integer $a \in \IZ_p$, we have $a^{p-1} = 1 + \cO(p)$, so defining $y = a^{p-1}-1$, we have $||y||_p<1$. Thus we can compute $\log_p(a^{p-1}) = \log_p(1+y)$ using the power series. By the usual multiplication identity, we then have $(p-1)\log_p(a) = \log_p(1+y)$. Therefore, the Iwasawa logarithm can be computed as
\begin{align} \notag
\log_p(a) \= \frac{\log_p(a^{p-1})}{p-1}~.
\end{align}
Continuity considerations allow one to define also other interesting special functions, such as the $p$-adic $\Gamma$- and $\zeta$-functions, which appear in the expression for the matrix $\mtU_p(\bm 0)$, discussed in \sref{sect:U(0)}. Here we just present their definitions, referring the reader to \cite{Candelas:2000fq,Candelas:2021tqt} for details. The $p$-adic $\Gamma$-function is obtained by $p$-adic extrapolation (which can be thought of as a $p$-adic analogue to analytic continuation) from the expression of $\Gamma_p(n)$ for non-negative integers $n$:
\begin{align} \notag
\Gamma_p(n) \= (-1)^n \prod_{\substack{k=1\\p \nmid k}}^{n-1} k~, \qquad n \in \IZ_{n \geq 0}~.
\end{align}
The $p$-adic zeta function can be likewise defined by extrapolation. There is a slight subtlety associated with the fact that the values of the zeta function $\zeta(s)$ for integers $s>0$ are believed to be transcendental, and thus they cannot be interpreted as $p$-adic numbers. However, for negative odd integers $s$, the values of $\zeta(s)$ are rational and can thus be interpolated. We define
\begin{align} \notag
\zeta_p(s) \= \frac{b_{s-1}}{s-1}~,
\end{align}
where $b_s$ are prime-dependent constants related to the Bernoulli numbers $B_n$ via
\begin{align} \notag
-\frac{1}{2k} (1-p^{2k-1}) B_{2k} \= -\frac{b_{-2k}}{2k}
\end{align}
for integers $k$. In particular, the zeta function value $\zeta_p(3)$ that we encounter in the main text is given by
\begin{align} \notag
\zeta_p(3) \= \frac{b_2}{2} \= - \frac{1}{2}\left( \Gamma_p'''(0)-\Gamma_p'(0)^3\right)~,
\end{align}
where the latter expression has been explicitly derived, for example, in \cite{Candelas:2021tqt}.
\newpage

\section{A Warm-Up Example: the Legendre Family of Elliptic Curves} \label{app:elliptic_curve}
\vskip-10pt
To get a feel for the kind of calculation we need to perform to find the zeta functions of multiparameter manifolds, we present here briefly the simplest example - that of an elliptic curve. To be specific, let us consider the Legendre family of elliptic curves $E_\lambda$ given, on an affine patch, by the~equation
\begin{align} \notag
y^2 \= x(x-1)(x-\lambda)~.
\end{align}
The canonical differential of this curve, corresponding to the (up to scaling) unique holomorphic $(1,0)$-form is given by
\begin{align} \notag
\Omega \= \frac{\dd x}{y} \= \frac{\dd x}{\sqrt{x(x-1)(x-\lambda)}}~.
\end{align}
Differentiating with respect to $\lambda$, we first get a form $\Omega' \in H^{(1,0)}(E_\lambda,\IC) \oplus H^{(0,1)}(E_\lambda,\IC)$ that together with $\Omega$ forms a basis for $H^1(E_\lambda,\IC)$, at least for almost every value of $\lambda$. Then it follows that the second derivative $\Omega''$ can be expressed in terms of $\Omega$ and $\Omega'$, at least up to an exact form. In fact a simple calculation reveals that
\begin{align} \notag
\lambda(\lambda-1) \frac{\dd^2}{\dd \lambda^2} \Omega + (2\lambda-1) \frac{\dd}{\dd \lambda} \Omega + \frac{1}{4} \Omega \= -\frac{1}{2} \dd \left(\frac{\sqrt{x(x-1)(x-\lambda)}}{(x-\lambda)^2}\right)~.
\end{align} 
Integrating over the two cycles in $H_1(E_\lambda,\IC)$, we find that the periods $\varpi_i$ satisfy the Picard--Fuchs equation, which can be written in terms of the logarithmic derivatives $\theta = \lambda \frac{\dd}{\dd \lambda}$ as
\begin{align} \notag
(\lambda-1)\theta^2 \varpi_i + \lambda \, \theta \varpi_i + \frac{\lambda}{4} \, \varpi_i \= 0~.
\end{align}
This is equivalent to the hypergeometric differential equation, and has as its solutions the elliptic integrals of the first kind $K(\lambda)$ and $K(1-\lambda)$. We wish to express the periods in the Frobenius basis, which is defined by requiring the asymptotics
\begin{align} \notag
\varpi^0(\lambda) = 1 + \cO(\lambda) \quad \text{and} \quad \varpi_0(\lambda) = \log(\lambda) \varpi^0 + \cO(\lambda)~.
\end{align}
With these conventions, the period vector corresponding to the holomorphic $(1,0)$-form can be expressed, in the Frobenius basis, as
\begin{align} \notag
\varpi \defineas \begin{pmatrix}
\varpi^0 \\
\varpi_0
\end{pmatrix} \= \begin{pmatrix}
\frac{2}{\pi} K(\lambda) \\[2pt]
-2 K(\lambda-1) + \frac{8 \log 2}{\pi} K(\lambda)
\end{pmatrix}~.
\end{align} 
Let us denote by $\{v_0,v^0\}$ the basis of $H^1(E_\lambda,\IC)$ in which $\Omega$ is given by
\begin{align} \notag
\Omega \= \varpi^0 v_0 + \varpi_0 v^0~.
\end{align}
In addition to this basis, we can take as the basis of the middle cohomology $H^1(E_\lambda,\IC)$ the span of $\Omega$ and $\theta \Omega$. The change-of-basis matrix $\mtE(\lambda)$ from the constant basis $\{ v_a, v^a \}$ of $H^1(E_\lambda,\IC)$ to the \textit{derivative basis} $\{ \Omega, \theta \Omega \}$ then takes the form
\begin{align} \notag
\mtE(\lambda) \= \begin{pmatrix}
\varpi^0 & \theta \varpi^0\\
\varpi_0 & \theta \varpi_0
\end{pmatrix}~.
\end{align}
The Picard--Fuchs equation can be written in the first-order form using this matrix as
\begin{align} \notag
\theta \mtE(\lambda) \= \mtE(\lambda)\mtB(\lambda)~, \qquad \text{where} \quad \mtB(\lambda) \= \begin{pmatrix}
0 & - \frac{\lambda}{4(\lambda-1)}\\
1 & - \frac{\lambda}{\lambda-1}
\end{pmatrix}~.
\end{align}
To find the characteristic polynomial $R(E_\lambda,T)$ defined in \eqref{eq:R(T)_determinant}, we need to find a matrix representing the action of the inverse Frobenius map $\Fr_{p}^{-1}$ \eqref{eq:Frobenius_map_Fr_definition} induced from the action $\Frob_p: \bm x \mapsto \bm x^p$ on the middle cohomology.

By considering the compatibility conditions \eqref{eq:Fr_Commutation}, it can be shown, completely analogously to \cite{Candelas:2021tqt} that the matrix $\mtU_p(\lambda)$ representing the action of $\Fr_{p}^{-1}$ in the derivative basis satisfies the following differential equation:
\begin{align} \label{eq:diff_eq_U_lambda}
\theta \mtU_p(\lambda) \= p \, \mtB(\lambda^p)^{-1} \mtU_p(\lambda) - \mtU_p(\lambda) \mtB(\lambda)~.
\end{align}
Let $\mtV_p(0)$ denote the matrix corresponding to this action in the large complex structure limit, in the constant basis. Then, it is easy to show that the matirx $\mtU(\lambda)$ is given by
\begin{align} \label{eq:U(0)V(0)_Elliptic_Curve}
\mtU_p(\lambda) \= \mtE(\lambda^p)^{-1}\mtV_p(0) \mtE(\lambda)~,
\end{align}
which is just the matrix $\mtV_p(0)$ expressed in the derivative basis in addition to which we have taken into account the fact that the Frobenius map acts on the scalar $\lambda$ as $\lambda \mapsto \lambda^p$.

The differential equation \eqref{eq:diff_eq_U_lambda} can be used to constrain the form of the matrix $\mtV_p(0)$. Taking the limit $\lambda \to 0$, we get the relation
\begin{align} \notag
p \, \mtB(0)^{-1} \mtU_p(0) - \mtU_p(0) \mtB(0) \= 0~, \qquad \text{with} \quad \mtB(0) = \begin{pmatrix}
0 & 0\\
1 & 0
\end{pmatrix}~.
\end{align}
This forces the matrix $\mtU_p(0)$ to take the lower-diagonal form
\begin{align} \notag
\mtU_p(0) \= u_p \begin{pmatrix}
1 & 0\\
\alpha_p & p
\end{pmatrix}~.
\end{align}
Plugging this form back to the equation \eqref{eq:U(0)V(0)_Elliptic_Curve}, one can show that in this case
\begin{align} \notag
\mtV_p(0) \= \mtU_p(0)~.
\end{align}
More conditions can be obtained from the fact \cite{Candelas:2021tqt} that the Frobenius map should be compatible with the symplectic product
\begin{align} \notag
(\alpha,\beta) \defineas \int_{E_\lambda} \alpha \wedge \beta~, \qquad \alpha,\beta \in H^1(E_\lambda,\IC)
\end{align}
in the sense that
\begin{align} \label{eq:Product_Compatability_Elliptic_Curve}
(\Fr_p \alpha, \Fr_p \beta) \= p \, (\alpha, \beta)~.
\end{align}
In the constant basis the matrix representing this product takes, up to an irrelevant overall constant of normalisation, the form
\begin{align} \notag
\sigma \= \begin{pmatrix}
\+0 & 1\\
-1 & 0
\end{pmatrix}~.
\end{align}
Written as a matrix equation, the condition \eqref{eq:Product_Compatability_Elliptic_Curve} becomes
\begin{align} \notag
V(0) \sigma V(0)^T \= p \sigma~,
\end{align}
which fixes $u_p =\pm 1$. 

This leaves the parameter $\alpha_p$ still free. We can fix this by appealing to the expectation (see for example \cite{Lauder2004a,Lauder2004b}) that when the coefficients of the power series in the matrix $\mtU_p(\lambda)$ are considered modulo $p^n$, the resulting matrix should be a matrix of rational functions. In practice, these rational functions have relatively low degrees, so we can test this expectation by computing the series in $\mtU_p(\lambda)$ to a high degree and seeing if we can find a polynomial $S_n(\lambda)$ such that multiplying $\mtU_p(\lambda)$ by a matrix that has the property that there exists an integer $K$ such that the coefficients of $\lambda^k$ for $k>K$ vanish $\text{mod } p^n$. 

For generic values of $\alpha_p$ it turns out that such a polynomial (of at least reasonably low degree) does not exists. However, we find that there is a unique value of $\alpha_p$ such that a polynomial $S_n(\lambda)$ can be found. In fact, we find that this polynomial is given by
\begin{align} \label{eq:U_denominator_Elliptic_Curve}
S_n(\lambda^p) \= (\lambda^p-1)^{n-2}~.
\end{align}
This vanishes exactly on the singular locus $\lambda=1$ of the Legendre family, which agrees with the expectation that the method described above works without modifications only for smooth elliptic~curves.

Let us consider, for instance the case $p=7$ and $n=7$. We find that the matrix $\mtU_p(\lambda)$, with coefficients of the series in $\lambda$ taken $\text{mod } 7^7$, becomes a rational matrix with the denominator \eqref{eq:U_denominator_Elliptic_Curve} with numerators being degree 31 polynomials in $\lambda$ if we choose 
\begin{align} \notag
\alpha_7 \= 620284 + \cO(7^7) \= \{0, 6, 2, 2, 6, 1, 5, \ldots\}~,
\end{align}
where $\cO(7^7)$ signifies that in this way we have found the value of $\alpha_7$ only modulo $7^7$. The second equality gives the $p$-adic digits of $\alpha_7$. Taking a higher $p$-adic accuracy, say $n=10$, would allow us to find a more $p$-adically accurate expression 
\begin{align} \notag
\alpha_7 \= 163681798 + \cO(7^{10}) \= \{0, 6, 2, 2, 6, 1, 5, 2, 0, 4,\ldots\}~.
\end{align}

For other values of $\alpha_7$, that is, for values that are not congruent to $620284$ modulo $7^7$, the numerator would have a degree at least $800$, which would strongly indicate that $\mtU_p(\lambda)$ is not a rational matrix. Conversely, existence of a solution for the value of $\alpha_7$ is remarkable, as there are \textit{a priori} hundreds of conditions that must be satisfied.
Specialising to a case where $\lambda = x \in \IF_{\! p}$, one might be tempted to think that to find the numerator of the zeta function $\zeta(E_x,T)$, it would be enough to evaluate the matrix $\mtU_p(\lambda)$ at $x$. However, we must take into account that in computing $\mtU_p(\lambda)$ and even defining the action of the inverse Frobenius map on the $p$-adic cohomology, we have treated $\mtU_p(\lambda)$ as a matrix with coefficients in the field of the $p$-adic numbers $\IQ_p$. Therefore, as discussed in \sref{sect:Weil_conjectures} we must use an embedding of $\IF_{\! p}$ into $\IQ_p$ given by the Teichmüller representatives $\Teich(\lambda)$ (see appendix \ref{app:p-adic_numbers}). 

To compute the numerator $R_p(E_x,T)$, we therefore substitute for $\lambda$ the $p$-adic expansion of $\Teich(x) \mod p^n$, and evaluate the matrix $\mtU_p(\Teich(x)) \mod p^n$. The characteristic polynomial of this matrix is the numerator $R_p(E_x,T)$ modulo $p^n$. It follows from the Weil conjectures that this polynomial is of the form
\begin{align} \notag
R_p(E_x,T) \= 1 - a_p T + p T^2~.
\end{align}
The only non-trivial coefficient $a$ can be expressed as a sum of the roots $a = \lambda_1 + \lambda_2$. Using the Riemann hypothesis, $|a|<2p^{1/2}$, so we know that the value of $a \! \mod p^2$ gives the exact value of~$a$. It is therefore enough to work modulo $p^2$.

\subsection{The choice of the signs $u_p$ and twists} \label{app:twists}
\vskip-10pt
For instance, if we take $\lambda = 5$ and $u_p=1$ for all $p$, we can find the values of $a_p$ for, say, the first 100 primes. The modularity theorem for elliptic curves defined over $\IQ$ implies that
these coefficients appear as Fourier coefficients of an ordinary modular form \cite{Wiles:1995ig,TaylorWiles}. In this case we find that the corresponding modular form is the one with the LMFDB \cite{LMFDB} label $\textbf{80.2.a.a}$ with its $q$-expansion is given by
\begin{align} \notag
f_{\textbf{80.2.a.a}} \= \sum_{n = 0}^\infty c_n q^n \= q + q^5 + 4 q^7 - 3 q^9 - 4 q^{11} - 2 q^{13} + 2 q^{17} - 4 q^{19} - 4 q^{23} + q^{25} + \cO(q^{29})
\end{align}
However, using the definition \eqref{eq:Zeta_function_definition} of the zeta function in terms of point counts, and counting points modulo $p$ on the Legendre curve with $\lambda = 5$ gives a zeta function corresponding to the modular form $\textbf{40.2.a.a}$ whose $q$-expansion differs from that of $\textbf{80.2.a.a}$ by a term-dependent sign:
\begin{align} \notag
    \wt f_{\textbf{40.2.a.a}} \=  \sum_{n = 0}^\infty \wt{c}_n q^n \=  q + q^5 - 4 q^7 - 3 q^9 + 4 q^{11} - 2 q^{13} + 2 q^{17} + 4 q^{19} + 4 q^{23} + q^{25} + \cO(q^{29})
\end{align}
To be specific, the difference between the coefficients of the modular forms is given by the Dirichlet character, which can be written in terms of the Kronecker symbol:
\begin{align} \label{eq:Legendre_family_twist_example_modular_form}
c_n \= \chi_{-1}(n) \wt{c}_n~, \qquad \text{with} \qquad \chi_{-1}(n)\= \left(\frac{-1}{n}\right)~.
\end{align}
In other words, we would obtain the correct zeta function by choosing the signs $u_p$ as
\begin{align} \notag
u_p \= \left(\frac{-1}{p}\right)~.
\end{align}
To understand why this is necessary, we note that the modular form $\textbf{80.2.a.a}$ which we find from the deformation method computation is related to the elliptic curve $\cE_5$ given by
\begin{align} \label{eq:Legendre_family_twist_curve}
Y^2 \= X^3-7X+6~,
\end{align}
whereas the Legendre family curve $E_5$ can be written as
\begin{align} \notag
y^2 \= x^3-7x-6~.
\end{align}
These two curves are isomorphic over $\IC$, as can be easily verified by computing their $j$-invariants or by explicitly finding the isomorphism
\begin{align} \label{eq:Legendre_family_twist_example_elliptic_curve}
Y = \ii y~, \qquad X = -x~.
\end{align}
From this we also see that the isomorphism is defined over $\IQ(\ii)$ and not over $\IQ$. Such a pair of curves are called \textit{twists} of each other.

To see why the modular forms associated to the two elliptic curves are related by \eqref{eq:Legendre_family_twist_example_elliptic_curve}, consider fixing $x \in \IF_p$. Then there exist exactly 2 points $(x,y) \in \IF_p^2$ on $E_5$ if $x^3-7x-6$ is a non-zero quadratic residue $\!\!\! \mod p$ so that its square roots exist in $\IF_p$. Corresponding to $x$ such that $x^3-7x-6=0$, there is exactly 1 point $(x,0)$ on $E_5$. Finally, if $x^3-7x-6$ is not a quadratic residue $\!\!\! \mod p$, there are no points of the form $(x,y)$ on $E_5$. This means that the number of points $N_p(E_5)$ can be expressed in terms of the Legendre symbol as
\begin{align} \notag
N_p(E_5) \= 1+ \sum_{x \in \IF_p} \left(1 + \left( \frac{x^3 -7x -6}{p} \right) \right) \= 1 + p + \sum_{x \in \IF_p}\left( \frac{x^3 -7x -6}{p} \right)~,
\end{align}
where the first term accounts for the point $(x:y:z) = (0:1:0)$ `at infinity' of the projectivised curve $E_5$. The modular form coefficients $c_p$ are related to the point counts $N_p(E_5)$ by
\begin{align} \notag
    c_p(E_5) \= p+1 - N_p(E_5)
\end{align}
so in terms of the modular form coefficients $c_p$, we have that
\begin{align} \notag
c_p \= \sum_{x \in \IF_p}\left( \frac{x^3 -7x -6}{p} \right)~.
\end{align}
Consider then the quadratic twist $\cE_5$ (see eq.~\eqref{eq:Legendre_family_twist_curve}) of $E_5$, so that the curves are related by the transformation~\eqref{eq:Legendre_family_twist_example_elliptic_curve}. The points on $\cE_5$ can be counted in a similar fashion to above, except that we must study whether 
\begin{align} \notag
   X^3-7X+6 \= -((-X)^3 -7(-X)-6) \= -(x^3 - 7x -6)
\end{align}
is a quadratic residue mod $p$ or not. This the origin of the relation between the $a_p(E_5)$ and $a_p(\cE_5)$. Using the fact that the Legendre symbol is multiplicative in the top argument, we have that
\begin{align} \notag
\left( \frac{X^3-7X+6}{p} \right) \= \left( \frac{-1}{p} \right) \left( \frac{x^3-7x-6}{p} \right)~.
\end{align}
so the above relation between which implies the relation \eqref{eq:Legendre_family_twist_example_modular_form} between the modular form coefficients $c_p$ and $\wt c_p$:
\begin{align} \notag
c_p \= \left( \frac{-1}{p} \right) \wt c_p \= \chi_{-1}(p) \wt c_p~.
\end{align}
Since we started with the periods of a family of elliptic curves defined over $\IC$, it stands to reason that with the choice $u_p =1$ for all $p$, we do not find the point counts of the particular rational elliptic curve, but rather find the point counts of a curve that is isomorphic over $\IC$. Some more examples are included in table~\eqref{tab:elliptic_curve_modular_forms}.

\begin{table}[H]
\renewcommand{\arraystretch}{1.1}
\begin{center}
\begin{tabular}{|c|c|c|c|} \hline
{\vrule height 14pt depth7pt width 0pt}$\lambda$ & $f$ & $\wt f$ & $\chi_d$\\ \hline \hline
$2$ & \textbf{32.2.a.a}  & \textbf{32.2.a.a} & $\chi_1$ \\ \hline
$3$ & \textbf{96.2.a.b}  & \textbf{96.2.a.a} & $\chi_{-1}$ \\ \hline 
$4$ & \textbf{24.2.a.a}  & \textbf{48.2.a.a} & $\chi_{-1}$ \\ \hline  
$5$ & \textbf{80.2.a.a}  & \textbf{40.2.a.a} & $\chi_{-1}$ \\ \hline  
$5/2$ & \textbf{960.2.a.f}  & \textbf{960.2.a.o} & $\chi_{-1}$ \\ \hline  
$8/5$ & \textbf{600.2.a.a}  & \textbf{1200.2.a.r} & $\chi_{-1}$ \\ \hline  
\end{tabular}
\vskip10pt
\capt{6.1in}{tab:elliptic_curve_modular_forms}{The parameter $\lambda$ together with the LMFDB label of the modular form $f$ found by using the coefficients $a_p$ obtained by computing the zeta function using the method outlined in this appendix and the modular form $\wt f$ obtained by counting points on the corresponding Legendre family curve. These two modular forms are related by a twist by the Dirichlet character given in the last column.}
\end{center}
\end{table}

\newpage

\section{Form of \texorpdfstring{$\mtU_p(0)$}{the matrix U(0)} from Commutation Relations} \label{app:U(0)}
\vskip-10pt
To see explicitly that we can write the matrix $\mtU_p(\bm 0)$ as
\begin{align} \notag
\mtU_p(\bm 0) = u_p \Lambda_p \left(\text{I} + \alpha_p^i \, \epsilon_i + \beta_i^{(p)} \, \mu^i + \left(\gamma_p + \frac{1}{3!}Y_{ijk} \alpha_p^i \alpha_p^j \alpha_p^k \right)\, \eta \right)~, \quad \Lambda_p = \text{diag}(1,p \, \bm 1,p^2 \, \bm 1,p^3)~,
\end{align}
we denote the components of $\mtU_p(\bm 0)$ by
\begin{align} \notag
\mtU_p(\bm 0) \= \begin{pmatrix}
u_0^0 & u_0^{1,i}  & u_0^{2,i} & u_0^3\\[3pt]
u_{1,i}^0 & u_{1,i}^{1,j} & u_{1,i}^{2,j} & u_{1,i}^{3}\\[3pt]
u_{2,i}^0 & u_{2,i}^{1,j} & u_{2,i}^{2,j} & u_{2,i}^{3} \\[3pt]
u_3^0 & u_{3}^{1,j} & u_{3}^{2,j} & u_{3}^{3}
\end{pmatrix}~.
\end{align}
From \eqref{eq:U0_Commutation} it is possible to obtain commutation relations of $\mtU_p(\bm 0)$ with matrices $\eta$ and~$\mu^i$: 
\begin{align} \label{eq:U0_Commutation2}
p^3 \eta \mtU_p(\bm 0) \= \mtU_p(\bm 0) \eta~, \qquad p^2 \mu^i \mtU_p(\bm 0) \= \mtU_p(\bm 0) \mu^i~.
\end{align}
The first of which is equivalent to requiring
\begin{align} \notag
u_{0}^{1,i} \= u_0^{2,i} \= u_0^3 \= 0~, \qquad u_0^3 \= u_{1,i}^3 \= u_{2,i}^3 \= 0~, \qquad p^3 u_0^0 \= u_3^3~. 
\end{align}
Substituting these conditions into the second equation in \eqref{eq:U0_Commutation2}, and contracting with $\wh Y_{ijk}$ where needed, the following conditions are found:
\begin{align} \notag
u_{1,i}^{1,j} \= p \, u_0^0 \, \delta_{i}^j ~, \quad u_{1,i}^{2,j} \= u_{1,i}^3 \= 0~, \quad u_{2,i}^{2,j} = p^2 \, u_0^0 \, \delta_i^j~, \quad u_{2,i}^3 \= 0~, \quad u_{3}^{2,i} \= 3 p^2 u_{1,i}^0~.
\end{align}
Finally, with these conditions, the original commutation relations \eqref{eq:U0_Commutation} are equivalent to 
\begin{align} \notag
u_{2,i}^{1,j} \= p u_{1,k}^0 Y_{ijk}~, \qquad u_3^{1,i} \= p u^0_{2,i}~.
\end{align}
Denoting the independent constants appearing here as
\begin{align} \notag
u_0^0 = u~, \qquad u^0_{i,1} = u p \alpha^i~, \qquad u^0_{2,i} = u p^2 \beta_i~, \qquad u_{3}^0 \= u p^3 \wh \gamma~,
\end{align}
we obtain the claimed result.

\newpage

\section{Univariate Hulek--Verrill Periods and Their Derivatives from Recurrences} \label{app:univariate_series}
\vskip-10pt
To compute the zeta functions of the five-parameter family of Hulek--Verrill manifolds, we could in principle use the five-parameter series expressions to compute the matrix $\mtE(\bm \varphi)$. However, these series are rather cumbersome which makes this method impractical for computing the series to a high~order.

As we are interested in series expressions on one-parameter lines in the moduli space, in particular on the symmetric lines $\varphi^i = \varphi$, we only need the periods and their logarithmic derivatives as series in one variable. However, having to first compute a five-variable series and then specialise to this line is a computationally a very expensive process, rendering again the computations practically impossible. Instead of having to do this, it is possible to derive recurrences for the periods and their derivatives directly as univariate series.\footnote{We thank Joseph McGovern for introducing this method to us and providing a Mathematica implementation.}

Let $\ft$ multiset of elements of $\{1,2,3,4,5\}$ (allowing duplicates) of cardinality $\leq 3$, and denote
\begin{align} \notag
\begin{split}
\theta_{\ft} &\=\prod_{i\in \ft}\theta_{i}~,\\
\theta_{\ft}\varpi^{a}\vert_{\varphi^{i} \to \varphi,\,\log \varphi \to 0}\=f_{\ft}^a~,& \qquad \theta_{\ft}\varpi_{a}\vert_{\varphi^{i} \to \varphi,\,\log \varphi \to 0}\=f_{\ft,a}~.
\end{split}
\end{align}
We seek an efficient way to compute a large number of terms in the power series $f_{\ft}^a$ and $f_{\ft,a}$. With these, we are able to compute the periods and their derivatives because the power series that multiply the logarithms in each are formed from known combinations of the above series. This problem can be broken up to computing the series expansions of the following sets:
\begin{equation} \notag
\begin{aligned}
A&\=\{f_{\ft,0},\, f_{\ft}^0\vert\text{ $\ft$ contains no repeated elements}\}~,\\[5pt]
B&\=\{f_{\ft,i}, \, f_{\ft}^i \vert\text{ $\ft$ contains no repeated elements and }i>0\}~,\\[5pt]
C&\=\{f_{\ft,a}, f_{\ft}^a \,\vert\text{ $\ft$ contains repeated elements and } a \geq 0\}~.
\end{aligned}
\end{equation}
In \cite{Verrill2004SumsOS}, Verrill gave a recursive method for computing the coefficients
\begin{align} \label{eq:HV_fundamental_coefficients}
c_n^{\bm \epsilon} \= \sum_{|\bm k| = n} \left( \frac{n!}{k_1! k_2! k_3! k_4! k_5!} \right)^2 \bm k^{\bm \delta}~,
\end{align}
where $\bm k = (k_1,\dots,k_5)$ is a five-component multi-index, and $\bm \delta$ is a five-component multi-index with $\delta_i \in \{0,1\}$. These appear as coefficients of the fundamental period $\varpi^0$ and its non-repeated logarithmic derivatives, which is the set $A$. The method in \cite{Verrill2004SumsOS} thus can be used to efficiently compute the functions in $A$.

Recurrences for the coefficients of the series in $B$ can be derived by considering the ordinary differential~equations
\begin{align} \label{eq:Picard_Fuchs_Line}
\cL^{\ft}(F) \= 0
\end{align}
satisfied by the functions $f_{\ft}^0=\theta_{\ft}\varpi^0$ on the $S_5$-symmetric line. These equations are obtained in the standard way from the recurrences for the functions in $A$ that can be computed by Verrill's~method.

About $\varphi=0$, there is a Frobenius basis of solutions to each of the equations \eqref{eq:Picard_Fuchs_Line} consisting of a single holomorphic solution, $f_{\ft}^0$, in addition to which there are solutions containing logarithms of $\varphi$. Let us denote the holomorphic part of each of these solutions by $g_{\ft,k}$, bearing in mind that the range of $k$ varies with $\bm t$. The coefficients of these functions satisfy almost the same recurrence relations as the corresponding functions in $A$, but with an inhomogeneity that can be computed in terms of functions $g_{\ft,l}$, with $l<k$. It can be shown that $f_{\ft}^a$ and $f_{\ft,a}$ are sums of functions $g_{\ft,k}$, $f_{\fr,a}$, and $f_{\fr}^a$, where $\fr \subsetneq \ft$. As an example of this principle, consider the function $f_{\{1\}}^{1}$, which has a series~expansion
\begin{equation} \notag
f_{\{1\}}^{1}\=2\sum_{n=0}^{\infty}\sum_{|\bm k| = n} \left( \frac{n!}{k_1! k_2! k_3! k_4! k_5!} \right)^2 k_{1}\left(H_{n}-H_{p_{1}}\right) \varphi^{n}~.
\end{equation}
The function $f_{\{1\}}^{0}$ has an expansion
\begin{equation} \label{eq:f10}
f_{\{1\}}^{0}\=\sum_{n=0}^{\infty}\sum_{|\bm k| = n} \left( \frac{n!}{k_1! k_2! k_3! k_4! k_5!} \right)^2 k_{1} \varphi^{n}~,
\end{equation}
and by applying the Frobenius method to this function, we can obtain a logarithmic solution to \eqref{eq:Picard_Fuchs_Line} with $\bm t=\{1\}$, which has a holomorphic part
\begin{equation} \notag
g_{\{1\},1}\=\partial_{\delta_{1}}\left(f_{\{1\}}^{0}\big\vert_{k_{i}\to k_{i}+\delta_{i},\;n\to n+\sum\delta_{i}}\right)\bigg\vert_{\delta_{i}\to0,\,\log\to0}\=f_{\{1\}}^{1}+f_{\{0\}}^{0}~.
\end{equation}
We see that it is possible compute $f_{\{1\}}^1$ in terms of two functions that we already defined recursively: $g_{\{1\},1}$, which satisfies the inhomogenous recurrence relation just derived, and $f_{\{0\}}^{0}$ which is a member of $A$. 

Recursion in $\ft$ allows for the functions in $B$ to be expressed in terms of solutions to a recurrence relation and functions in $A$, which we can already compute.

To better illustrate the recurrences for the functions $g$, consider $\varpi^{1} =  f_{\emptyset}^0 \log \varphi + f_{\emptyset}^1$. By substituting this into the differential equation \eqref{eq:Picard_Fuchs_Line} with $\ft=\emptyset$, we obtain an inhomogeneous differential equation for the functions $f_{\emptyset}^1$:
\begin{align} \notag
\cL^{\emptyset}(f_{\emptyset}^1) \= -\cL^{\emptyset}(\log \varphi \, f_{\emptyset}^0)~.
\end{align}
From this equation, we can read off an inhomogeneous recurrence relation satisfied by the coefficients $d_{n}$ appearing in the expansion
\begin{align} \notag
f_{\emptyset}^1 \= \sum_{n=1}^\infty d_n \varphi^n~.
\end{align}
The inhomogeneity involves the coefficients of $f_{\emptyset}^{0}$, which is a member of $A$ and can be computed quickly. This solves the problem for the functions in $B$.

Finally, we turn to the functions in $C$. Luckily, there are identities that give these functions in terms of functions belonging to $A$ or $B$. For example, consider the derivatives $\theta_1^2 \theta_2 \varpi^0$. From the series expansion \eqref{eq:Fundamental_Period_Series}, it follows that
\begin{align} \notag
\theta_1^2 \theta_2 \varpi^0 &\= \sum_{n=0}^\infty \sum_{|\bm{k}| = n} \left(\frac{n!}{k_1!\cdots k_5!} \right)^{2} k_1^2 k_2 \, \bm \varphi^{n}\\[5pt] 
&\= \sum_{n=0}^\infty n^2 \sum_{|\bm{k}| = n} \left(\frac{(n-1)!}{(k_1-1)! k_2! \cdots k_5!} \right)^{2} k_2 \, \varphi^n \=  \sum_{n=0}^\infty n^2 c_{n-1}^{(2)} \varphi^n~,
\end{align}
where the coefficients $c_{n-1}^{(2)}$ are those appearing in the series expansion of $\theta_2 \varpi^0$. It is possible to find an exhaustive list of such identities, one for each function in $C$, thus finishing the problem of finding a fast method of computing the periods and their derivatives on a line in the moduli space.

This problem is made simpler by working on the line with the full $S_{5}$ symmetry, where not all of the functions $f_{\ft}^a$ and $f_{\ft,a}$ are independent. For instance $f_{\{1,2,3\}}^0=f_{\{3,4,5\}}^0$ on this line. However, the method described above can be used to work on any line $\varphi^{i}=s^i \varphi$, where $s^{i}$ are constants. The main difference is that there are in full generality ten periods $\varpi^{i}$ and $\varpi_{i}$, and so 12 functions and their numerous derivatives to consider. The series expansions of all of these can be computed with the above method, but the recurrence relations and intermediate differential equations \eqref{eq:Picard_Fuchs_Line} become more complicated the less symmetry one has.

\newpage

\section{\texorpdfstring{\texttt{CY3Zeta}}{CY3Zeta}, a Mathematica Package for Computing Zeta Functions of Calabi--Yau Threefolds} \label{app:CY3Zeta}
\vskip-10pt
To make the computations using the methods developed in this paper more accessible to a wider audience, we present a \textit{Mathematica} package \texttt{CY3Zeta} which contains implementations of many algorithms described in the paper. The aim of the package is to make these as user-friendly as possible, and work for any Calabi--Yau threefold with sufficiently few complex structure parameters to make the computations feasible. As a result, the implementation provided in the package could often be slightly improved on case-by-case basis, for instance by taking into account symmetries of the manifolds in question.

\subsection{Downloading and installing}
\vskip-10pt
The package can be downloaded from \url{https://github.com/PyryKuusela/CY3Zeta}. It comes with two files. The file \mmtd{CY3Zeta.wl} contains the package itself and file \mmtd{CY3Zeta\_Examples.nb} contains instructions and examples.

To install the package, in Mathematica front end, go to menu \mmtd{File}$\rightarrow$\mmtd{Install...}. Then in the resulting dialog choose \mmtd{Package} as \mmtd{Type of Item to Install}, and as \mmtd{Source}, choose \mmtd{From File...}, navigate to the directory containing \mmtd{CY3Zeta.wl}, and open it. After this, choose either to install the package for current user or all users.

Alternatively the file \mmtd{CY3Zeta.wl} can be manually placed to the directory \mmtd{\$UserBaseDirectory}.
\subsection{Setup and options}
\vskip-10pt
The package can be loaded by using
\begin{mmaCell}{Input}
	 <<\mmaDef{CY3Zeta.wl}`
\end{mmaCell}
After this, one needs to specify some data related to the Calabi--Yau manifold $X_{\bm \varphi}$ whose zeta function is to be computed: the number of complex structure parameters of $X_{\bm \varphi}$, the order to which the Taylor series expansions should be computed, the values of the coefficients $Y_{ijk}$ and $\wh Y_{ijk}$, as well as the equations specifying the singular loci. These parameters are specified by using the following functions
\begin{optDefnN}[label=zSetNParams]{\mmtd{zSetNParams}}
	\mmtd{zSetNParams[NParams]}\\
	Sets the number of complex structure parameters for which the following computations are to be performed.\\
	\textbf{Arguments}\\
	\mmtd{NParams} is the number of complex structure parameters ($=h^{1,2}$) of the manifold $X_{\bm \varphi}$.
\end{optDefnN}
\begin{optDefnN}[label=zSetNMax]{\mmtd{zSetNMax}}
	\mmtd{zSetNMax[NMax]}\\
	Sets the maximal order to which the power series are evaluated during the following computations. \\
	\textbf{Arguments}\\
	\mmtd{NMax} is the maximal order to which the power series are to be evaluated.
\end{optDefnN}
\begin{optDefnN}[label=zSetY]{\mmtd{zSetY}}
	\mmtd{zSetY[YRules]}\\
	Sets values of the triple intersection numbers $Y_{ijk}$ of the mirror manifold of $X_{\bm \varphi}$. The $Y_{ijk}$ are assumed to be symmetric. \\
	\textbf{Arguments}\\
	\mmtd{YRules} is a list of rules (in the form \mmtd{Y[i,j,k]->yval}) giving the values of the independent triple intersection numbers $Y_{ijk}$.
\end{optDefnN}
\begin{optDefnN}[label=zSetYhat]{\mmtd{zSetYhat}}
	\mmtd{zSetYhat[YhatRules]}\\
	Sets values of the `inverse' triple intersection numbers $\wh Y_{ijk}$ (see \eqref{eq:Yhat_definition}) of the mirror manifold of $X_{\bm \varphi}$. The $\wh Y_{ijk}$ are assumed to be symmetric in the first two indices. To modify this behaviour, set \mmtd{\$zYhatSymmetryRules=\{\}}.\\
	\textbf{Arguments}\\
	\mmtd{YhatRules} is a list of rules (in the form \mmtd{Yhat[i,j,k]->yval}) giving the values of the independent $\wh Y^{ijk}$.
\end{optDefnN}
\begin{optDefnN}[label=zSetConifoldLocus]{\mmtd{zSetConifoldLocus}}
	\mmtd{zSetConifoldLocus[D]}\\
	Specifies the polynomial $\Delta$ (see \sref{sect:form_of_U(varphi)}) defining the conifold locus of the manifold $X_{\bm \varphi}$. The complex structure moduli space coordinates are \mmtd{$\phi$[1],\dots,$\phi$[NParams]}.\\
	\textbf{Arguments}\\
	\mmtd{D} is a polynomial whose vanishing locus is the conifold locus of $X_{\bm \varphi}$.
\end{optDefnN}
\begin{optDefnN}[label=zSetOtherSingularLocus]{\mmtd{zSetOtherSingularLocus}}
	\mmtd{zSetOtherSingularLocus[Y]}\\
	Specifies the polynomial $\cY$ (see \sref{sect:form_of_U(varphi)}) defining moduli space locus of the manifold $X_{\bm \varphi}$, where $X_{\bm \varphi}$ is has a singularity not of conifold or large complex structure type.\\
	\textbf{Arguments}\\
	\mmtd{Y} is a polynomial whose vanishing locus is the singular locus of $X_{\bm \varphi}$, without the conifold and large complex structure singularities.
\end{optDefnN}
For example, to study the example presented in \sref{sect:non-symmetric_split_example}, we can first specify that we are studying a two-parameter model, and that we wish to perform the computations to 200 terms in the series:
\begin{mmaCell}{Input}
	 \mmaDef{zSetNParameters}[2]
	 \mmaDef{zSetNMax}[200]
\end{mmaCell}
After this, we input the independent triple intersection numbers $Y_{ijk}$ and their inverses $\wh Y^{ijk}$.
\begin{mmaCell}{Input}
	 \mmaDef{zSetY}[\{Y[1,1,1]->0,Y[1,1,2]->0,Y[2,1,2]->4,Y[2,2,2]->5\}]
	 \mmaDef{zSetYhat}[\{Yhat[1,2,1]->-5/32,Yhat[1,2,2]->1/8,Yhat[2,2,1]->1/4,
	  Yhat[2,2,2]->0,Yhat[1,1,1]->0,Yhat[1,1,2]->0\}]
\end{mmaCell}
\begin{mmaCell}{Output}
  \{Y[1,1,1]->0,Y[1,1,2]->0,Y[2,2,1]->4,Y[2,2,2]->5\}
\end{mmaCell}
\begin{mmaCell}{Output}
	 \mmmd{\Big\{}Yhat[1,2,1]->-\mmaFrac{5}{32},Yhat[1,2,2]->1/8,Yhat[2,2,1]->1/4,
	 Yhat[2,2,2]->0,Yhat[1,1,1]->0,Yhat[1,1,2]->0\mmmd{\Big\}}
\end{mmaCell}
Finally, the conifold locus and the rest of the singular locus of $X_{\bm \varphi}$, disregarding the large complex structure singularities, are specified using \mmtd{zSetConifoldLocus} and \mmtd{zSetOtherSingularLocus}.
\begin{mmaCell}{Input}
	 \mmaDef{zSetConifoldLocus}\mmmd{\Big[}65536\mmaSup{\mmmd{\phi}[2]}{2} - \mmaSup{(\mmmd{\phi}[1]-1)}{5} -
	  \mmmd{\phi}[2](512+2816\mmmd{\phi}[1]-320\mmaSup{\mmmd{\phi}[1]}{2}+144\mmaSup{\mmmd{\phi}[1]}{3}-27\mmaSup{\mmmd{\phi}[1]}{4}) \mmmd{\Big]}
	 \mmaDef{zSetOtherSingularLocus}[1]
\end{mmaCell}
\begin{mmaCell}{Output}
	 65536\mmaSup{\mmmd{\phi}[2]}{2} - \mmaSup{(\mmmd{\phi}[1]-1)}{5} - \mmmd{\phi}[2](512+2816\mmmd{\phi}[1]-320\mmaSup{\mmmd{\phi}[1]}{2}+144\mmaSup{\mmmd{\phi}[1]}{3}-27\mmaSup{\mmmd{\phi}[1]}{4})
\end{mmaCell}
\begin{mmaCell}{Output}
	 1
\end{mmaCell}
\subsection{Period coefficients}
\vskip-10pt
After specifying the data related to the Calabi--Yau manifold, one needs to input the coefficients $c_{a_1,\dots,a_m}$ that specify the functions $f, f^i, \wt f_i$ and $\wt f$ (defined in \eqref{eq:period_vector_Frob_basis_expansion} and \eqref{eq:tilded_f_definition}) via
\begin{align} \notag
f \= \sum_{a_i=0}^\infty c_{a_1,\dots,a_m} \varphi_1^{a_1} \dots \varphi_m^{a_m}~, \qquad f^i \= \sum_{a_i=0}^\infty c^i_{a_1,\dots,a_m} \varphi_1^{a_1} \dots \varphi_m^{a_m}~,
\end{align}
and similar relations for $\wt f^i$ and $\wt f$. The coefficients $c_{a_1,\dots,a_m}$ corresponding to $f$ are denoted by \mmtd{zc[0,\{\},\{a1,\dots,am\}]}, the coefficients of $f^i$ are denoted by \mmtd{zc[1,\{i\},\{a1,\dots,am\}]}, whereas the coefficients of $\wt f^i$ and $\wt f$ are denoted by \mmtd{zc[2,\{i\},\{a1,\dots,am\}]} and \mmtd{zc[3,\{\},\{a1,\dots,am\}]}, respectively.
\begin{objDefnN}[label=zc]{\mmtd{zc}}
	\mmtd{zc[s,\{\},\{a1,\dots,am\}]}\\
	Gives the coefficient of $\varphi_1^{a_1}\dots\varphi_m^{a_m}$ in $f$ or $\wt f$, depending on whether \mmtd{s} equals 0 or 3.\\
    \mmtd{zc[s,\{i\},\{a1,\dots,am\}]}\\
	Gives the coefficient of $\varphi_1^{a_1}\dots\varphi_m^{a_m}$ in $f^i$ or $\wt f^i$, depending on whether \mmtd{s} equals 1 or 2.
\end{objDefnN}
The periods can be derived from the fundamental period using the expansion \eqref{eq:all-period_epsilon_expansion}. This can be done automatically by using the function \mmtd{zPeriodsFromFundamental}.
\begin{funcDefnN}[label=zPeriodsFromFundamental]{\mmtd{zPeriodsFromFundamental}}
	\mmtd{zPeriodsFromFundamental[fundPeriodCoeff,\{k1,\dots,km\}]}\\
	Computes the periods $\varpi^i, \varpi_i$ and $\varpi_0$ from the fundamental period coefficients using the expansion \eqref{eq:all-period_epsilon_expansion}.\\
	\textbf{Arguments}\\
	\mmtd{fundPeriodCoeff} is the coefficient of $\varphi_1^{k_1}\dots \varphi_m^{k_m}$ of the fundamental period $\varpi^0$.\\
    \mmtd{k1, \dots, km} are the variables $k_1,\dots,k_m$ that give the powers of $\varphi_i$ corresponding to the coefficient \mmtd{fundPeriodCoeff}.
\end{funcDefnN}
For instance, recalling that the periods of the split quintic studied in section \ref{sect:non-symmetric_split_example} are given by \eqref{eq:split_quintic_fundamental_period}, the periods can be specified using this function by calling
\begin{mmaCell}{Input}
  \mmaDef{zPeriodsFromFundamental} \!\!\mmmd{\Bigg[}\mmaFrac{(m+n)! (m+4n)!}{\mmaSup{(m)!}{2} \mmaSup{(n)!}{5}},\{m,n\} \!\!\mmmd{\Bigg]}
\end{mmaCell}
However, this uses the in-built Mathematica functions to simplify the derived expressions for the coefficients, and as such can be very slow. In practice, it is often more advisable to simplify the expressions by hand, and use these for faster evaluation times, although usually the recurrence relations analogous to \eqref{eq:c^0_recurrence} are the fastest way of obtaining the coefficients. For instance, to define the coefficients \mmtd{zc[0,{m,n}]} giving the coefficient of $\varphi_1^m \varphi_2^n$ in the fundamental period, we can define
\begin{mmaCell}{Input}
	 \mmaDef{zc}[0,\{\},\{\mmaPat{m\_},\mmaPat{n\_}\}] := \mmaDef{zc}[0,\{\},\{\mmaPat{m},\mmaPat{n}\}] = \mmaFrac{(\mmaPat{m}+\mmaPat{n})! (\mmaPat{m}+4\mmaPat{n})!}{\mmaSup{(\mmaPat{m})!}{2} \mmaSup{(\mmaPat{n})!}{5}}
\end{mmaCell}
The period $\varpi_1$ can similarly be specified as
\begin{mmaCell}{Input}
	 \mmaDef{zc}[1,\{1\},\{\mmaPat{m\_},\mmaPat{n\_}\}] := \mmaDef{zc}[1,\{1\},\{\mmaPat{m},\mmaPat{n}\}] = \mmaFrac{(\mmaPat{m}+\mmaPat{n})! (\mmaPat{m}+4\mmaPat{n})!}{\mmaSup{(\mmaPat{m})!}{2} \mmaSup{(\mmaPat{n})!}{5}} 
  
  (\mmaDef{HarmonicNumber}[\mmaPat{m}+\mmaPat{n}]+\mmaDef{HarmonicNumber}[\mmaPat{m}+4\mmaPat{n}]-2\mmaDef{HarmonicNumber}[\mmaPat{m}])
\end{mmaCell}
After specifying the coefficients \mmtd{zc}, the logarithm-free period vectors $\wt{\vartheta_a \varpi}$, and $\wt{\vartheta^a \varpi}$ defined in \eqref{eq:logarithm_free_period_vectors} can be accessed from the variables \mmtd{$\omega$t[]}, \mmtd{$\theta\omega$t[i]}, \mmtd{$\theta$2$\omega$t[i]}, and \mmtd{$\theta$3$\omega$t[]}, where \mmtd{i} ranges from \mmtd{1} to \mmtd{NParams}.
\begin{objDefnN}[label=zPeriods]{\mmtd{$\omega$t,} \mmtd{$\theta\omega$t,} \mmtd{$\theta$2$\omega$t,} \mmtd{$\theta$3$\omega$t}}
	\mmtd{$\omega$t[]}\\
	Gives the logarithm-free period vector $\wt{\vartheta_0\varpi}(\bm \varphi)$.\\
	\mmtd{$\theta\omega$t[i]}\\
	Gives the logarithm-free period vector $\wt{\vartheta_i\varpi}(\bm \varphi)$ involving the first derivatives.\\
	\mmtd{$\theta$2$\omega$t[i]}\\
	Gives the logarithm-free period vector $\wt{\vartheta^i\varpi}(\bm \varphi)$ involving the second derivatives.\\
	\mmtd{$\theta$3$\omega$t[]}\\
	Gives the logarithm-free period vector $\wt{\vartheta^0\varpi}(\bm \varphi)$ involving the third derivatives..	
\end{objDefnN}
The periods are given as series in \mmtd{$\phi$[i]} and \mmtd{$\lambda$}, where \mmtd{$\lambda$} keeps track of the overall degree. For example, the fundamental period is given, to the second order, by
\begin{mmaCell}{Input}
	 \mmmd{\omega}\mmaDef{t}[][[1]]+\mmaDef{O}\mmaSup{[\mmm{\lambda}]}{3}
\end{mmaCell}
\begin{mmaCell}{Output}
	 1+(\mmaDef{\(\phi\)}[1]+24\mmaDef{\(\phi\)}[2])\mmaDef{\(\lambda\)} + \mmmd{\big(}\mmaSup{\mmmd{\phi}[1]}{2}+240\mmmd{\phi}[1]\mmmd{\phi}[2]+2520\mmaSup{\mmmd{\phi}[2]}{2}\mmmd{\big)}\mmaSup{\mmmd{\lambda}}{2} + \mmaDef{O}\mmaSup{[\mmmd{\lambda}]}{3}
\end{mmaCell}
\subsection{Reading/writing to/from a file}
\vskip-10pt
As the evaluation of periods as series in multiple variables tends to cause a significant bottleneck in the method presented in this paper, \mmtd{CY3Zeta} includes simple functionality for saving the period expansions as plain text files and reading the expressions from the files. First a directory for saving the text files must be created. After that, to tell \mmtd{CY3Zeta} to use that folder for saving/loading, one can use \mmtd{zSetDirectory}.
\begin{optDefnN}[label=zSetDirectory]{\mmtd{zSetDirectory}}
	\mmtd{zSetDirectory[path]}\\
	Specifies the directory where the text files containing the period coefficients are stored.\\
	\textbf{Arguments}\\
	\mmtd{path} is a string containing the path of the directory relative to the \mmtd{Directory[]}.
\end{optDefnN}
Then the logarithm-free period vectors \mmtd{$\omega$t[]}, \mmtd{$\theta\omega$t[i]}, \mmtd{$\theta$2$\omega$t[i]}, and \mmtd{$\theta$3$\omega$t[]} can be saved into their corresponding \mmtd{.txt} files using the following function.
\begin{funcDefnN}[label=zSavePeriodsToFile]{\mmtd{zPeriodsToFile}}
	\mmtd{zPeriodsToFile[]}\\
	Saves the logarithm-free period vectors stored into the variables \mmtd{$\omega$t[]}, \mmtd{$\theta\omega$t[i]}, \mmtd{$\theta$2$\omega$t[i]}, and \mmtd{$\theta$3$\omega$t[]} to \mmtd{.txt} files named \mmtd{wtilde\_Coeffs}, \mmtd{thwtilde\_Coeffs\_i}, \mmtd{th2wtilde\_Coeffs\_i}, and \mmtd{th3wtilde\_Coeffs}, where \mmtd{i} ranges from \mmtd{1} to \mmtd{NParams}. The files are located in the directory specified by \mmtd{zSetDirectory}. \\
	\textbf{Arguments}\\
	None.
\end{funcDefnN}
The saved expressions can be read from the files and stored into the logarithm-free period vectors \mmtd{$\omega$t[]}, \mmtd{$\theta\omega$t[i]}, \mmtd{$\theta$2$\omega$t[i]}, and \mmtd{$\theta$3$\omega$t[]} by using the function \mmtd{zPeriodsFromFile}.
\begin{funcDefnN}[label=zPeriodsFromFile]{\mmtd{zPeriodsFromFile}}
	\mmtd{zPeriodsFromFile[]}\\
	Reads the series expressions for the logarithm-free periods from \mmtd{.txt} files to which they have been saved and stores the expressions to the variables \mmtd{$\omega$t[]}$,\dots,$\mmtd{$\theta$3$\omega$t[]} representing these vectors. \\
	\textbf{Arguments}\\
	None.
\end{funcDefnN}
We can save the period computed above (and the other periods) to their corresponding files by first specifying a directory. In this case we use \mmtd{Directory[]/Split\_Quintic}.
\begin{mmaCell}{Input}
	 \mmaDef{zSetDirectory}["Split\_Quintic"]
	 \mmaDef{zPeriodsToFile}[]
\end{mmaCell}
After this, we can clear the definitions of the period vectors, and check that reading them from the files\footnote{The text files containing the periods to 200 terms can also be found at \url{https://github.com/PyryKuusela/CY3Zeta/releases}} gives the same result as the expression given above.
\begin{mmaCell}{Input}
	\mmaDef{Clear}[\mmmd{\omega}\mmaDef{t},\mmmd{\theta\omega}\mmaDef{t},\mmmd{\theta}2\mmmd{\omega}\mmaDef{t},\mmmd{\theta}3\mmmd{\omega}\mmaDef{t}]
	 \mmaDef{zPeriodsFromFile}[]
	[\mmmd{\omega}\mmaDef{t}[]+\mmaDef{0}\mmaSup{[\mmm{\lambda}]}{3}
\end{mmaCell}
\begin{mmaCell}{Output}
	 1+(\mmaDef{\(\phi\)}[1]+24\mmaDef{\(\phi\)}[2])\mmaDef{\(\lambda\)} + \mmmd{\big(}\mmaSup{\mmmd{\phi}[1]}{2}+240\mmmd{\phi}[1]\mmmd{\phi}[2]+2520\mmaSup{\mmmd{\phi}[2]}{2}\mmmd{\big)}\mmaSup{\mmmd{\lambda}}{2} + \mmaDef{O}\mmaSup{[\mmmd{\lambda}]}{3}
\end{mmaCell}
\subsection{The matrix \texorpdfstring{$\mtE$}{E} and its inverse}
\vskip-10pt
After the coefficients \mmtd{zc} determining the periods have been specified or the period vectors $\wt \varpi$ have been read from a file, one can compute the matrices $\wt \mtE(\bm \varphi)$ and $\wt \mtE^{-1}(\bm \varphi)$. As discussed in \sref{sect:inversion_of_E}, to compute $\wt \mtE^{-1}$, the matrix $W$ (see \eqref{eq:W_matrix_definition}) must first be found. An often convenient method for finding this is to compute the inner products $(\vartheta \Omega, \vartheta \Omega)$ as series to high enough accuracy. One can then take a generic ansatz for the denominator of the rational matrix $W$. If the periods have been computed to high enough accuracy, one should be able to solve for the denominator by requiring that the matrix $W$ be rational. This procedure is implemented as the function \mmtd{zFindW}.
\begin{funcDefnN}[label=zFindW]{\mmtd{zFindW}}
	\mmtd{zFindW[\{deg1,\dots,degm\},NMax,NumDeg]}\\
	Looks for the rational matrix $W$ by evaluating it as a series and solving for the denominator using a generic ansatz. \\
	\textbf{Arguments}\\
	\mmtd{deg1,\dots,degm} is a list giving the degree of the ansatz for the denominator of $W$ in the coordinates \mmtd{$\phi$[1],\dots,$\phi$[NParams]}. \\
	\mmtd{NMax} is number of terms to which the series used for finding the denominator should be computed.\\
	\mmtd{NumDeg} is number of terms to which the series used for finding the numerator should be computed.
\end{funcDefnN}
For example, to compute the matrix $W$ of the mirror of the non-symmetric split of the quintic of \ref{sect:non-symmetric_split_example}, we first can try to use a linear ansatz for the denominator. We expect the denominator and the numerator to be relatively simple, so we compute the series used to find the denominator to degree 50. We also expect that the numerator is of degree less than 20:
\begin{mmaCell}{Input}
	\mmaDef{\mmaDef{zFindW}}[\{1,1\},50,20]
\end{mmaCell}
However, it turns out that the degree of the ansatz (or the degree to which the series are evaluated) is too low, and we get a warning, and the output is an empty list indicating that no solution was found:
\begin{mmaCell}{Output}
	\mmaDef{No solution to the given accuracy}\\
	 \mmaDef{\{\}}
\end{mmaCell}
Increasing the degrees of the ansatz to $\deg(\varphi_1,\varphi_2)=(3,5)$ gives a solution. However, in this case the solution for the denominator is not unique, so we get a warning:
\begin{mmaCell}{Input}
	\mmaDef{\mmaDef{zFindW}}[\{5,3\},50,20]
\end{mmaCell}
\begin{mmaCell}{Output}
	\mmtd{There are free variables - a denominator of lower degree may exist.}\\
\end{mmaCell}
Although the solution found by \mmtd{zFindW} might not be of the simplest form, it is a valid solution and the matrix $W$ is stored in the variable \mmtd{zW}.
\begin{objDefnN}[label=zW]{\mmtd{zW}}
	\mmtd{zW}\\
	Gives the matrix $W$ defined in \eqref{eq:W_matrix_definition}.
\end{objDefnN}
In this case, it turns out that the simplest denominator is of degree $\deg(\varphi_1,\varphi_2)=(2,5)$, and indeed, running
\begin{mmaCell}{Input}
	\mmaDef{\mmaDef{zFindW}}[\{5,2\},50,20]
\end{mmaCell}
\begin{mmaCell}{Output}
	 
\end{mmaCell}
does not result in any warnings, indicating that the solution has been found successfully.

If the matrix $W$ is known in advance or is not found by using the function \mmtd{zFindW}, one can set it manually by using the function \mmtd{zSetW}.
\begin{funcDefnN}[label=zSetW]{\mmtd{zSetW}}
	\mmtd{zSetW[WMat]}\\
	Specifies the the matrix $W$. \\
	\textbf{Arguments}\\
	\mmtd{WMat} is the matrix $W$.
\end{funcDefnN}
After the matrix $W$ has been found, one can compute the matrices $\wt \mtE(\bm \varphi)$ and $\wt \mtE^{-1}(\bm \varphi)$. This is done by running the function \mmtd{zComputeEMatrices}.
\begin{funcDefnN}[label=zComputeEMatrices]{\mmtd{zComputeEMatrices}}
	\mmtd{zComputeEMatrices[]}\\
	Computes the matrices $\wt \mtE(\bm \varphi)$ and $\wt \mtE^{-1}(\bm \varphi)$ and stores them in internal variables. \\
	\textbf{Arguments}\\
	None.
\end{funcDefnN}
\subsection{Finding the coefficients \texorpdfstring{$\alpha^i$}{alpha} and \texorpdfstring{$\hat \gamma$}{gamma}}
\vskip-10pt
If the coefficients $\alpha^i$ and $\wh \gamma$, defining the matrix $\mtU_p(\bm 0)$ are not known, they can be solved for numerically, order by order in $p$, by requiring that the series in the matrix $S_n(\bm \varphi^p)\mtU_p(\bm \varphi)$ terminate to the specified order in $\bm \varphi$. In all of the examples we have studied, it is actually enough to check this for $S_n(\varphi^p,\dots,\varphi^p) \tr \left[\mtU_p(\varphi,\dots,\varphi) \right]$. This numerical method is implemented by \mmtd{zFindU0Constants}.
\begin{funcDefnN}[label=zFindU0Constants]{\mmtd{zFindU0Constants}}
	\mmtd{zFindU0Constants[p,acc,maxdeg]}\\
	Computes coefficients $\alpha^i$ and $\wh \gamma$ which appear in the expression \eqref{eq:U(0)_general} for $\mtU_p(\bm 0)$. \\
	\textbf{Arguments}\\
	\mmtd{p} is the prime $p$ for which the matrix $\mtU_p(\bm 0)$ is computed.\\
	\mmtd{acc} is the $p$-adic target accuracy to which the function aims to compute the constants. However, a lower accuracy solution may be returned, if higher-accuracy solution is not found.
	\mmtd{maxdeg} is the maximum degree of the series $S_n(\bm \varphi^p)\mtU_p(\bm \varphi)$ for the series to be considered terminating. If the degree of $S_n(\bm \varphi^p)\mtU_p(\bm \varphi)$ is higher than \mmtd{maxdeg} for a particular set of constants $\alpha^i$ and $\wh \gamma$, then the series is considered non-terminating and the values of $\alpha^i$ and $\wh \gamma$ are considered not to give a solution.
\end{funcDefnN}
The output of the function is a pair \mmtd{\{\{$\alpha_1$->val1,\dots,$\alpha_{\text{NParams}}$->valNparams,$\gamma$hat->val0\},acc\}}, where the first entry is a list of rules that give the values of $\alpha^i$ and $\wh \gamma$, and the second entry is the accuracy to which they have been found. Note that this accuracy can be lower than that specified as an input, if no higher accuracy solution is found, for example due to the order to which the periods have been computed. The value of \mmtd{maxdeg} is to be chosen such that vanishing of the terms in the series expansion of $S_n(\varphi^p,\dots,\varphi^p) \tr \left[\mtU_p(\varphi,\dots,\varphi) \right]$ which are of order higher than \mmtd{maxdeg} should provide enough independent equations to uniquely fix the coefficients that appear in the $p$-adic expansions of $\alpha^i$ and $\wh \gamma$ modulo~\mmtd{$p^{\text{acc}}$}.

By using this function, we can, for instance, easily verify that the coefficients are indeed in this case given by \eqref{eq:alphagamma_coefficients_split_quintic}. For primes $p=7,11$, running the function gives
\begin{mmaCell}{Input}
   \mmaDef{zComputeEMatrices}[]
   \mmaDef{zFindU0Constants}[7,6,170]
   \mmaDef{zFindU0Constants}[11,6,170]
\end{mmaCell}
\begin{mmaCell}{Output}
   \{\mmaSub{\mmmd{\alpha}}{1}->0,\mmaSub{\mmmd{\alpha}}{2}->0,\mmmd{\gamma}hat->77\}
\end{mmaCell}
\begin{mmaCell}{Output}
   \{\mmaSub{\mmmd{\alpha}}{1}->0,\mmaSub{\mmmd{\alpha}}{2}->0,\mmmd{\gamma}hat->722\}
\end{mmaCell}
One can verify that the values of $\wh \gamma$ given by \mmtd{zFindU0Constants} are indeed equal to $-168\zeta_p(3)$ to the accuracy $p^3$ so that the quantity $p^3 \wh \gamma$ appearing in $\mtU_p(0)$ agrees to the accuracy $p^6$ as expected. The $p$-adic zeta function is given numerically by the function \mmtd{pzeta3}.
\begin{funcDefnN}[label=pzeta3]{\mmtd{pzeta3}}
	\mmtd{pzeta3[p,acc]}\\
	Gives the $p$-adic zeta function $\zeta_p(3)$ to the $p$-adic accuracy $p^{\text{\mmtd{acc}}}$. \\
	\textbf{Arguments}\\
	\mmtd{p} is the prime $p$ for which the zeta function is computed.\\
    \mmtd{acc} is the $p$-adic accuracy to which the zeta function is computed.
\end{funcDefnN}
\begin{mmaCell}{Input}
   \mmaDef{Mod}[-168 pzeta3[7,3],\mmaSup{7}{3}]
   \mmaDef{Mod}[-168 pzeta3[11,3],\mmaSup{11}{3}]
\end{mmaCell}
\begin{mmaCell}{Output}
   77
   722
\end{mmaCell}
To use these constants in later computations, they must be saved in the variables \mmtd{$\alpha_i$} and \mmtd{$\gamma$hat}. 
\begin{objDefnN}[label=alphagamma]{\mmtd{$\alpha_i$,$\gamma$hat}}
	\mmtd{$\alpha_i$[p]}\\
	Stores the coefficients $\alpha_i$ appearing in the matrix $\mtU_p(\bm \varphi)$.\\
 	\mmtd{$\gamma$hat[p]}\\
	Stores the coefficient $\wh \gamma$ appearing in the matrix $\mtU_p(\bm \varphi)$.
\end{objDefnN}
\begin{mmaCell}{Input}
  \mmaSub{\mmmd{\alpha}}{1}[7]=\mmaSub{\mmmd{\alpha}}{2}[7]=0;    
  \mmmd{\gamma}\mmaDef{hat}[7]=77;
  \mmaSub{\mmmd{\alpha}}{1}[11]=\mmaSub{\mmmd{\alpha}}{2}[11]=0;    
  \mmmd{\gamma}\mmaDef{hat}[11]=722;  
\end{mmaCell}
\subsection{The matrix \texorpdfstring{$\mtU_p(\bm \varphi)$}{U} and the polynomials \texorpdfstring{$R_p(X_{\bm \varphi},T)$}{R}}
\vskip-10pt
Once the matrices $\mtE(\bm \varphi)$ and $\mtE^{-1}(\bm \varphi)$ have been computed and the coefficients $\alpha_i$ and $\gamma$ found, one can study the matrix $\mtU_p(\bm \varphi)$ and the polynomials $R_p(X_{\bm \varphi},T)$. The matrix $\mtU_p(\bm \varphi)$ can be obtained in three forms: as a matrix of series, as a matrix of rational functions, and as a numerical matrix, where the rational functions have been evaluated at a Teichmüller representative of a point $(\varphi_1,\dots,\varphi_m) \in \IZ^m$ in the moduli space, evaluated to a specified $p$-adic accuracy.
\begin{funcDefnN}[label=zUSeries]{\mmtd{zUSeries}}
	\mmtd{zUSeries[p]}\\
	Gives the matrix $\mtU_p(\bm \varphi)$ as a matrix of Taylor series. \\
	\textbf{Arguments}\\
	\mmtd{p} is the prime $p$ for which the matrix is computed.
\end{funcDefnN}
\begin{funcDefnN}[label=zURational]{\mmtd{zURational}}
	\mmtd{zURational[p,padicacc]}\\
	Gives the matrix $\mtU_p(\bm \varphi)$ as a matrix of rational functions in the coordinates \mmtd{$\phi$[i]}. \\
	\textbf{Arguments}\\
	\mmtd{p} is the prime $p$ for which the matrix is computed.\\
	\mmtd{padicacc} is the $p$-adic accuracy to which the matrix is computed, i.e. the coefficients in the series appearing in the matrix $\mtU_p(\bm \varphi)$ are treated$\!\!\mod p^{\text{\mmtd{padicacc}}}$ when computing the rational functions.	
\end{funcDefnN}
\begin{funcDefnN}[label=zUNumeric]{\mmtd{zUNumeric}}
	\mmtd{zUNumeric[\{$\phi$[1],\dots,$\phi$[NParams]\},p,padicacc]}\\
	Gives the matrix $\mtU_p(\bm \varphi)$ as a matrix of integers, given to the specified $p$-adic accuracy.\\
	\textbf{Arguments}\\
	\mmtd{$\phi$[1],\dots,$\phi$[NParams]} is a list of integers at whose Teichmüller representatives the matrix $\mtU_p(\bm \varphi)$ is evaluated.\\
	\mmtd{p} is the prime $p$ for which the matrix is computed.\\
	\mmtd{padicacc} is the $p$-adic accuracy to which the matrix is computed, i.e. the entries of the matrix $\mtU_p(\bm \varphi)$ are treated$\!\!\mod p^{\text{\mmtd{padicacc}}}$.
\end{funcDefnN}
One can compute the rational matrix $\mtU_7(\varphi_1,\varphi_2)$ to accuracy $\cO(7^6)$. We check that the series in the numerator of the matrix terminate, and indeed, the highest-order term is of order 104, well under 200. Note that we have above set the values for $\alpha_i$ and $\gamma$, which are needed to complete these computations.
\begin{mmaCell}{Input}
	 \mmaDef{Max}[\mmaDef{Exponent}[\mmaDef{Numerator}[\mmaDef{zURational}[7,6]]/.\mmm{\phi}[\mmaPat{i\_}]:>\mmm{\lambda},\mmm{\lambda}]]
\end{mmaCell}
\begin{mmaCell}{Output}
	 104
\end{mmaCell}
Consider then the point $(\varphi_1,\varphi_2)=(2,1)$, which corresponds to a smooth manifold $X_{\bm \varphi}/\IF_7$. We can compute the corresponding matrix $\mtU_p(2,1)$ to accuracy $p^5$.
\begin{mmaCell}{Input}
	 \mmaDef{zUNumeric}[\{2,1\},7,5]
\end{mmaCell}
\begin{mmaCell}{Output}
	 \{\{8507, 10224, 22, 13637, 8741, 7632\}, \{10206, 7693, 12208, 9275, 
	 1673, 168\}, \{8302, 1799, 16527, 9415, 2289, 14539\}, \{10731, 12544, 
	 7987, 14994, 15925, 6713\}, \{12201, 3234, 3822, 4949, 15043, 3283\},
	 \{2744, 7546, 4459, 4116, 11319, 4459\}\}
\end{mmaCell}
The data contained in these matrices can be used to compute the polynomials $R_p(X_{\bm \varphi},T)$, which can be most conveniently done with the function \mmtd{zR}.
\begin{funcDefnN}[label=zR]{\mmtd{zR}}
	\mmtd{zR[\{$\phi$[1],\dots,$\phi$[NParams]\},p,padicacc]}\\
	Gives the characteristic polynomial $R_p(X_{\bm \varphi},T)$ of the matrix $\mtU_p(\bm \varphi)$.\\
	\textbf{Arguments}\\
	\mmtd{$\phi$[1],\dots,$\phi$[NParams]} is a list of integers at whose Teichmüller representatives the matrix $\mtU_p(\bm \varphi)$ is evaluated.\\
	\mmtd{p} is the prime $p$ for which the matrix is computed.\\
	\mmtd{padiacc} is the $p$-adic accuracy to which the matrix is computed, i.e. the entries of the matrix $\mtU_p(\bm \varphi)$ are treated$\!\!\mod p^{\text{\mmtd{padicacc}}}$.
\end{funcDefnN}
The characteristic polynomial $R_p(X_{\bm \varphi},T)$ of $\mtU_7(2,1)$ is given by
\begin{mmaCell}{Input}
	 \mmaDef{zR}[\{2,1\},7,5]
\end{mmaCell}
\begin{mmaCell}{Output}
	 1 + 5T - 3234\mmaSup{T}{3} + 588245\mmaSup{T}{5} + 40353607\mmaSup{T}{6}
\end{mmaCell}
The individual coefficients of $T^i$ in $R_p(X_{\bm \varphi},T)$ can be accessed with the command \mmtd{zRCoefficient}.
\begin{funcDefnN}[label=zRCoefficient]{\mmtd{zRCoefficient}}
	\mmtd{zR[i,\{$\phi$[1],\dots,$\phi$[NParams]\},p,padicacc]}\\
	Gives the coefficient of $T^i$ in the characteristic polynomial $R_p(X_{\bm \varphi},T)$ of the matrix $\mtU_p(\bm \varphi)$.\\
	\textbf{Arguments}\\
	\mmtd{i} is a power of $T$ whose coefficient in the polynomial $R_p(X_{\bm \varphi},T)$ is to be computed.\\
	\mmtd{$\phi$[1],\dots,$\phi$[NParams]} is a list of integers at whose Teichmüller representatives the matrix $\mtU_p(\bm \varphi)$ is evaluated.\\
	\mmtd{p} prime $p$ for which the matrix is computed.\\
	\mmtd{padiacc} is the $p$-adic accuracy to which the matrix is computed, i.e. the entries of the matrix $\mtU_p(\bm \varphi)$ are treated$\!\!\mod p^{\text{\mmtd{padicacc}}}$.
\end{funcDefnN}
For instance, the coefficient of $T^3$ in the characteristic polynomial of $\mtU_7(2,1)$ can be computed as
\begin{mmaCell}{Input}
	 \mmaDef{zRCoefficient}[3,\{2,1\},7,5]
\end{mmaCell}
\begin{mmaCell}{Output}
	 -3234
\end{mmaCell}
Note that the above functions may not give correct results when the point $\bm \varphi$ corresponds to a manifold $X_{\bm \varphi}/\IF_{\! p}$ with a(n apparent) singularity. Existence and the type of the singularity can be checked with the function \mmtd{zSingularityType}.
\begin{funcDefnN}[label=zSingularityType]{\mmtd{zSingularityType}}
	\mmtd{zSingularityType[\{$\phi$[1],\dots,$\phi$[NParams]\},p]}\\
	Gives the list of singularity types of the manifold $X_{\bm \varphi}/\IF_{\! p}$ corresponding to the point $(\Teich (\varphi_1), \dots, \Teich (\varphi_m))$ in the complex structure moduli space.\\
	\textbf{Arguments}\\
	\mmtd{$\phi$[1],\dots,$\phi$[NParams]} is a list of integer coordinates in the moduli space, specifying the manifold $X_{\bm \varphi}$.\\
	\mmtd{p} is the prime $p$ giving the number of elements in the finite field $\IF_{\! p}$ over which $X_{\bm \varphi}$ in considered to be defined.
\end{funcDefnN}
In the example we have been using thus far, the point $(1,1)$ as an apparent singularity, $(1,2)$ is smooth, $(2,5)$ is a conifold, and $(5,1)$ is both an apparent and a conifold singularity.
\begin{mmaCell}{Input}
	 \mmaDef{zSingularityType}[\{1,1\},7]
	 \mmaDef{zSingularityType}[\{2,1\},7]
	 \mmaDef{zSingularityType}[\{5,2\},7]
	 \mmaDef{zSingularityType}[\{1,5\},7]	 	 	 
\end{mmaCell}
\begin{mmaCell}{Output}
	 \{\mmtd{apparent}\}
\end{mmaCell}
\begin{mmaCell}{Output}
	 \{\}
\end{mmaCell}
\begin{mmaCell}{Output}
	 \{\mmtd{conifold}\}
\end{mmaCell}
\begin{mmaCell}{Output}
	 \{\mmtd{apparent},\mmtd{conifold}\}
\end{mmaCell}

\newpage
\newpage
\bibliographystyle{JHEP}
\bibliography{MZF}

\end{document}